\documentclass[10pt]{article}
\usepackage[left=0.8in,top=0.8in,right=0.8in,bottom=0.8in]{geometry}

\usepackage{graphicx}
\usepackage{amsmath}
\usepackage{amssymb}
\usepackage{amsthm}
\usepackage{latexsym}
\usepackage{bm}
\usepackage{array,supertabular}
\usepackage{color}
\usepackage{framed}
\usepackage{setspace}
\usepackage{fancyhdr}
\usepackage{stmaryrd}
\usepackage{mathrsfs}
\usepackage{mdframed}
\usepackage{url}
\usepackage{multirow}
\usepackage{algorithm}
\usepackage{algpseudocode}
\usepackage{pifont}
\usepackage{longtable}
\usepackage{booktabs}
\usepackage{adjustbox}
\usepackage{xr-hyper} 
\usepackage[hidelinks]{hyperref}
\usepackage{cleveref}

\newcolumntype{P}[1]{>{\centering\arraybackslash}p{#1}}

\SetSymbolFont{stmry}{bold}{U}{stmry}{m}{n}

\newtheorem{proposition}{Proposition}
\newtheorem{remark}{Remark}

\numberwithin{equation}{section}

\begin{document}
\title{Half-explicit Runge-Kutta integrators for variational multiscale turbulence modeling: Toward higher-order accuracy in space and time}
\author{Yujie Sun, Chi Ding, and Ju Liu$^{\textup{*}}$\\
\textit{\small Department of Mechanics and Aerospace Engineering,}\\
\textit{\small Southern University of Science and Technology,}\\
\textit{\small 1088 Xueyuan Avenue, Shenzhen, Guangdong 518055, China}\\
$^{*}$ \small \textit{E-mail address:} liuj36@sustech.edu.cn
}
\date{}
\maketitle

\section*{Abstract}
The residual-based variational multiscale (VMS) formulation has achieved remarkable success in large-eddy simulation of turbulent flows. However, its temporal discretization has largely remained limited to second-order implicit schemes. The present work aims at advancing this direction through the introduction of Runge-Kutta (RK) schemes within the VMS framework in a mathematically consistent manner. Guided by the Rothe method, the half-explicit RK scheme is employed as its accuracy is theoretically guaranteed for index-2  differential-algebraic equations. Owing to the explicit treatment of the nonlinear term, the resulting spatial problem exhibits a structure analogous to that of the Darcy equation. Following the philosophy of the VMS analysis, a subgrid-scale model is derived without invoking linearization based on perturbation series and related assumptions. The analysis further reveals that the parameter in the subgrid model is independent of the spatial mesh size. Fourier analysis demonstrates that the Rothe method, compared with the conventional vertical method of lines, provides improved dissipation and dispersion properties and exhibits a larger stability region for convection-dominated regimes. In the Taylor-Green vortex benchmark, the proposed schemes demonstrate superior performance as a large-eddy simulation model, achieving higher fidelity in predicting the kinetic energy evolution, energy spectra, and vortex structures than the conventional VMS formulation. Simulations of the open cavity flow further show that the proposed schemes can accurately capture the periodic limit cycle caused by the supercritical Hopf bifurcation, confirming its effectiveness and fidelity for highly sensitive flow instability problems.	
\vspace{5mm}

\noindent \textbf{Keywords:}  Variational multiscale formulation, Half-explicit Runge-Kutta methods, Higher-order schemes, Large-eddy simulation,  Dispersion-dissipation analysis, Taylor-Green vortex, Limit cycle oscillations

\section{Introduction}
Over the past decades, the variational multiscale (VMS) formulation has established itself as an effective framework for the numerical modeling of the Navier-Stokes (NS) equations. Within this framework, a variety of large-eddy simulation (LES) models have been formulated. Early efforts focused on the fine-scale viscosity modeling, an approach analogous to conventional eddy-viscosity closures \cite{Hughes2001,Hughes2001a,Gravemeier2005,Oberai2016,Lederer2023,Temellini2025}. More recently, the residual-based approach was advocated \cite{Bazilevs2007,Masud2011,Liu2012,Oberai2014,Meliga2019}, which has been increasingly favored largely because it retains strong consistency for all resolved scales and thus tends to deliver higher accuracy. Improved turbulence statistical results have been reported, in comparison with classical LES models based on filtered equations \cite{Hughes2001,Gravemeier2011,Chen2022}.

Numerical schemes interact with the subgrid-scale model in practical LES calculations, and it has long been recognized that the truncation error of low-order schemes contaminates the resolved flow structures \cite{Kravchenko1997,Kokkinakis2015}. In the VMS framework, the numerical method is not just a vehicle for solving the equations, but in fact defines the model. The advent of robust higher-order, higher-continuity spline basis functions has further advanced the capability of VMS-based models in achieving higher accuracy for turbulent flow modeling \cite{Bazilevs2007,Akkerman2008,Evans2023}. Compared with conventional low-order bases, splines enjoy superior dispersion and dissipation characteristics, especially for large wavenumbers, and  their spectral-like approximation property enables the accurate resolution of a broad range of scales. Furthermore, spline technology facilitates the design of structure-preserving schemes \cite{Evans2013,Evans2023,Van2017} and offers a seamless means of handling complex geometries within the framework of isogeometric analysis. These attributes make it rather attractive for LES.

While considerable efforts have been devoted to designing spectrally accurate spatial discretizations for LES, the temporal discretization is equally critical \cite{Artiano2025,Moin1998}. Inadequate temporal resolution may compromise the spectral fidelity of the LES or even laminarize the flow \cite{Choi1994}. Historically, the VMS formulation has been predominantly discretized by fully-implicit schemes \cite{Bazilevs2007,Masud2011,Liu2012,Oberai2014,Meliga2019,Guasch2013,Colomes2015,Akkerman2008,Evans2023,Colomes2016a,Sun2024}. Among these, the generalized-$\alpha$ scheme emerges as the most popular choice, likely due to its controllable dissipation on the high-frequency modes without compromising accuracy or stability \cite{Jansen2000}. As an implicit scheme, the generalized-$\alpha$ scheme places primary emphasis on stability. This reflects the design philosophy inherited from structural dynamics \cite{Chung1993}, where robustness against spurious oscillations is paramount. In turbulence simulations, sufficient temporal accuracy is demanded to resolve the turnover time of the smallest resolved eddies \cite{Moin1998}, as otherwise vortex dynamics are smeared out and the energy cascade gets distorted. Consequently, third- and higher-order temporal schemes are increasingly favored in turbulence simulations, ensuring that the temporal accuracy keeps pace with the spatial resolution \cite{Wang2006,Goc2021,Lysenko2012,Charlette2002,Nagarajan2003}.

There is thus strong motivation to develop time integration schemes within the VMS framework that extend beyond second-order accuracy, ensuring temporal accuracy is commensurate with the spatial discretization. One such attempt is an implicit-explicit scheme that employs the backward differentiation formulas for the time derivative and the Newton-Gregory extrapolation for the nonlinear convection term \cite{Forti2015}. It is expected to achieve third-order temporal accuracy, although the authors did not verify that. Explicit Runge-Kutta (RK) schemes are particularly popular in flow problems \cite{Moin1998,Jameson1981,Pirozzoli2011}. A segregated RK scheme has been designed for the NS system, allowing implicit and explicit treatments of different terms within the RK framework \cite{Colomes2016,Colomes2017}. However, the discretization was restricted to inf-sup stable elements with a local-projection-type stabilization. Indeed, the introduction of the residual-based modeling terms inevitably alter the index-2 differential-algebraic equation (DAE) structure, thereby undermining the setting on which the conventional RK schemes rely. For general implicit ordinary differential equations (ODEs), both the available numerical scheme and related numerical analysis are rather limited \cite{Hairer1996}. Moreover, the presence of the time derivatives in the residual renders the resulting mass matrix non-symmetric. In \cite[p.~208]{Jansen2023}, the authors proposed an explicit time marching strategy by using small time steps and a lagged form to treat the time derivative in the residual. Rigorously speaking, that approach inevitably suffers from a consistency issue. The present work aims to develop a systematic higher-order time integration scheme for the residual-based VMS formulations. In particular, we favor explicit time stepping methods, because they intrinsically demand less data movement and achieve higher algorithm intensity than the implicit methods. This makes them well-suited for modern heterogeneous computing architectures \cite{Goc2021,Karam2021,Keyes2013}. This study establishes the numerical foundation for spatiotemporal high-resolution turbulence simulations. The core methodologies employed involve the Rothe method, the half-explicit Runge-Kutta (HERK) method, and VMS analysis for the Darcy problem. In what follows, we provide a concise review of these concepts.

\paragraph{The Rothe method}
Two primary strategies exist for the numerical treatment of partial differential equations, distinguished by the order in which temporal and spatial discretizations are performed. As the classical approach, the \textit{vertical method of lines} first discretizes the spatial domain, reducing the problem to a system of ODEs, which are then integrated using an appropriate time-stepping scheme. In contrast, the \textit{horizontal method of lines}, also known as the Rothe method, begins with temporal discretization, transforming the transient problem into a sequence of steady-state problems that are subsequently discretized in space at each time step. Introduced by Rothe in 1930, this approach was originally developed to show the existence of differential equation solutions \cite{Roubicek2005}. For standard problems with fixed discretizations, changing the order of discretizations does not alter the resulting discrete scheme. However, in the context of VMS formulations, the Rothe method offers distinct advantages. Each time step reduces to a steady-state problem, which provides the equations for multiscale analysis. In the one-dimensional advection-diffusion equation, the Rothe method yields nodally exact solutions regardless of the time-step size and exhibits greater robustness than the VMS formulations derived via the vertical method of lines \cite{Asensio2007,Harari2004,Harari2007,Henao2010}. This suggests that the numerical instabilities often observed in conventional semi-discrete formulations may be intrinsic to the vertical method of lines. 

By reducing each time increment to a stationary problem, the Rothe method offers a consistent setting for residual-based VMS analysis and a systematic foundation for designing subgrid-scale terms. Crucially, it removes the transient coupling that complicates RK schemes and thereby addresses the difficulty that has prevented residual-based VMS models from being integrated into the RK framework. In addition, the resulting subgrid-scale models are structurally simpler, as they are independent of the transient terms. These observations indicate that the Rothe method holds strong potential for the NS equations, enabling residual-based VMS models to be applied in a manner that is both consistent and robust.

\paragraph{The RK methods for the incompressible NS equations}
Based on the Rothe method, one may focus on the time integration of the NS system before modeling the subgrid scale. Due to the differential-algebraic nature of the incompressible NS equations, the time integration requires extra care. The linear multistep methods, in particular the backward differentiation formulas, are often favored because they retain the same order of accuracy as for ordinary differential equations \cite{Forti2015,Hairer1996,Loetstedt1986}. In contrast, RK methods enjoy strong stability, are self-starting, and can be conveniently implemented for time adaptivity. Yet, they need to be carefully designed to avoid order reduction in the DAE context. Early works often employed the projection or fractional step methods to decouple velocity and pressure, apply an RK scheme to the momentum equations, and project the velocity to the divergence-free field at the end of each time step \cite{Kim1985,Le1991,Zheng2006}. It is later recognized that such an approach compromises the formal order of accuracy, especially for the pressure. The application of implicit RK to index-2 DAEs has been analyzed \cite{Hairer1996,Sanderse2013}. With stiffly accurate schemes, the incompressibility constraint is satisfied at the end of the time step, and the methods are L-stable. Moreover, implicit RK schemes generally possess favorable stability properties and, when properly designed, can be symplectic, energy-conserving, and time reversible \cite{Sanderse2013}. Nevertheless, order reduction may still occur, and a variety of remedies have been designed to improve their accuracy \cite{Hairer1996,Cai2025}. A further limitation is that implicit RK schemes generally require solving a coupled nonlinear system involving all stages, leading to a prohibitive computational cost \cite{Southworth2022}.

A distinct line of research is the HERK schemes, where the differential equations are treated explicitly and the algebraic constraint is enforced implicitly at each stage. Originally introduced by Brasey and Hairer for general index-2 DAEs \cite{Brasey1993}, HERK has been shown to retain full order accuracy for the differential and algebraic variables, provided that the classical order conditions are supplemented with additional order conditions for the index-2 DAE structure \cite{Hairer1996,Brasey1993,Hairer2006}. Regarding the incompressible NS equations on stationary meshes, the algebraic variable enters the momentum equations through a linear term, and the constraint equation itself is linear. This particular structure ensures that the additional DAE order conditions are automatically satisfied, at least up to fourth-order schemes \cite{Sanderse2012}. The accuracy of the algebraic variable, namely the pressure, can be preserved by solving the hidden constraint equation at each time step. Alternatively, Sanderse and Koren discussed strategies that achieve higher order accuracy for pressure without solving the hidden constraint \cite{Sanderse2012}. Moreover, in high-Reynolds-number flows, the advantages of implicit methods, primarily their ability to accommodate larger time steps, are often outweighed by the computational cost of solving nonlinear systems for velocity and pressure. Recent numerical experiments demonstrate that, compared to the implicit RK and linear multistep methods, HERK exhibits superior overall efficiency for incompressible flow problems \cite{Cai2025}. Owing to their desirable accuracy and flexibility, the HERK schemes have become an attractive discretization strategy for flow problems, including turbulence simulations \cite{Lehmkuhl2019} and flows in unbounded domains using the immersed boundary method \cite{Liska2017}. Building on the HERK framework, subsequent efforts sought to further improve efficiency by employing algebraic approximation techniques for the pressure-like variable, thereby avoiding the Poisson solves at the intermediate stages \cite{Aithal2020,Aithal2023,Iwatsu2018,Karam2021,Karam2023}. Another related line of development has led to segregated RK schemes, which extends the HERK framework by allowing flexible implicit and explicit treatment of different terms in the momentum equations. This approach offers a balance between stability and efficiency while enabling higher-order accuracy for the pressure \cite{Colomes2016,Colomes2017}.

\paragraph{The VMS formulation and the Darcy problem}
Applying the Rothe method and HERK schemes leads to a steady-state problem at each RK stage. The resulting problem exhibits the form of the Darcy equations, in which the effective permeability explicitly depends on the time-step size. Owing to the saddle-point nature, the problem requires either an inf-sup stable element pair or an enhancement of the variational problem through subgrid-scale modeling. Early work in this direction introduced a stabilized formulation for the Darcy equations, accompanied by a rigorous convergence analysis \cite{Masud2002,Hughes2006}, enabling the use of equal-order interpolations. A particularly appealing feature is that the parameter in the subgrid model is constant, independent of the discretization parameter. Subsequently, a multiscale analysis was carried out for the Darcy equations, deriving the stabilized formulation through a bubble enrichment strategy \cite{Nakshatrala2006}. In parallel, the residual-based VMS analysis was performed, using both the algebraic subgrid-scale method and the orthogonal subgrid-scale method \cite{Badia2009, Badia2010}. Based on a heuristic spectral analysis, the authors derived stabilization parameters that depend on a characteristic length scale. Four distinct designs of the stabilization parameter were proposed based on the primal, dual, or intermediate variational settings, each with distinct stability and convergence properties. Notably, that analysis also permits the use of piecewise constant pressure approximations for the Darcy problem. Overall, those studies provide valuable theoretical and practical insights that inform the numerical scheme design pursued in the present work.

\subsection*{Outline}
In the remainder, we introduce the HERK schemes based on a generic DAE system and summarize their order conditions. Following the Rothe method, we apply the HERK method to the weak form of the NS equations, which are then spatially discretized within the VMS framework. Since the nonlinear terms are treated explicitly, the VMS analysis is more straightforward, relies on fewer modeling assumptions, and avoids the need for perturbation-based linearization procedures. Through bubble enrichment, a consistent and convenient fully discrete formulation is obtained. The stability and spectral characteristics of the proposed method are examined by Fourier analysis, revealing several distinctive properties relative to conventional schemes. Due to the HERK schemes, the solver in all stages can effectively reuse the block matrices through a carefully designed algorithm, leading to an efficient overall solver design. Finally, the performance of the method is assessed through a series of benchmark problems.

\section{Strong- and weak-form problems}
\label{sec:NS_equations}
We begin by formulating the strong- and weak-form problems of the incompressible NS equations. Let $\Omega \subset \mathbb R^3$ be an open bounded domain with Lipschitz boundary $\Gamma := \partial \Omega$. The time interval of interest is denoted by $(0,T) \subset \mathbb R$ with $T>0$. Given the density $\rho$, dynamic viscosity $\mu$, and the body force per unit volume $\bm f : \Omega \rightarrow \mathbb R^3$, the initial/boundary-value problem aims to determine the velocity $\bm v : \Omega \rightarrow \mathbb R^{3}$ and pressure $p : \Omega \rightarrow \mathbb R$ that satisfy
\begin{align}
\label{eq:ns_mom}
\rho \dot{\bm v} + \rho (\bm v \cdot \nabla )\bm v - \nabla \cdot \bm \sigma - \rho \bm f &= \bm 0,  && \mbox{ in } \Omega \times (0,T), \displaybreak[2] \\
\label{eq:ns_mass}
\nabla \cdot \bm v &= 0, && \mbox{ in } \Omega \times (0, T),
\end{align}
where $\bm \sigma := 2 \mu \bm{\varepsilon}(\bm v) - p \bm I$ is the Cauchy stress and $\bm \varepsilon(\bm v):= 1/2\left(\nabla \bm{v} + \nabla \bm{v}^{T} \right)$ is the rate of strain. In this work, $\bm I$ denotes the rank-two identity tensor. We also assume the density and viscosity are both positive constants. Equations \eqref{eq:ns_mom}-\eqref{eq:ns_mass} characterize the balance of the linear momentum and the incompressibility constraint. The initial condition is specified as $\bm v(\cdot, 0) = \bm v_0(\cdot)$ on $\overline{\Omega}$, with a prescribed divergence-free velocity field $\bm v_0: \overline{\Omega} \rightarrow \mathbb R^3$. We consider a non-overlapping subdivision of the boundary, that is, $\overline{\Gamma} = \overline{\Gamma_g \cup \Gamma_h}$ and $\emptyset = \Gamma_g \cap \Gamma_h$. Denote the unit outward normal vector to the boundary $\Gamma$ by $\bm n$. Given the Dirichlet data $\bm g$ and boundary traction $\bm h$, the boundary conditions are prescribed as
\begin{align}
\label{eq:dirichlet_bc}
\bm v = \bm g && \mbox{ on } \Gamma_{g} \times (0,T), \displaybreak[2] \\
\label{eq:neumann_bc}
\bm \sigma \bm n = \bm h  && \mbox{ on } \Gamma_{h} \times (0,T).
\end{align}
The condition \eqref{eq:dirichlet_bc} prescribes the velocity on the Dirichlet boundary $\Gamma_{g}$, while the condition \eqref{eq:neumann_bc} specifies the total traction on the Neumann boundary $\Gamma_{h}$. The system \eqref{eq:ns_mom}-\eqref{eq:neumann_bc} constitutes the strong-form problem for the incompressible NS equations.

To establish the variational formulation, or the weak-form problem, we introduce $(\cdot, \cdot)$ and $(\cdot, \cdot)_{\Gamma_h}$ as the $L^2$ inner products over the domain $\Omega$ and the boundary $\Gamma_h$, respectively. The space of square integrable functions is denoted as $L^2(\Omega)$, and $H^1(\Omega)$ denotes the Sobolev space of functions whose distributional derivatives up to order one are square integrable. We introduce the trial solution and test function spaces as
\begin{align*}
\mathcal S := \left\lbrace \bm v \in H^1(\Omega) : \bm v\middle|_{\Gamma_g} = \bm g \right\rbrace \quad \mbox{and} \quad
\mathcal V := \left\lbrace \bm w \in H^1(\Omega) : \bm w\middle|_{\Gamma_g} = \bm 0 \right\rbrace,
 \end{align*}
respectively. Given $\bm g \in H^{1/2}(\Gamma_g)$, the regularity of the two function spaces allows the integrals $(\cdot, \cdot)$ and $(\cdot, \cdot)_{\Gamma_h}$ to be well-defined. The variational formulation for the problem \eqref{eq:ns_mom}-\eqref{eq:neumann_bc} is then stated as follows. Find $\bm v: [0,T] \rightarrow \mathcal S$ and $p : [0,T] \rightarrow L^2(\Omega)$ such that $\forall \lbrace \bm w, q \rbrace \in \mathcal V \times L^{2}(\Omega)$ and $t \in (0,T)$,
\begin{align}
\label{eq:weak_mom}
0 &= \big(\bm w, \rho \dot{\bm v} \big) + \mathbb C(\bm w; \bm v, \bm v) - \big(\nabla \cdot \bm w, p \big) + \big( \bm {\varepsilon(\bm w)}, 2 \mu \bm {\varepsilon}(\bm {v}) \big) - \big( \bm w, \rho \bm{f} \big) - \big(\bm w, \bm {h} \big)_{\Gamma_h}, \displaybreak[2] \\
\label{eq:weak_mass}
0 &= \big( q, \nabla \cdot \bm{v} \big),
\end{align}
where $\mathbb C(\bm w; \bm v_1, \bm v_2) := \big( \bm w, \rho (\bm v_1 \cdot \nabla) \bm v_2 \big)$ is the convective trilinear form. The initial condition for the weak-form problem is given by $\big(\bm w, \bm v \big) \big|_{t=0} = \big(\bm w, \bm v_0 \big)$.

\begin{remark}
Alternative variational formulations of the incompressible NS equations can also be constructed. For example, the trilinear form $\mathbb C$ can be defined in the skew-symmetric form, which leads to better conservation property of the kinetic energy and enjoys superior de-aliasing property \cite{Kravchenko1997}. On the other hand, there is a conservative trilinear form that emphasizes the momentum conservation \cite{Evans2023}.
\end{remark}

\section{Half-explicit Runge-Kutta methods}
In this study, the Rothe method \cite{Rothe1930}, also known as the horizontal method of lines, is adopted as the guiding principle for the design of the overall numerical formulation. In previous applications of the Rothe method, the generalized trapezoidal rule was applied to transport problems \cite{Henao2010,Harari2007}. In this study, with a goal of pursuing higher-order temporal accuracy, we employ the HERK scheme for the incompressible NS equations. In the following, we first review the basic settings of the HERK method and then apply it to the weak-form problem of the NS equations. In the time stepping schemes, the time interval $(0, T)$ is partitioned into $N_{\mathrm{ts}}$ subintervals, each associated with a time-step size $\Delta t_n := t_{n+1} - t_n$.

\subsection{Conceptural framework: HERK method for an index-2 DAE system}
\label{sec:HERK-DAE}
To illustrate the basic idea of the HERK method, we begin with a prototypical ordinary differential-algebraic system
\begin{align}
\label{eq:DAE-example}
\dot{\bm y}(t) &= \mathcal F(t, \bm y(t), \bm z(t)), \displaybreak[2] \\
\label{eq:DAE-example-constraint}
\bm 0  &= \mathcal G(\bm y(t)),
\end{align}
where $\bm y(t) \in \mathbb R^n$ is the differential variable, $\bm z(t) \in \mathbb R^m$ is the algebraic variable that acts as a Lagrange multiplier for enforcing the algebraic constraint. The function $\mathcal F : \mathbb R \times \mathbb R^n \times \mathbb R^m \rightarrow \mathbb R^n$ governs the evolution of $\bm y$, while the constraint function $\mathcal G : \mathbb R^n \rightarrow \mathbb R^m$ defines the submanifold on which the dynamics evolve. Both $\mathcal F$ and $\mathcal G$ can be nonlinear in general. If $\mathcal G_{,\bm y} \mathcal F_{,\bm z}$ is non-singular in the neighborhood of the solution, the problem \eqref{eq:DAE-example}-\eqref{eq:DAE-example-constraint} is referred to as an index-2 differential-algebraic system. For such a system, the initial condition needs to satisfy $\bm 0=\mathcal G(\bm y_0)$ and $\bm 0 = \mathcal G_{,\bm y}(\bm y_0) \mathcal F(t_0, \bm y_0, \bm z_0)$, which are collectively referred to as the consistency of the initial condition. The second condition, also known as the \textit{hidden constraint} \cite{Hairer1996,Ascher1998}, enforces that the instantaneous velocity $\dot{\bm y}_0$ is tangent to the constraint manifold $\lbrace \bm y \in \mathbb R^{n} \mid \mathcal G(\bm y) = \bm 0 \rbrace$. Geometrically, this guarantees that the solution starting on the manifold remains on it throughout the evolution. The consistency of the initial condition is necessary for the local existence of a unique solution to the differential-algebraic system \eqref{eq:DAE-example}-\eqref{eq:DAE-example-constraint} \cite{Brasey1993}.

\begin{table}[h!]
	\centering
	\renewcommand{\arraystretch}{1.5}
	\begin{tabular}{c|cccc}
		$c_1$ & $a_{11}$ & $a_{12}$ & $\cdots$ & $a_{1s}$ \\
		$c_2$ & $a_{21}$ & $a_{22}$ & $\cdots$ & $a_{2s}$ \\
		$\vdots$ & $\vdots$ & $\vdots$ & $\ddots$ & $\vdots$ \\
		$c_s$ & $a_{s1}$ & $a_{s2}$ & $\cdots$ & $a_{ss}$ \\ \hline
		& $b_1$ & $b_2$ & $\cdots$ & $b_s$
	\end{tabular}
	\hspace{10mm}
	\begin{tabular}{c|cccc}
		$0$ & $0$ & $0$ & $\cdots$ & $0$ \\
		$c_2$ & $a_{21}$ & $0$ & $\cdots$ & $0$ \\
		$\vdots$ & $\vdots$ & $\vdots$ & $\ddots$ & $\vdots$ \\
		$c_s$ & $a_{s1}$ & $a_{s2}$ & $\cdots$ & $0$ \\ \hline
		& $b_1$ & $b_2$ & $\cdots$ & $b_s$
	\end{tabular}
	\caption{Butcher tableaux for a general (left) and an explicit (right) RK scheme, where $a_{ij}$ are the coefficients, and $b_i$  are the weights for the final combination. Explicit schemes are characterized by $a_{ij}=0$ for $i \leq j$.}
	\label{tab:explicit_RK_butcher}
\end{table}

In this context, the HERK method applies an explicit RK scheme to the differential equation followed by solving the constraint equation at each stage. An $s$-stage scheme is specified by a set of algorithmic coefficients: $a_{ij}$ are the stage coefficients, $b_i$ are the weights used to compute the final solution, and $c_i$ are the node values and indicate the evaluation time instances of each intermediate stage. Conventionally, the node values satisfy $c_i = \sum_{j=1}^{i-1}a_{ij}$. The coefficients are typically presented in a compact form using the Butcher tableau (see Table \ref{tab:explicit_RK_butcher}). With those, the scheme is stated as follows. Given a solution of the previous time step $\bm y_n$, the first stage solution is initialized as $\bm y_{n(1)} = \bm y_n$. For stages $i=2,\cdots, s$, the  solutions $\bm y_{n(i)}$ of the intermediate stages are determined by solving
\begin{align}
\label{eq:HERK_int_F}
\bm y_{n(i)} =& \bm y_n + \Delta t_n \sum_{j=1}^{i-1} a_{ij} \mathcal F(t_n + c_j \Delta t_n, \bm y_{n(j)}, \bm z_{n(j)}), \displaybreak[2] \\
\label{eq:HERK_int_G}
\bm 0 =& \mathcal G(\bm y_{n(i)}),
\end{align}
where the subscript $(i)$ indicates the solution at the $i$-th intermediate stage. In practice, one inserts \eqref{eq:HERK_int_F} into the constraint equations \eqref{eq:HERK_int_G} and obtains the nonlinear algebraic equations for $\bm z_{n(i)}$. For an index-2 problem, the existence of $\bm z_{n(i)}$ is guaranteed, provided $a_{i\:i-1} \neq 0$. Once $\bm z_{n(i)}$ is determined, the solution $\bm y_{n(i)}$ is obtained in an explicit manner according to \eqref{eq:HERK_int_F}. The algebraic variables $\bm z_{n(i)}$ thus play the role of enforcing the value of the solution to remain on the constraint manifold. After all $s$ intermediate stages are completed, the solution at the next time step, $\bm y_{n+1}$, is obtained by solving
\begin{align}
\label{eq:herk_final_F}
\bm y_{n+1} =& \bm y_n + \Delta t_n \sum_{i=1}^{s}b_{i} \mathcal F(t_n + c_i \Delta t_n, \bm y_{n(j)}, \bm z_{n(j)}), \displaybreak[2] \\
\label{eq:herk_final_G}
\bm 0 =& \mathcal G(\bm y_{n+1}).
\end{align}
A feature of the above algorithm is that only the constraint equations associated with $\mathcal G$ need to be solved implicitly, reducing the computational cost from $m+n$, as in fully implicit schemes, to $m$. At the final stage, one obtains $\bm y_{n+1}$ together with $\bm z_{n(s)}$. An accurate approximation for the algebraic variables at the next time step can be made by considering the hidden constraint at $t_{n+1}$, i.e.,
\begin{align}
\label{eq:herk_hidden_for_z_n_plus_1}
\mathcal G_{,\bm y}(\bm y_{n+1}) \mathcal F(t_{n+1}, \bm y_{n+1}, \bm z_{n+1}) = \bm 0.
\end{align}
The resulting solution $\bm z_{n+1}$ has the same order of accuracy as the differential variable $\bm y_{n+1}$. The main convergence property of this scheme is summarized as follows, whose proof can be found in \cite{Brasey1993}.
\begin{proposition}
If $\mathcal G_{,\bm y}\mathcal F_{,\bm z}$ is non-singular, the consistency of the initial condition is satisfied, the local error of the scheme is of order $p+1$, the coefficient $b_s \neq 0$, and $a_{i\: i-1} \neq 0$ for $i=2, \cdots, s$, the method has a convergence order of $p$, that is, $\bm y_{n+1} - \bm y(t_{n+1}) = \mathcal O(\Delta t^p)$ for $t_{n+1} - t_0 = (n+1)\Delta t \leq T$. If the algebraic variable is determined according to \eqref{eq:herk_hidden_for_z_n_plus_1}, its order of convergence is also of order $p$.
\end{proposition}

The local error of the RK schemes is determined by matching the Taylor expansions of the exact and numerical solutions, leading to the classical Butcher conditions. The order conditions can be systematically derived by employing the rooted trees \cite{Ascher1998}. For a scheme to achieve third-order accuracy for ODEs, the coefficients need to satisfy
\begin{align}
\label{eq:Butcher-order-cond-123}
\sum_{i=1}^{s} b_i = 1, \quad \sum_{i=1}^{s} b_i c_i = \frac12, \quad \sum_{i=1}^{s} b_i c_i^2 = \frac13, \quad \sum_{i,j=1}^{s} b_i a_{ij} c_j = \frac16.
\end{align}
To reach fourth-order accuracy, four additional order conditions arise as follows,
\begin{align}
\label{eq:Butcher-order-cond-4}
\sum_{i=1}^{s}b_i c_i^3 = \frac14, \quad \sum_{i,j=1}^{s}b_i a_{ij}c_j^2 = \frac{1}{12}, \quad \sum_{i,j=1}^{s} b_i c_i a_{ij}c_j = \frac18, \quad \sum_{i,j,k=1}^{s}b_i a_{ij}a_{jk}c_k = \frac{1}{24}.
\end{align}
For index-2 DAEs, the situation is more subtle. Additional order conditions arise due to the coupling between the differential and algebraic variables \cite{Brasey1993,Hairer2006}. Because of these, the set of feasible coefficients is more restricted than for ODEs. There exists a unique three-stage HERK method of order three, while no four-stage HERK method can attain order four \cite{Brasey1993}. In an important special case, the function $\mathcal F$ is linear in the algebraic variable $\bm z$ and the constraint $\mathcal G$ is linear, so that $\mathcal F_{,\bm z}$ and $\mathcal G_{,\bm y}$ are both constant. The additional trees in the local order analysis get simplified substantially. Under these conditions, the extra algebraic order conditions vanish or reduce to the Butcher conditions, at least up to fourth-order accuracy \cite{Sanderse2012}. Moreover, since the algebraic variable $\bm z_{n+1}$ does not enter into the evolution for the differential variables, its accuracy does not influence the convergence order of $\bm y_{n+1}$. However, if the final stage solution $\bm z_{n(s)}$ is used as an approximation of $\bm z(t_{n+1})$, extra order conditions need to be imposed to achieve the desired order of accuracy \cite{Hairer2006}.

\subsection{HERK method for the weak-form problem}
\label{sec:HERK_transient_algorithm}
Here, we apply the HERK method to the weak-form problem \eqref{eq:weak_mom}-\eqref{eq:weak_mass}. At time $t_n$, the HERK method advances the solution by computing the velocity at $s$ intermediate stages, which are then combined with the weights $\lbrace b_i \rbrace$ to obtain the velocity $\bm{v}_{n+1}$ at the next time step. At the $i\text{-}$th intermediate stage, the external body force and boundary traction are evaluated as $\bm{f}_{n(i)} := \bm{f}(t_n + c_i \Delta t_n)$ and $\bm{h}_{n(i)} := \bm{h}(t_n+c_i \Delta t_n)$, for $i=1,\cdots, s$. The solution procedure is stated as follows.

\begin{enumerate}
\item Initialization: 
Given the velocity $\bm{v}_n$ at time $t_n$, the velocity at the stage $i=1$ is evaluated as $\bm{v}_{n(1)} = \bm{v}_n$.
	
\item Intermediate stage:
For $i = 2, \dots, s$, determine the intermediate velocity $\bm{v}_{n(i)} \in \mathcal{S}$ and pressure $p_{n(i-1)} \in L^2(\Omega)$ by solving
\begin{align}
	\label{eq:herk_weak_mom_sub}
	\Big(\bm w, \frac{\rho}{\Delta t} {\bm{v}}_{n(i)} \Big) - {a}_{i\:i-1}\big(\nabla \cdot \bm w, p_{n(i-1)} \big) =& \Big(\bm w, \frac{\rho}{\Delta t_n} \bm {v}_n \Big) -\sum_{j=1}^{i-1} {a}_{ij}  \Big[ \mathbb C(\bm w; \bm {v}_{n(j)}, \bm {v}_{n(j)}) + \big(\bm {\varepsilon}(\bm w), 2 \mu \bm {\varepsilon}(\bm {v}_{n(j)})\big) \nonumber \displaybreak[2] \\  
	&  - \big( \bm w, \rho \bm{f}_{n(j)} \big) - \big(\bm w, \bm {h}_{n(j)} \big)_{\Gamma_h} \Big] + \sum_{j=1}^{i-2} {a}_{ij}\big(\nabla \cdot \bm w, p_{n(j)} \big), \displaybreak[2] \\
	\label{eq:herk_weak_mass_sub}
	\big( q, \nabla \cdot {\bm{v}}_{n(i)} \big) =& 0,
\end{align}
for $\forall \lbrace \bm w, q \rbrace \in \mathcal V \times L^{2}(\Omega)$.

\item Final stage:
The velocity $\bm{v}_{n+1}$ and $p_{n(s)}$ are obtained by solving the variational problem
\begin{align}
\label{eq:herk_weak_mom_final}
\Big(\bm w, \frac{\rho}{\Delta t_n} \bm {v}_{n+1} \Big) - {b}_s\big(\nabla \cdot \bm w, p_{n(s)} \big) =& \Big(\bm w, \frac{\rho}{\Delta t_n} \bm {v}_n \Big) - \sum_{i=1}^{s} {b}_{i} \Big[ \mathbb C(\bm w; \bm {v}_{n(i)}, \bm {v}_{n(i)}) + \big(\bm {\varepsilon}(\bm w), 2 \mu \bm {\varepsilon}(\bm {v}_{n(i)}) \big) \nonumber \displaybreak[2]  \\ 
&- \big( \bm w, \rho \bm{f}_{n(i)} \big) - \big(\bm w, \bm {h}_{n(i)} \big)_{\Gamma_h} \Big] 
+ \sum_{i=1}^{s-1} {b}_{i}\big(\nabla \cdot \bm w, p_{n(i)} \big), \displaybreak[2]  \\
\label{eq:herk_weak_mass_final}
\big( q, \nabla \cdot \bm{v}_{n+1} \big) =& 0,
\end{align}
for $\forall \lbrace \bm w, q \rbrace \in \mathcal V \times L^{2}(\Omega)$.
	
\item Time advancement: Set $t_n \leftarrow t_{n+1}$ and $\bm{v}_n \leftarrow \bm{v}_{n+1}$ and return to step 1.
\end{enumerate}

\begin{remark}
The proposed algorithm enforces the divergence-free constraint at all intermediate and final stages, thereby ensuring incompressibility throughout the time integration process. The satisfaction of the divergence-free condition at intermediate stages is crucial for achieving higher-order temporal accuracy of the velocity. In contrast, several approaches inspired by the pioneering work of Le and Moin \cite{Le1991} have sought to reduce the computational cost by avoiding the Poisson equation solve at intermediate RK stages through pressure extrapolation or relaxed divergence enforcement \cite{Karam2021,Aithal2020, Aithal2023, Karam2023}.
\end{remark}

It can be found that in HERK scheme, advancing the solution from time $t_n$ to $t_{n+1}$ requires only the velocity $\bm{v}_n$, and the pressure $p_n$ is not involved. The pressure plays the role as a Lagrange multiplier to enforce incompressibility and does not propagate in time. As a result, no initial condition is required for the pressure, and the pressure at time $t_{n+1}$ remains undetermined immediately after computing $\bm{v}_{n+1}$. In order to determine the pressure $p_{n+1}$, one may solve an additional set of equations, corresponding to the hidden constraint at time $t_{n+1}$. Let $\dot{\bm{v}}_{n+1}$ denote the time derivative of the velocity at $t_{n+1}$. The formulation is stated as follows. Determine $\dot{\bm v}_{n+1} \in \dot{\mathcal S}$ and $p_{n+1} \in L^{2}(\Omega)$ such that
\begin{align}
\label{eq:herk_weak_pres_v}
\big( \bm{w}, \rho \dot{\bm{v}}_{n+1} \big) - \big( \nabla \cdot \bm{w}, p_{n+1} \big) =& 
- \mathbb{C}(\bm{w}; \bm{v}_{n+1}, \bm{v}_{n+1})
- \big( \bm{\varepsilon}(\bm{w}), 2\mu \bm{\varepsilon}(\bm{v}_{n+1}) \big) + \big( \bm{w}, \rho \bm{f}_{n+1} \big) + \big( \bm{w}, \bm{h}_{n+1} \big)_{\Gamma_h}, \displaybreak[2]  \\ 
\label{eq:herk_weak_pres_p}
\big(q, \nabla \cdot \dot{\bm{v}}_{n+1} \big) =& 0,
\end{align}
for $\forall \lbrace \bm w, q \rbrace \in \mathcal V \times L^{2}(\Omega)$. In the above, the trial solution space for $\dot{\bm v}_{n+1}$ is given by $\dot{\mathcal S} := \left\lbrace \dot{\bm v} \in H^1(\Omega) : \dot{\bm v}\middle|_{\Gamma_g} = \dot{\bm g}\right\rbrace$. It is worth noting that the incompressible NS equations belong to the aforementioned special case, in which the additional algebraic order conditions reduce to the classical Butcher conditions \eqref{eq:Butcher-order-cond-123}-\eqref{eq:Butcher-order-cond-4}. Consequently, the velocity can achieve up to fourth-order accuracy using the RK schemes designed for ODEs. Although the pressure acts as an instantaneous variable and does not directly influence the temporal accuracy of the velocity in the HERK schemes, its own accuracy often remains critical. Alternative strategies exist for determining $p_{n+1}$ without solving the additional equations. One may approximate $p_{n+1}$ by the final stage solution $p_{n(s)}$ and expect higher-order accuracy if additional algebraic order conditions are imposed. Alternatively, one may reconstruct a higher-order accurate pressure from solutions of intermediate stages through averaging strategies. In contrast, with time-independent inflow, a simpler pressure formulation can be designed, and readers may refer to \cite{Sanderse2012} for a detailed discussion. In the present study, we adopt the approach that directly solves the hidden constraint for the pressure (i.e., \eqref{eq:herk_weak_pres_v}-\eqref{eq:herk_weak_pres_p}), since this step can be performed independently and deferred to a post-processing stage. This treatment offers the flexibility in employing arbitrary RK schemes without concern for additional algebraic order constraints or inflow boundary conditions.

\section{Variational multiscale formulation}
\label{sec:Variational multiscale formulation}
Noticing the structural similarity between the variational equations at the intermediate stage \eqref{eq:herk_weak_mom_sub}-\eqref{eq:herk_weak_mass_sub} and at the final stage \eqref{eq:herk_weak_mom_final}-\eqref{eq:herk_weak_mass_final}, we introduce a `shifted' velocity vector
\begin{align}
\bm u^T =
\begin{pmatrix}
\bm u_{1}, &
\cdots &
\bm u_{s-1}, &
\bm u_{s}
\end{pmatrix}
:=
\begin{pmatrix}
\bm v_{n(2)}, &
\cdots &
\bm v_{n(s)}, &
\bm v_{n+1}
\end{pmatrix},
\end{align}
and a `shifted' RK coefficient matrix $\bm \alpha \in \mathbb R^{s\times s}$ with components given by
\begin{align*}
\bm{\alpha} =
\begin{pmatrix}
\alpha_{11} & 0  & \cdots & 0 \\
\vdots & \ddots &   & \vdots \\
\alpha_{s-1\: 1} & \cdots & \alpha_{s-1\:s-1} & 0 \\
\alpha_{s1}  & \cdots & \alpha_{s\:s-1} & \alpha_{ss}
\end{pmatrix}
:=
\begin{pmatrix}
a_{21} & 0      & \cdots & 0 \\
\vdots & \ddots &        & \vdots \\
a_{s1} & \cdots & a_{s\:s-1} & 0 \\
b_1    & \cdots & b_{s-1} & b_s
\end{pmatrix}
\end{align*}
to streamline the notation for the VMS analysis. Notice that in HERK, $\bm v_{n(1)}$ is set to be $\bm v_n$ and is thus not included in the shifted velocity vectour. In the shifted coefficient matrix $\bm \alpha$, the stage coefficients and the weights of the final stage are arranged in a unified compact form.

Let the stage-wise trial and test functions be denoted as $\bm{U}_i = \lbrace \bm{u}_i, p_{n(i)} \rbrace$, and $\bm{W} = \lbrace \bm{w}, q \rbrace$, respectively. With these notations, the variational problem for the intermediate and final stages can be written in a unified form
\begin{align}
	\label{eq:unified_weak}
	B_i(\bm{W}, \bm{U}_i) = L_i(\bm{W}),
\end{align}
with the bilinear and linear forms defined as
\begin{align}
	\label{eq:unified_B}
	B_i(\bm{W}, \bm{U}_i) :=& \big( \bm{w}, \frac{\rho}{\Delta t_n} \bm{u}_i \big) 
	- \alpha_{ii} \big( \nabla \cdot \bm{w}, p_{n(i)} \big) 
	+ \alpha_{ii} \big( q, \nabla \cdot \bm{u}_i \big), \displaybreak[2] \\
	\label{eq:unified_L}
	L_i(\bm{W}) :=& \big( \bm{w}, \frac{\rho}{\Delta t_n} \bm{v}_n \big)
	- \sum_{j=1}^{i} \alpha_{ij} \Big[ \mathbb{C}(\bm{w}; \bm{v}_{n(j)}, \bm{v}_{n(j)}) + \big( \bm{\varepsilon}(\bm{w}), 2 \mu \bm{\varepsilon}(\bm{v}_{n(j)} \big) \nonumber \displaybreak[2] \\
	& \quad -  \big(\bm{w}, \rho \bm{f}_{n(j)} \big) - \big(\bm{w}, \bm{h}_{n(j)})_{\Gamma_h} \big) \Big] + \sum_{j=1}^{i-1} \alpha_{ij} \big( \nabla \cdot \bm{w}, p_{n(j)} \big),
\end{align}
for $i=1,\cdots, s$. We note that the continuity equation term in \eqref{eq:unified_B} is scaled by a factor $\alpha_{ii}$ compared with the formulation introduced earlier. This modification does not affect consistency but facilitates the definition of a stability norm in the subsequent analysis. The linear form $L_i(\bm{W})$ in \eqref{eq:unified_L} collects all explicit contributions of the previous stages, including the convective, viscous, pressure, body force, and boundary terms. The variational problem \eqref{eq:unified_weak} corresponds to the intermediate stage when $i=1,\cdots, s-1$ and corresponds to the final stage problem when $i=s$.  By applying integration-by-parts for the variational problem \eqref{eq:unified_weak}, one obtains the corresponding strong-form problem, which can be written abstractly as
\begin{align*}
	\mathscr L_i \bm U_i = \mathcal F_i,
\end{align*}
together with appropriate boundary conditions. The differential operator takes the explicit form
\begin{align}
	\mathscr L_i \bm U_i = 
	\begin{pmatrix}
		\frac{\rho}{\Delta t_n} \bm u_i + \alpha_{ii} \nabla p_{n(i)} \\
		\alpha_{ii} \nabla \cdot \bm u_i
	\end{pmatrix}.
\end{align}
We note that the operator $\mathscr L_i$ is structurally analogous to the Darcy operator, with the permeability replaced by the time-stepping factor $\rho/\Delta t_n$, and both the gradient and divergence operators are scaled by the RK coefficient $\alpha_{ii}$. Despite these, it preserves the characteristic saddle-point structure of a Darcy problem, coupling the velocity and pressure through the gradient and divergence terms. This structural resemblance is crucial for the design of numerical schemes. In the following section, we construct a residual-based VMS formulation for incompressible flows with the mathematical construction grounded in this Darcy-type differential operator. Importantly, the Darcy-type operator facilitates a convenient treatment of multiscale coupling, which in turn enables us to obtain an effective LES model without relying on excessive numerical modeling assumptions. The derivation differs from the VMS formulation developed under fully implicit integration frameworks \cite{Bazilevs2007,Masud2011,Codina2007,Hughes2018,Scovazzi2004}.

\subsection{Multiscale decomposition}
\label{sec:Multiscale formulation}
In this section, we perform a multiscale analysis of the NS equations that have been discretized by the HERK method in time. The objective is to derive a spatial formulation that systematically accounts for the impact from the subgrid scales. Within the residual-based framework, various strategies have been proposed for modeling the fine scales. A widely used algebraic approach assumes the fine-scale velocity to be proportional to the residual of the resolved scales \cite{Bazilevs2007}, yielding an inexpensive and effective model. An alternative strategy enforces orthogonality of subgrid scales to the coarse scales \cite{Codina2002}, which enjoys improved dissipative behavior by preventing contamination of the resolved modes \cite{Guasch2013,Colomes2015}. A further refinement treats subgrid scales as time dependent, evolving them via ordinary differential equations \cite{Codina2007}. This not only allows for backscatter of the energy \cite{Principe2010} but also improves the robustness of the model \cite{Colomes2015}. In this work, we derive the fine-scale model based on a bubble enrichment strategy. The resulting formulation leads naturally to an algebraic subgrid-scale model, offering a clear theoretical foundation for coupling HERK time integration with the VMS framework.
 
 To this end, we introduce direct-sum decomposition of the trial solution space $\mathcal S$ and test function space $\mathcal V$ as coarse-scale and fine-scale subspaces, that is, $\mathcal S = \bar{\mathcal S} \oplus \mathcal S^{\prime}$ and $\mathcal V = \bar{\mathcal V} \oplus \mathcal V^{\prime}$. The coarse-scale space, denoted by an overbar, is assumed to be finite dimensional and is also referred to as the resolved scale. In contrast, the fine-scale component, denoted by a prime and sometimes referred to as the subgrid scale, belongs to an infinite dimensional function space. Given a projector, the coarse- and fine-scale components are uniquely defined \cite{Hughes2007}. This implies that the trial solution $\bm U_i$, for $i=1,\cdots, s$, can be uniquely decomposed as 
\begin{align*}
\bm U_i = \bar{\bm U}_i + \bm U^{\prime}_i, \qquad \bar{\bm U}_i \in \bar{\mathcal S} \quad \mbox{and} \quad \bm U^{\prime}_i \in \mathcal S^{\prime}.
\end{align*}
Analogously, the test function admits the decomposition 
\begin{align*}
\bm W = \bar{\bm W} + \bm W^{\prime}, \qquad \bar{\bm W} \in \bar{\mathcal V} \quad \mbox{and} \quad \bm W^{\prime} \in \mathcal V^{\prime}.
\end{align*}

To simplify our discussion, we assume that no fine-scale pressure contributions are present \cite{Gravemeier2005,Nakshatrala2006}, meaning $\bm U^{\prime}_i$ only involves the fine-scale velocity $\bm u_i^{\prime}$. Moreover, it is typical to assume that the fine-scale velocity vanishes along the Dirichlet boundary, i.e., $\bm{u}_i^{\prime}=\bm {w}^{\prime}=\bm 0$ on $\Gamma_g$, which in turn suggests that the coarse-scale velocity $\bar{\bm u}_i$ satisfies the essential boundary condition \cite{Hughes1998}. With the scale decomposition, we may write the weak-form problem \eqref{eq:unified_weak} at the $i$-th stage in terms of the coarse-scale and fine-scale equations by virtue of the linear independence of $\bar{\bm W}$ and $\bm W^{\prime}$,
\begin{align}
\label{eq:darcy_vms_coarse}
B_i(\bar{\bm{W}},\bar{\bm{U}}_i) + B_i(\bar{\bm{W}},{\bm{U}^{\prime}_i}) &= L_i(\bar{\bm{W}}), \displaybreak[2] \\
\label{eq:darcy_vms_fine}
B_i({\bm{W}^{\prime}},\bar{\bm{U}}_i) + B_i({\bm{W}^{\prime}},{\bm{U}^{\prime}_i}) &= L_i({\bm{W}^{\prime}}).
\end{align}
The original variational equation \eqref{eq:unified_weak} achieves the above decomposition thanks to the bilinearity of $B_i(\cdot, \cdot)$ and linearity of $L_i(\cdot)$. The bilinear forms $B_i(\bar{\bm{W}},{\bm{U}'_i})$ and $B_i({\bm{W}'},\bar{\bm{U}}_i)$ on the left-hand side of \eqref{eq:darcy_vms_coarse} and \eqref{eq:darcy_vms_fine}, respectively, represent the scale-coupling effects. We remark that, in conventional multiscale analyses of nonlinear problems, the weak-form problem is nonlinear with respect to the trial solution, which necessitates the use of a perturbation-series expansion to generate a hierarchy of linearized subproblems for further analysis. Such procedures inherently rely on the assumption that coarse-scale residual is sufficiently ``small" \cite{Bazilevs2007,Scovazzi2004,Liu2018}.

The fine-scale problem \eqref{eq:darcy_vms_fine} can be reorganized by moving the scale-coupling term to the right-hand side as
\begin{align}
\label{eq:darcy_vms_fine2}
B_i( \bm W^{\prime}, \bm U_i^{\prime} ) = L_i(\bm W^{\prime} \big) - B_i\big( \bm W^{\prime}, \bar{\bm U}_i ).
\end{align}
Since the pressure is assumed not to involve a fine-scale component, the left-hand side of \eqref{eq:darcy_vms_fine2} can be explicitly written as
\begin{align*}
B_i( \bm W^{\prime}, \bm U_i^{\prime} ) = \big( \bm{w}^{\prime}, \frac{\rho}{\Delta t_n} \bm{u}^{\prime}_i \big) + \alpha_{ii} \big( q^{\prime}, \nabla \cdot \bm{u}^{\prime}_i \big).
\end{align*}
In the meantime, the right-hand side of \eqref{eq:darcy_vms_fine2} can be represented as
\begin{align*}
L_i(\bm W^{\prime} \big) - B_i\big( \bm W^{\prime}, \bar{\bm U}_i ) = \big( \bm{w}^{\prime}, \mathcal R_i \big) - \alpha_{ii} \big( q^{\prime}, \nabla \cdot \bar{\bm{u}}_i \big),
\end{align*}
with the residual $\mathcal R_i$ defined as
\begin{align}
\label{eq:darcy_vms_coarse_residual}
\mathcal R_i := \frac{\rho}{\Delta t_n} ( \bm{v}_n - \bar{\bm u}_i) - \sum_{j=1}^{i} \alpha_{ij} \Big( \rho (\bm{v}_{n(j)} \cdot \nabla) \bm{v}_{n(j)} - \nabla \cdot 2 \mu \bm{\varepsilon}(\bm{v}_{n(j)}) + \nabla p_{n(j)} - \rho \bm{f}_{n(j)} \Big).
\end{align}
It represents the residual of the momentum balance equation written in terms of the previous step and stage components. Collecting the terms associated with the test function $\bm w^{\prime}$ leads to
\begin{align}
\label{eq:darcy_vms_fine_eqn_simplified}
\big( \bm{w}^{\prime}, \frac{\rho}{\Delta t_n} \bm{u}^{\prime}_i \big) = \big( \bm{w}^{\prime}, \mathcal R_i \big).
\end{align}
It is evident that the left-hand side of \eqref{eq:darcy_vms_fine_eqn_simplified} involves no differential operator, and the solution of this problem is in fact the $L^2$-projection of $\mathcal R_i$ onto the fine-scale space $\mathcal S^{\prime}$. Denoting the projector by $\mathcal P^{\prime}$, the solution of the fine-scale equation \eqref{eq:darcy_vms_fine_eqn_simplified} can be represented as
\begin{align}
\label{eq:darcy_vms_fine_sol}
\bm u^{\prime}_i = \frac{\Delta t_n}{\rho} \mathcal P^{\prime}\left( \mathcal R_i \right),
\end{align}
and it is driven by the resolved scales through the residual $\mathcal R_i$. Inserting the fine-scale velocity \eqref{eq:darcy_vms_fine_sol} into the coarse-scale equation \eqref{eq:darcy_vms_coarse}, we obtain
\begin{align}
\label{eq:darcy_vms_coarse_reformulation}
B_i(\bar{\bm{W}},\bar{\bm{U}}_i) + \big( \frac{\rho}{\Delta t_n}\bar{\bm w}-\alpha_{ii}\nabla \bar{q}, \frac{\Delta t_n}{\rho} \mathcal P^{\prime}\big( \mathcal R_i \big) \big) = L_i(\bar{\bm{W}}).
\end{align}
Since the residual depends only on the solutions of the previous stages, the coarse-scale velocity $\bar{\bm u}_i$, and the pressure $p_{n(i)}$, the equation \eqref{eq:darcy_vms_coarse_reformulation} is closed for the coarse-scale unknowns. As is customary in VMS formulations, we regard the solution $\bar{\bm U}_i$ of the problem \eqref{eq:darcy_vms_coarse_reformulation} as the approximation to $\bm U_i$ and do not attempt to use $\bar{\bm U}_i + \bm U^{\prime}_i$ to represent $\bm U_i$. The reason is that the fine-scale component $\bm U^{\prime}_i$ is often provided by a numerical model, which may not always be sufficiently accurate \cite{Oberai2016}. In our setting, using $\bar{\bm U}_i + \bm U^{\prime}_i$ to represent the solutions complicates the numerical implementation, since the HERK scheme requires multiple solutions of the previous stages and previous time step. In our experience, the gain in accuracy is marginal.

For the moment, we only assume that there is no subgrid-scale pressure and the fine-scale velocity vanishes on the Dirichlet boundary. No additional modeling assumption has been invoked in the above derivation, and in this sense, the equation \eqref{eq:darcy_vms_coarse_reformulation} represents an exact formulation for the coarse-scale problem, with the influence of the subgrid scales taken into account. In the conventional approach, the fine-scale modeling necessitates truncating a perturbation series as a linearization procedure \cite{Bazilevs2007,Scovazzi2004}. Such a procedure inherently depends on the convergence of the perturbation-series expansion, rendering it somewhat ad hoc. Here, the explicit treatment of the nonlinear terms renders the VMS analysis more straightforward, which we regard as an appealing feature of the approach.


\subsection{Approximation of the projector via bubble enrichment}
\label{sec:VMS formulation}
It is computationally infeasible to determine the fine-scale component $\bm u^{\prime}_i$ through \eqref{eq:darcy_vms_fine_eqn_simplified} as it is projecting the residual onto an infinite-dimensional space. From dimensional analysis, we may conclude that $\mathcal P^{\prime}(\mathcal R_i)$ has the dimension of velocity, or the action of $\mathcal P^{\prime}$ is dimensionless. Yet, this observation alone is not sufficient, and a more explicit model for projection $\mathcal P^{\prime}$ is needed to make the formulation \eqref{eq:darcy_vms_coarse_reformulation} computationally tractable. To this end, we consider a discretization of the bounded domain $\Omega$ into $n_{\mathrm{el}}$ non-overlapping regions $\Omega^e$ with element boundaries $\Gamma^e := \partial \Omega^{e}$, for $e=1,\cdots,n_{\mathrm{el}}$. The fine-scale problem \eqref{eq:darcy_vms_fine_eqn_simplified} can then be simplified by assuming that $\mathcal S^{\prime}$ is the space of bubbles, i.e., $\mathcal S^{\prime} = \oplus_{e=1}^{n_{\mathrm{el}}} H_0^1(\Omega^{e})$. Consequently, the fine-scale velocity can be expressed as
\begin{align*}
\bm u^{\prime}_i = \sum_{e=1}^{n_{\mathrm{el}}} b_{e} \bm u^{\prime}_{i\:e}, \quad \mbox{for} \quad i = 1, \cdots, s,
\end{align*}
where $b_{e} \in H_0^1(\Omega^{e})$ denotes a bubble basis function supported on the element $\Omega^{e}$, and $\bm u^{\prime}_{i\:e}$ is the associated unknown coefficient. With this choice, the problem \eqref{eq:darcy_vms_fine_eqn_simplified} can be localized to individual elements. In particular, choosing the test function $\bm w^{\prime} = b_{e} \bm e_m$, with $\bm e_m$ being the $m$-th Cartesian basis vector, one obtains
\begin{align*}
\frac{\rho}{\Delta t_n}\big( b_{e}, b_{e} \big)_{\Omega_e} u^{\prime}_{i\:e \: m} =  \big( b_{e}, \mathcal R_{i\:m}\big)_{\Omega_e}.
\end{align*}
In the above, $u^{\prime}_{i\:e \: m}$ and $R_{i\:m}$ are the $m$-th Cartesian components of $\bm u^{\prime}_{i\:e}$ and $R_{i}$, respectively. Solving the coefficients from the above equation, we obtain an expression for the fine-scale component in element $\Omega^{e}$ as
\begin{align*}
\bm u^{\prime}_{i} =  \frac{\Delta t_n}{\rho} P_e^{\prime}\big( \mathcal R_i \big) \quad \mbox{and} \quad P_e^{\prime}\big( \mathcal R_i \big) = \big( b_{e}, b_{e} \big)_{\Omega_e}^{-1}\big( b_{e}, \mathcal R_i \big)_{\Omega_e} b_{e}.
\end{align*}
This procedure can be performed element by element, yielding the fine-scale velocity $\bm u^{\prime}$ and thus completing the specification of \eqref{eq:darcy_vms_coarse_reformulation}. The design of the bubble functions $b_e$ directly defines the projection $\mathcal P_e^{\prime}$, and hence the effectiveness of the numerical model hinges on this choice. It is recognized that residual-free bubbles provide the most accurate representation of the subgrid scale. However, determining the residual-free bubble elementwise is generally nontrivial \cite{Gravemeier2005,Gravemeier2004}. A practical alternative is to adopt polynomial bubble functions. The resulting formula is not necessarily the optimal choice and can be viewed as an approximation to the residual-free bubble. Nevertheless, consistent good results have been reported with polynomial bubbles \cite{Masud2011,Masud2011a,Mourad2007}.

To further simplify the model, we  seek an approximation that reduces the fine-scale effect to an elementwise algebraic relation and preserves the essential stabilization properties. Let $\bar{\mathcal R}_i := (1, \mathcal R_i)/|\Omega^{e}|$ be the average of the residual, and we have $(b^{e}, \mathcal R_i) = (b^{e}, \bar{\mathcal R}_i) + (b^{e}, \mathcal R_i - \bar{\mathcal R}_i)$. If the fluctuation $\mathcal R_i - \bar{\mathcal R}_i$ is sufficiently small, the average $\bar{\mathcal R}_i$ serves as a good approximation of $\mathcal R_i$, and we have
\begin{align*}
\bm u^{\prime}_{i} = \tau_{b} \frac{\Delta t_n}{\rho}  \bar{\mathcal R}_i, \quad \mbox{with} \quad \tau_{b} = \big( b_{e}, b_{e} \big)_{\Omega_e}^{-1}\big( b_{e}, 1\big)_{\Omega_e} b_{e}.
\end{align*}
We may then define the elementwise average of $\tau_{b}$ as
\begin{align*}
\tau := \frac{(\tau_{b}, 1)}{|\Omega_e|} = \frac{\big( b_{e}, 1\big)^2_{\Omega_e}}{|\Omega_e|\big( b_{e}, b_{e} \big)_{\Omega_e}}.
\end{align*}
If we introduce that $\mu := \big( b_{e}, 1\big)_{\Omega_e}/|\Omega_e|$ and $\sigma^2 := \big( b_{e}-\mu, b_{e} - \mu \big)_{\Omega_e}/|\Omega_e|$ as the mean and fluctuation of the residual-free bubble function in the element $\Omega_e$, respectively, the averaged value can be represented as  $\tau = 1/(1+\sigma^2/\mu^2) \in (0,1)$. If the residual-free bubble function approaches a constant function, the fluctuation $\sigma \rightarrow 0$, and $\tau$ tends to $1$; if the mean of the residual-free bubble function is zero, $\tau$ is zero. Although $\tau$ can in principle be computed from the bubble function via the above expression, such an implementation is still cumbersome in practice. Instead, we exploit the fact that $\tau \in (0,1)$ and adopt the pragmatic choice 
\begin{align}
\label{eq:tau_equal_1_2}
\tau = \frac12 \quad \mbox{and} \quad \bm u^{\prime}_{i} = \frac{\Delta t_n}{2\rho} \mathcal R_i, \quad \mbox{for} \quad i = 1, \cdots, s,
\end{align}
as our model for the fine-scale velocity. In the above algebraic model, the fine-scale contribution is scaled by $\Delta t_n/2\rho$, and notably, it does not depend on any spatial discretization parameter. This feature distinguishes the present formulation from many classical subgrid-scale model designs.

\begin{remark}
The stabilized formulation and multiscale analysis of the Darcy problem constitutes a fundamental basis for the present work. One of the earliest stabilized formulation was proposed by Masud and Hughes \cite{Masud2002}, in which they introduced the constant $1/2$ as the stabilization parameter. Their work not only provided a rigorous convergence analysis for general interpolation pairs but also motivates the approach adopted in the present study.
\end{remark}

Substituting the fine-scale model into the coarse-scale equation \eqref{eq:darcy_vms_coarse} leads to
\begin{align}
\label{eq:darcy_vms_coarse_closed}
B_i(\bar{\bm{W}},\bar{\bm{U}}_i) + \big( \frac{\rho}{\Delta t_n}\bar{\bm w}-\alpha_{ii}\nabla \bar{q}, \frac{\Delta t_n}{2\rho} \mathcal R_i \big) = L_i(\bar{\bm{W}}), \quad \mbox{for} \quad i = 1, \cdots, s.
\end{align}
The above represents the coarse-scale formulation at the intermediate and final stages of the HERK method within the VMS framework. Due to the residual-driven nature, it maintains consistency with the original strong-form problem. The formulation \eqref{eq:darcy_vms_coarse_closed} can be written explicitly by substituting the residual \eqref{eq:darcy_vms_coarse_residual}, the bilinear form \eqref{eq:unified_B}, and the linear form \eqref{eq:unified_L} into \eqref{eq:darcy_vms_coarse_closed}. The resulting formulation can be represented as
\begin{align}
\label{eq:herk_vms_coarse_closed}
B^{\mathrm{VMS}}_i(\bar{\bm{W}},\bar{\bm{U}}_i) = L^{\mathrm{VMS}}_i(\bar{\bm{W}}),
\end{align}
where
\begin{align}
	\label{eq:herk_vms_coarse_closed_B}
	B^{\mathrm{VMS}}_i(\bar{\bm{W}},\bar{\bm{U}}_i) :=
	&\big(\bar{\bm w}, \frac{\rho}{\Delta t_n} \bar{\bm {u}}_{i} \big) - {\alpha}_{ii} \big(\nabla \cdot \bar{\bm w}, \bar{p}_{n(i)} \big) + {\alpha}_{ii}\big( \bar{q}, \nabla \cdot \bar{\bm {u}}_{i} \big) + \big(\bar{\bm w}, \frac{\rho}{\Delta t_n} \bm {u}^{\prime}_{i} \big) - {\alpha}_{ii}\big( \nabla \bar{q}, \bm u^{\prime}_{i} \big), \displaybreak[2] \\
	\label{eq:herk_vms_coarse_closed_L}
	L^{\mathrm{VMS}}_i(\bar{\bm{W}}) := 
	& \big(\bar{\bm w}, \frac{\rho}{\Delta t_n} \bm v_n \big) +  \sum_{j=1}^{i-1}{\alpha}_{ij}\big( \nabla \cdot \bar{\bm w}, p_{n(j)} \big) -\sum_{j=1}^{i} {\alpha}_{ij} \Big[\big(\bm {\varepsilon}(\bar{\bm{w}}), 2 \mu \bm {\varepsilon}(\bm {v}_{n(j)})\big) \nonumber \displaybreak[2] \\
	& - \big( \bar{\bm w}, \rho \bm{f}_{n(j)} \big) - \big(\bar{\bm w}, \bm {h}_{n(j)} \big)_{\Gamma_h} + \mathbb C(\bar{\bm w}; \bm v_{n(j)}, \bm v_{n(j)}) - \big( \nabla \bar{\bm w}, \rho \bm{v}^{\prime}_{n(j)} \otimes \bm v_{n(j)} \big)  \nonumber \displaybreak[2] \\
	&  + \big( \nabla \bm v_{n(j)}, \rho \bar{\bm w} \otimes \bm{v}^{\prime}_{n(j)} \big) - \big( \nabla \bar{\bm w}, \rho \bm{v}^{\prime}_{n(j)} \otimes \bm{v}^{\prime}_{n(j)} \big) \Big],
\end{align}
and
\begin{align}
	\label{eq:herk_vms_fine}
	\bm {u}^{\prime}_{i} := - \frac{\Delta t_n}{2\rho} \Big(\rho\frac{\bar{\bm u}_{i} - \bm{v}_n}{\Delta t_n} +  \sum_{j=1}^{i} {\alpha}_{ij} \big( \rho \bm{v}_{n(j)} \cdot \nabla \bm{v}_{n(j)} - \nabla \cdot 2\mu \bm{\varepsilon}(\bm{v}_{n(j)})  + \nabla p_{n(j)} - \rho \bm {f}_{n(j)} \big) \Big),
\end{align}
for $i=1,\cdots, s$. The last three terms in \eqref{eq:herk_vms_coarse_closed_L} arise from the multiscale decomposition of $\mathbb C(\bar{\bm w}; {\bm v}_{n(j)}, {\bm v}_{n(j)})$, where the symbol $\otimes$ denotes the dyadic product of vectors. The first two terms are referred to as the cross-stress terms, whereas the third term is referred to as the Reynolds stress term.

The stability of the newly derived VMS formulation can be demonstrated directly. Substituting the test functions into the bilinear form $B^{\mathrm{VMS}}_i\left(\cdot \:, \cdot\right)$, one obtains a stability norm for the $i$-th stage as
\begin{align}
	\label{eq:Bvms_stability}
	 \left|\!\left|\!\left| \bar{\bm{W}} \right|\!\right|\!\right|_{i}^{2} := B^{\mathrm{VMS}}_i(\bar{\bm{W}},\bar{\bm{W}}) &=
	\Big(\bar{\bm w}, \frac{\rho}{\Delta t_n} \bar{\bm w} \Big) - {\alpha}_{ii} \big(\nabla \cdot \bar{\bm w}, \bar{q} \big) + {\alpha}_{ii}\big( \bar{q}, \nabla \cdot \bar{\bm w} \big) - \frac{1}{2}\Big(\bar{\bm w}, \frac{\rho}{\Delta t_n}\bar{\bm w} \Big) \nonumber \displaybreak[2] \\ 
	&\quad - \frac{1}{2}{\alpha}_{ii}\big(\bar{\bm w}, \nabla{\bar{q}} \big) + \frac{1}{2}{\alpha}_{ii}\big( \nabla \bar{q}, \bar{\bm w} \big) + \frac{1}{2}\Big({\alpha}_{ii}\nabla \bar{q}, \frac{\Delta t_n}{\rho}{\alpha}_{ii}\nabla \bar{q} \Big) \nonumber \displaybreak[2] \\ 
	&= \frac{1}{2}\Big(\bar{\bm w}, \frac{\rho}{\Delta t_n}\bar{\bm w} \Big) + \frac{1}{2}\Big({\alpha}_{ii}\nabla \bar{q}, \frac{\Delta t_n}{\rho}{\alpha}_{ii}\nabla \bar{q} \Big) \nonumber \displaybreak[2] \\ 
	&= \frac{1}{2} \Big(  \frac{\rho}{\Delta t_n} \|  \bar{\bm w} \|^2  + 
	\frac{\Delta t_n}{\rho} \alpha^2_{ii} \| \nabla \bar{q} \|^2  \Big).
\end{align}
Since the subgrid-scale model does not depend on the spatial mesh parameter, the norms $\left|\!\left|\!\left| \cdot \right|\!\right|\!\right|_{i}$, for $i=1,\cdots, s$, are consequently independent of the spatial discretization. With respect to this norm, the VMS formulation is unconditionally stable for any continuous velocity-pressure interpolation. It can be observed that $\tau=1/2$ is a balanced option, providing stability of the velocity in the $L^2$-norm and the stability of the pressure in the $H^1$-seminorm. The impact of the choice of $\tau$ is further revealed in the forthcoming study of Fourier analysis. With the above development, we have obtained a fully discrete scheme whose spatial discretization is stable based on the above norm. Consequently, the numerical stability of our formulation primarily lies in the interplay between the HERK scheme and the VMS formulation. In the next section, we combine the VMS formulation with a prototype problem to examine the numerical properties, which offers further insights into the proposed discrete scheme.

\begin{remark}
While the pressure subgrid scale is not explicitly considered, the formulation can be directly extended by including pressure subscales \cite{Masud2002}. Incorporating a pressure model leads to the grad-div stabilization and an extra term in the stability norm that provides additional control over the velocity divergence \cite{Badia2010}. Yet, this term typically yields more pronounced improvements in mass conservation when applied to inf-sup stable elements than to equal-order velocity-pressure interpolations \cite{Colomes2015,Colomes2016a}. 
\end{remark}

\begin{remark}
The VMS analysis for the Darcy problem was initially explored to provide a rationale for the stabilized formulation \cite{Nakshatrala2006}. This direction was later pursued by Badia and Codina, who carried out a more rigorous analysis \cite{Badia2010}. By means of Fourier analysis, they identified four distinct multiscale models, each characterized with different stabilization parameters and numerical properties. For a related analysis in the context of the coupled Darcy-Stokes problem, readers are referred to \cite{Badia2009}. 
\end{remark}

\begin{remark}
Given the velocity $\bm v_{n+1}$, the hidden constraint \eqref{eq:herk_weak_pres_v}-\eqref{eq:herk_weak_pres_p} is solved, whose numerical scheme is designed following the same manner as described above. The corresponding VMS formulation for $\dot{\bar{\bm {v}}}_{n+1}$ and $\bar{p}_{n+1}$ is stated as
\begin{align}
\label{eq:herk_vms_pres_coarse_closed}
&\big(\bar{\bm w}, \rho\dot{\bar{\bm {v}}}_{n+1} \big) - \big(\nabla \cdot \bar{\bm w}, \bar{p}_{n+1} \big) + \big( \bar{q}, \nabla \cdot \dot{\bar{\bm v}}_{n+1} \big) + \big(\bar{\bm w}, \rho\dot{\bm {v}}^{\prime}_{n+1} \big) - \big( \nabla \bar{q}, \dot{\bm v}^{\prime}_{n+1} \big) \nonumber \displaybreak[2] \\
=& \big( \bm{w}, \rho \bm{f}_{n+1} \big) + \big( \bm{w}, \bm{h}_{n+1} \big)_{\Gamma_h} - \big( \bm{\varepsilon}(\bm{w}), 2\mu \bm{\varepsilon}(\bm{v}_{n+1}) \big) - \mathbb C(\bar{\bm w}; \bm v_{n+1}, \bm v_{n+1}),
\end{align}
where the fine-scale term is modeled as
\begin{align}
\label{eq:herk_vms_pres_fine}
\dot{\bm {v}}^{\prime}_{n+1} &= -\frac{\tau}{\rho} \Big(\rho\dot{\bar{\bm v}}_{n+1} + \rho \bm{v}_{n+1} \cdot \nabla \bm{v}_{n+1} - \nabla \cdot \left( 2\mu \varepsilon(\bm{v}_{n+1}) \right) + \nabla \bar{p}_{n+1} - \rho \bm {f}_{n+1} \Big).
\end{align}
\end{remark}

\section{Fourier analysis}
\label{sec:Fourier_analysis}
In the incompressible NS equations, the pressure appears as an instantaneous algebraic variable that enforces the divergence-free constraint. Since it is treated implicitly, its impact on the spectral and stability properties of the overall scheme is indirect. For this reason, it is more tractable and offers sufficient insight to analyze the spectral behavior of a proposed method through the canonical advection-diffusion equation, which captures the essential wave-like and diffusive mechanisms \cite{Alhawwary2018,Dettmer2003,Shakib1991a,Sagaut2023}. We focus on the one-dimensional advection-diffusion equation for a scalar field $\phi(x,t)$ over a periodic domain $(0, L)$. We use $\phi_{,x}$ and $\phi_{,xx}$ to denote, respectively, the first- and second-order spatial derivatives of $\phi$, and the governing equation reads
\begin{align}
\label{eq:advec-diffu-eq}
\dot{\phi} + a  \phi_{,x} - \kappa  \phi_{,xx} = 0, \quad x \in (0,L) \quad \mbox{and} \quad t > 0,
\end{align}
subject to the initial condition $\phi(x,0)=\phi_{0}(x)$ and periodic boundary conditions. Here, $a>0$ is the prescribed constant flow velocity, and $\kappa > 0$ is the constant diffusivity. The present analysis simultaneously examines the characteristics of the spatial and temporal discretizations. Specifically, we assume that the exact solution of \eqref{eq:advec-diffu-eq} admits a Fourier series expansion, with a typical mode of the form $ \phi = \exp(\nu t + \iota K x) $. Here, $K$ is the spatial wavenumber, $\nu = -\xi + \iota \omega$ governs the temporal evolution of the mode, $\iota$ is the imaginary unit (i.e., $\iota^2 = -1$), $\xi$ denotes the damping coefficient, and $\omega$ is the frequency. Substituting this ansatz into \eqref{eq:advec-diffu-eq} yields the relations $ \xi = \kappa K^2 $ and $ \omega = -a K $, which characterize, respectively, the dissipation and dispersion of each Fourier mode of the continuous problem. 

We compare two strategies that integrate explicit RK schemes with the VMS formulation. In the RK$(s,p)$-VMS approach, following the horizontal method of lines, an explicit $s$-stage, $p$-th-order RK discretization is first applied to generate a sequence of steady problems, which are then discretized in space using the VMS formulation. In the VMS-RK$(s,p)$ approach, the VMS formulation is first applied to obtain a stabilized semi-discrete problem, which is then advanced in time by an RK scheme. Because the subgrid-scale models enter the discrete scheme in different manners in the two cases, the resulting fully discrete schemes are distinct and exhibit different spectral characteristics. The analysis below examines and compares their stability and spectral properties. We mention that the VMS semi-discrete problem with implicit time stepping, including the generalized-$\alpha$ scheme and space-time formulation, has been analyzed in prior studies \cite{Dettmer2003,Shakib1991a}.

\subsection{Fully discrete schemes}
\label{sec:fully discrete form of RK-VMS scheme}
To obtain a compact and unified representation of both the intermediate and final stages of the RK scheme, we follow the notation introduced in Section \ref{sec:Variational multiscale formulation} and analogously define the shifted scalar variables as
\begin{align*}
\begin{pmatrix}
\psi_{1}, & \cdots & \psi_{s-1}, & \psi_{s}
\end{pmatrix}
:=
\begin{pmatrix}
\phi_{n(2)}, & \cdots & \phi_{n(s)}, & \phi_{n+1}
\end{pmatrix}.
\end{align*}
Analogous to the notation introduced in Section \ref{sec:Multiscale formulation}, we use an overbar and a prime to denote the coarse- and fine-scale components, respectively. In the following, we present the fully discrete schemes based on the two considered approaches. In the derivation of the stencil, we consider a uniform spatiotemporal mesh, with the spatial mesh size denoted by $\Delta x$ and time-step size denoted by $\Delta t$. The subsequent analysis of stability and spectral properties is based on the stencils using $C^1$-continuous, quadratic B-splines.

\paragraph{RK$(s,p)$-VMS} The fully discrete RK$(s,p)$-VMS formulation is summarized as follows. Let $\bar{\mathcal S}$ and $\bar{\mathcal V}$ denote the trial and weighting function spaces for the coarse-scale components, respectively. For stage $i=1,\dots,s$, find $\bar{\psi}_{i} \in \bar{\mathcal S}$ such that,  
\begin{align}
\label{eq:advec-diffu-rkvms}
\big( \bar{w}, \bar{\psi}_{i} \big)+\big( \bar{w}, \psi_{i}^{\prime} \big) = \big( \bar{w}, \phi_{n} \big) + \Delta t \sum_{j=1}^{i} \alpha_{ij} \Big[&-\big( \bar{w}, a  \phi_{n(j),x} \big)-\big( \bar{w}_{,x}, \kappa \phi_{n(j),x} \big) \nonumber\\
& +\big( \bar{w}_{,x}, a  \phi_{n(j)}^{\prime} \big) +\big( \bar{w}_{,xx}, \kappa  \phi_{n(j)}^{\prime} \big) \Big],
\end{align}
for $\forall \bar{w} \in \bar{\mathcal V}$. The fine-scale component $\psi_{i}^{\prime}$ of the $i$-th stage is given by 
\begin{align}
\label{eq:advec-diffu-fine-rkvms}
\psi_{i}^{\prime} = -\tau^{*} \Big( \bar{\psi}_{i} - \phi_{n} - \Delta t \sum_{j=1}^{i} \alpha_{ij}	\big( -a  \phi_{n(j),x} + \kappa  \phi_{n(j),xx} \big) \Big).
\end{align}
The above equations are to be solved sequentially for stages $i=1,\dots,s$. It is worth noting that at the first stage, the coarse-scale variable is initialized as $\bar{\phi}_{n(1)} = \phi_n$, which implies that the corresponding fine-scale component vanishes, i.e., $\phi_{n(1)}^{\prime} = 0$. The parameter $\tau^{*}$ plays the same role as the parameter $\tau$ introduced in Section \ref{sec:Multiscale formulation}, and it can be interpreted as an approximation of the $L^2$-projection of the coarse-scale residual onto the fine-scale subspace. A similar derivation based on bubble enrichment reveals that $\tau^{*} \in (0,1)$ is dimensionless. Upon substituting the quadratic B-spline basis functions, this algebraic system leads to a five-point stencil.

\paragraph{VMS-RK$(s,p)$} 
The semi-discrete formulation of the VMS-RK$(s,p)$ approach reads: find $\bar{\phi} \in \bar{\mathcal{S}}$ such that
\begin{align}
	\label{eq:advec-diffu-vmsrk-semi}
	\big(\bar{w}, \dot{\bar{\phi}}\big)+\big(\bar{w}, a{\bar\phi}_{,x}\big)+\big(\bar{w}_{,x}, \kappa{\bar\phi}_{,x}\big) - \big(\bar{w}_{,x}, a\phi^{\prime} \big) - \big(\bar{w}_{,xx}, \kappa\phi^{\prime} \big) = 0,
\end{align}
for all $\bar{w} \in \bar{\mathcal{V}}$, where the fine-scale component $\phi^{\prime}$ is modeled as
\begin{align*}
\phi^{\prime} = -\tau_{\diamond}\Big( \dot{\bar{\phi}} + a \bar{\phi}_{,x} - \kappa \bar{\phi}_{,xx}\Big),
\end{align*}
with the parameter $\tau_{\diamond}$ defined as \cite{Bazilevs2007,Shakib1991b}
\begin{align}
	\label{eq:tau-advec-diffu-vms-rk}
	\tau_{\diamond}=\Big(\big(\frac{2}{\Delta t}\big)^2+\big(\frac{2a}{\Delta x}\big)^2+9\big(\frac{4\kappa}{{\Delta x}^2}\big)^2\Big)^{-1/2}.
\end{align}
The asymptotic behavior of $\tau_{\diamond}$ can be summarized as follows. For sufficiently small time steps, i.e., $\Delta t \ll \Delta x$, $\tau_{\diamond} = \mathcal O(\Delta t)$; when $\Delta x  \ll \Delta t$, one has $\tau_{\diamond} = \mathcal O(\Delta x)$ in the advection-dominated regime and $\tau_{\diamond}= \mathcal O(\Delta x^2)$ in the diffusion-dominated regime \cite{Shakib1991b}. The semi-discrete system \eqref{eq:advec-diffu-vmsrk-semi} can be integrated in time using an explicit RK scheme, leading to the fully discrete formulation: find $\bar{\psi}_{i} \in \bar{\mathcal S}$ such that,
\begin{align}
	\label{eq:advec-diffu-vmsrk}
	\big(\bar{w}+\tau_{\diamond} a \bar{w}_{,x}+\tau_{\diamond}\kappa \bar{w}_{,xx}, \bar{\psi}_i\big) 
	=& \big(\bar{w}+\tau_{\diamond} a \bar{w}_{,x}+\tau_{\diamond}\kappa \bar{w}_{,xx}, \bar{\phi}_n\big) \nonumber \displaybreak[2] \\
	&+\sum_{j=1}^{i} \alpha_{ij}\Big[\big(-a \Delta t \bar{w}-\tau_{\diamond} a^2 \Delta t \bar{w}_{,x}-\tau_{\diamond}\kappa a \Delta t \bar{w}_{,xx} -\kappa \Delta t \bar{w}_{,x}, \bar{\phi}_{n(j),x}\big)  \nonumber \displaybreak[2] \\
	& +\big(a \tau_{\diamond}\kappa \Delta t \bar{w}_{,x}+\tau_{\diamond}\kappa^2 \Delta t \bar{w}_{,xx}, \bar{\phi}_{n(j),xx}\big)\Big]
\end{align}
for $\forall \bar{w} \in \bar{\mathcal V}$. Upon spatial discretization with quadratic B-spline basis functions, one obtains the corresponding stencil. A direct comparison of \eqref{eq:advec-diffu-rkvms} and \eqref{eq:advec-diffu-vmsrk} already reveals that reversing the order of RK method and the VMS formulation yields distinct fully discrete schemes.

\subsection{Stability and spectral properties}
To perform the Fourier analysis, the discrete solution at the time step $t_n$ can be represented in terms of the spatial basis functions, where the coefficient associated with the $A$-th basis function takes the form $\bar{\phi}_{A;n}  = \exp(\nu^{h } n \Delta t + \iota K A h)$. Here, $\nu^{h } = -\xi^{h } + \iota \omega^{h }$ denotes the discrete counterpart of $\nu$, with $\xi^{h}$ and $\omega^{h}$ representing the algorithmic damping and algorithmic frequency, respectively. The amplification factor is then defined as  
\begin{align}
\label{eq:amplification_factor_def}
\zeta^{h } := \frac{\bar{\phi}_{A;n+1} }{\bar{\phi}_{A;n} } = \exp(\nu^{h } \Delta t).
\end{align}  
For RK schemes, one may also define the stage-wise amplification factors as
\begin{align}
\label{eq:stage_amplification_factor_def}
\zeta_i^{h } := \frac{\bar{\phi}_{A;n(i)} }{\bar{\phi}_{A;n} }, \quad \mbox{for} \quad i = 1,\dots,s.
\end{align}  
The amplification factor $\zeta^{h}$ can be constructed through a weighted combination of the stage contributions $\lbrace \zeta_i^{h } \rbrace_{i=1}^{s}$. To characterize the spectral behavior, we introduce the following nondimensional parameters,
\begin{align*}
K^* := \Delta x K, \quad a^* := a \frac{\Delta t}{\Delta x}, \quad \kappa^* := \kappa \frac{\Delta t}{\Delta x^{2}}.
\end{align*}
Here, $K^*$ represents the nondimensional wavenumber, with $K^* = \pi$ corresponding to the Nyquist limit; the nondimensional advective speed $a^* = a \Delta t / \Delta x$ plays the role of the Courant number; the nondimensional diffusivity $\kappa^* = \kappa \Delta t / \Delta x^{2}$ leads to the mesh P\'{e}clet number through the ratio $a^*/2\kappa^*$. With these nondimensional parameters in place, we next derive the amplification factors in closed form.

\subsubsection{Expressions of the amplification factors}
\paragraph{RK$(s,p)$-VMS}
Substituting the discrete Fourier mode representation into the stencil leads to a recursive expression for the stage-wise amplification factor. For the first two stages, one obtains
\begin{align}
\label{eq:stage-wise-amplification-advec-diffu-rk-vms-zeta-1-and-2}
	\zeta^{h}_1 = 1 \quad \mbox{and} \quad
	\zeta^{h}_2 = 1 + \left( \lambda_1 + \lambda_2 \right) \alpha_{11}.
\end{align}
For $i \ge 3$, the stage-wise amplification factor is given by
\begin{align}
	\label{eq:stage-wise-amplification-advec-diffu-rk-vms}
	\zeta^{h}_i
	= 1+\lambda_1\sum_{j=1}^{i-1}\alpha_{i-1\,j}
	+\lambda_2\sum_{j=1}^{i-1}\alpha_{i-1\,j}\zeta^{h}_j
	+\lambda_3\sum_{j=2}^{i-1}\Big(
	\alpha_{i-1\,j}\sum_{k=1}^{j-1}\alpha_{j-1\,k}\zeta^{h}_k
	\Big).
\end{align}
The overall amplification factor associated with the RK$(s,p)$-VMS scheme then reads
\begin{align}
	\label{eq:overall-amplification-advec-diffu-rk-vms}
	\zeta^{h } &= 1 + \lambda_1 \sum_{i=1}^{s} \alpha_{si}
	+ \lambda_2 \sum_{i=1}^{s} \alpha_{si}\zeta^h_i
	+ \lambda_3 \sum_{i=2}^{s} \Big(\alpha_{si} \sum_{j=1}^{i-1} \alpha_{i-1\,j}\zeta^{h }_j\Big),
\end{align}
in which the coefficients are given by
\begin{align*}
	\lambda_1 &= \frac{1}{ \mathcal{C} }\big(20\tau^*\kappa^*\cos(2K^*) + 40\tau^*\kappa^*\cos(K^*) - 60\tau^* \kappa^* 
	+ \mathrm{\iota} \left( -5\tau^*a^*\sin(2K^*) - 50\tau^*a^*\sin(K^*) \right) \big),  \displaybreak[2] \\ 
	\lambda_2 &= \frac{1}{ \mathcal{C} } \big( (-40\tau^*\kappa^* + 20\kappa^*)\cos(2K^*) + (-80\tau^*\kappa^* + 40\kappa^*)\cos(K^*) + 120\tau^*\kappa^* - 60\kappa^*  \nonumber \displaybreak[2] \\ 
	& \quad + \mathrm{\iota}  (10\tau^* a^* - 5 a^*)\sin(2K^*) + (100\tau^* a^* - 50 a^*)\sin(K^*) \big), \displaybreak[2] \\ 
	\lambda_3 &= \frac{1}{ \mathcal{C} } \big((20\tau^*{a^*}^2 + 120\tau^*{\kappa^*}^2)\cos(2K^*) 
	+ (40\tau^*{a^*}^2 - 480\tau^*{\kappa^*}^2)\cos(K^*) + 360\tau^*{\kappa^*}^2 - 60\tau^*{a^*}^2 \nonumber \displaybreak[2] \\ 
	& \quad + \mathrm{\iota} ( -120\tau^*\kappa^* a^*\sin(2K^*) + 240\tau^*\kappa^* a^*\sin(K^*) ) \big), \displaybreak[2] \\ 
	\mathcal{C} &= (1-\tau^*)\cos(2K^*) + 26(1-\tau^*)\cos(K^*) + 33(1-\tau^*).
\end{align*}
In the subsequent analysis, we restrict our attention to explicit RK schemes with stage number $s$ equal to the order of accuracy $p$, which holds for $p \leq 4$ due to the Butcher barrier \cite{Ascher1998}. We observe from \eqref{eq:stage-wise-amplification-advec-diffu-rk-vms}-\eqref{eq:overall-amplification-advec-diffu-rk-vms} that the amplification factors are defined recursively in terms of those from preceding stages. Starting with $\zeta^{h }_1$ and $\zeta^{h}_2$ given in \eqref{eq:stage-wise-amplification-advec-diffu-rk-vms-zeta-1-and-2}, one may obtain $\zeta^h_3$ by direct substitution into \eqref{eq:stage-wise-amplification-advec-diffu-rk-vms}, and this process repeats stage by stage until the overall amplification factor \eqref{eq:overall-amplification-advec-diffu-rk-vms} is recovered. By invoking the order conditions \eqref{eq:Butcher-order-cond-123}-\eqref{eq:Butcher-order-cond-4} of explicit RK schemes, these amplification factors can be simplified, and the resulting expressions are summarized in Table \ref{tab:rk-vms-amplification} for schemes up to $p=4$.
\begin{table}[htbp]
\centering
\renewcommand{\arraystretch}{1.5}
\caption{Amplification factors $\zeta^h$ for RK$(s,p)$-VMS schemes.}
\label{tab:rk-vms-amplification}
\begin{tabular}{ll}
	\hline
	Scheme & Amplification factor $\zeta^h$ \\
	\hline
	RK$(1,1)$-VMS &
	$1 + \lambda_1 + \lambda_2$ \\	
	RK$(2,2)$-VMS &
	$1 + \lambda_1 + \lambda_2 
	+ \frac{1}{2}\lambda_1 \lambda_2 
	+ \frac{1}{2}\lambda_2^2 
	+ \frac{1}{2}\lambda_3$ \\
	RK$(3,3)$-VMS &
	$1 + \lambda_1 + \lambda_2 
	+ \frac{1}{2}\lambda_1 \lambda_2 
	+ \frac{1}{2}\lambda_2^2 
	+ \frac{1}{2}\lambda_3 
	+ \frac{1}{6}\lambda_1 \lambda_3 
	+ \frac{1}{6}\lambda_1 \lambda_2^2 
	+ \frac{1}{3}\lambda_2 \lambda_3 
	+ \frac{1}{6}\lambda_2^3$ \\
	RK$(4,4)$-VMS &
	$1 + \lambda_1 + \lambda_2 
	+ \frac{1}{2}\lambda_1 \lambda_2 
	+ \frac{1}{2}\lambda_2^2 
	+ \frac{1}{2}\lambda_3 
	+ \frac{1}{6}\lambda_1 \lambda_3 
	+ \frac{1}{6}\lambda_1 \lambda_2^2 
	+ \frac{1}{3}\lambda_2 \lambda_3 
	+ \frac{1}{6}\lambda_2^3$ \\
	& $+ \frac{1}{24}\lambda_1 \lambda_2^3 
	+ \frac{1}{24}\lambda_2^4 
	+ \frac{1}{24}\lambda_3^2 
	+ \frac{1}{8}\lambda_2^2 \lambda_3 
	+ \frac{1}{12}\lambda_1 \lambda_2 \lambda_3$ \\
	\hline
\end{tabular}
\end{table}

\paragraph{VMS-RK$(s,p)$} An analogous procedure applied to the VMS-RK$(s,p)$ scheme leads to the amplification factors expressed as
\begin{align*}
\zeta^{h}_i=1+\gamma \sum_{j=1}^{i-1}\alpha_{i-1\:j}\zeta^{h}_j, \qquad
\zeta^{h}=1+\gamma \sum_{i=1}^{s}\alpha_{si}\zeta^{h}_i,
\end{align*}
with
\begin{align*}
\gamma =& \frac{1}{\mathcal{D}}\big((20\tau_{\diamond}^*{a^*}^2+20\kappa^*+120\tau_{\diamond}^*{\kappa^*}^2)\cos(2K^*)+(40\tau_{\diamond}^*{a^*}^2+40\kappa^*-480\tau_{\diamond}^*{\kappa^*}^2)\cos(K^*) \nonumber \displaybreak[2] \\
&\quad-60\tau_{\diamond}^*{a^*}^2-60\kappa^*+360\tau_{\diamond}^*{\kappa^*}^2 - \mathrm{\iota}\big( (5a^*+120\tau_{\diamond}^*\kappa^*a^*)\sin(2K^*) +(50a^*-240\tau_{\diamond}^*\kappa^*a^*)\sin(K^*)\big)\big), \displaybreak[2] \\
\mathcal{D} =& (1 + 20\tau_{\diamond}^*\kappa^*)\cos(2K^*) 
+ (26 + 40\tau_{\diamond}^*\kappa^*)\cos(K^*) 
+ 33 - 60\tau_{\diamond}^*\kappa^* - \mathrm{\iota}\big( 5\tau_{\diamond}^*a^*\sin(2K^*) 
+ 50\tau_{\diamond}^*a^*\sin(K^*) \big).
\end{align*}
Here, $\tau_{\diamond}^* :=\tau_{\diamond}/\Delta t = (4+4a^{*2}+144 \kappa^{*2})^{-1/2}$ stands for the dimensionless counterpart of the parameter $\tau_{\diamond}$. For the case of $s=p$, the overall amplification factors of the schemes up to fourth order are summarized in Table \ref{tab:vms-rk-amplification}. As expected, when $s=p$, the amplification factor for the VMS-RK$(s,p)$ scheme coincides with the $p$-th-order Taylor expansion of $\exp({\gamma})$ \cite{Sagaut2023}. 

\begin{table}[htbp]
\centering
\renewcommand{\arraystretch}{1.5}
\caption{Amplification factors for VMS-RK$(s,p)$ schemes.}
\label{tab:vms-rk-amplification}
\begin{tabular}{ll}
	\hline
	Scheme & Amplification factor $\zeta^h$ \\
	\hline
	VMS-RK$(1,1)$ & 
	$1 + \gamma$ \\
	VMS-RK$(2,2)$ & 
	$1 + \gamma + \frac{1}{2}\gamma^2$ \\
	VMS-RK$(3,3)$ & 
	$1 + \gamma + \frac{1}{2}\gamma^2 + \frac{1}{6}\gamma^3$ \\
	VMS-RK$(4,4)$ & 
	$1 + \gamma + \frac{1}{2}\gamma^2 + \frac{1}{6}\gamma^3 + \frac{1}{24}\gamma^4$ \\
	\hline
\end{tabular}
\end{table}

\subsubsection{Stability}
\begin{figure}[htbp]
\centering
\setlength{\tabcolsep}{2pt}
\begin{tabular}{@{}p{0.5cm}cccc@{}}
\raisebox{1\height}{\rotatebox{90}{$\tau^*=0.1$}} & 
\includegraphics[width=0.22\linewidth, trim=0 30 0 32, clip]{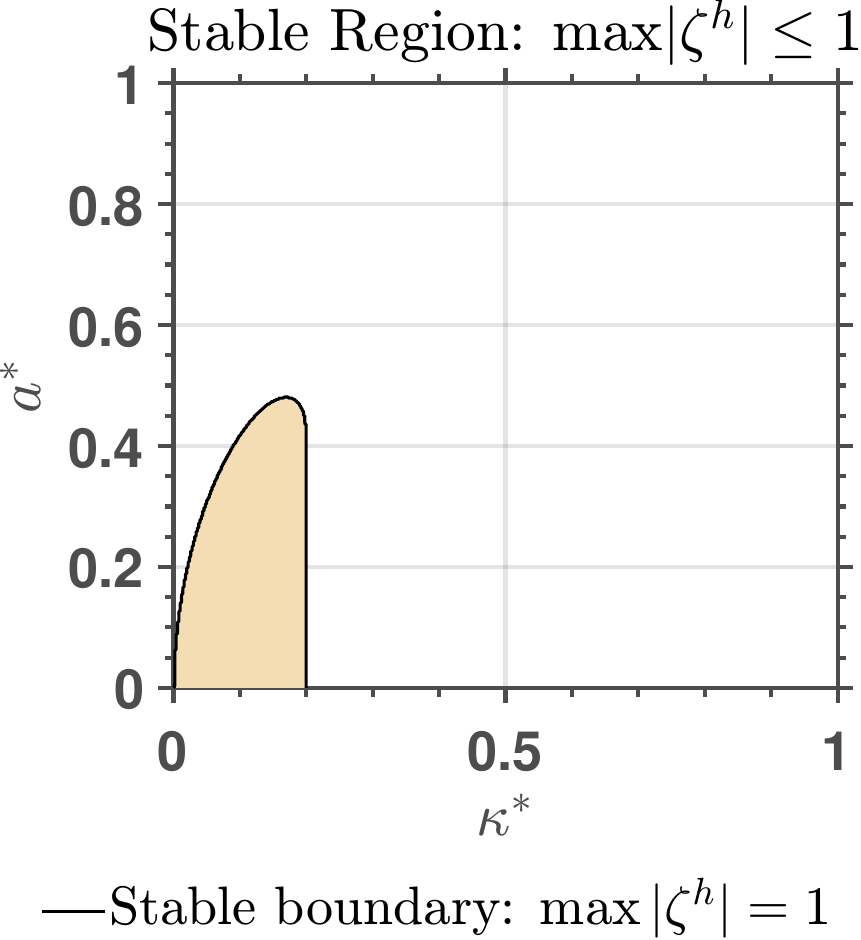} &
\includegraphics[width=0.22\linewidth, trim=0 30 0 32, clip]{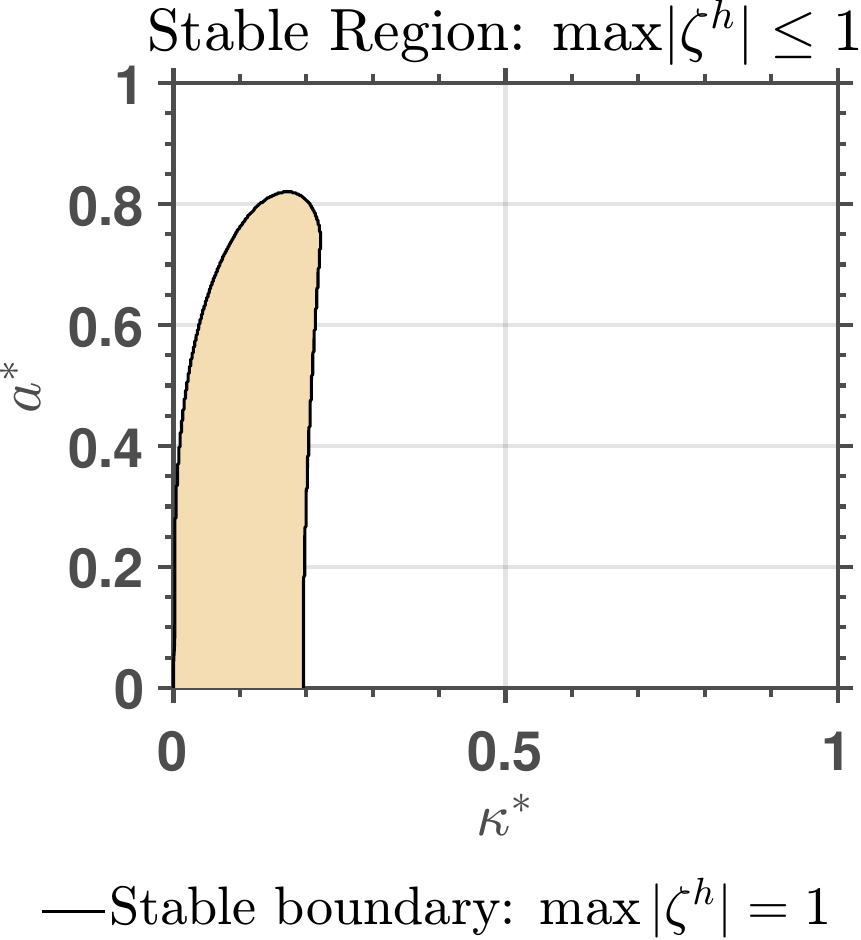} &
\includegraphics[width=0.22\linewidth, trim=0 30 0 32, clip]{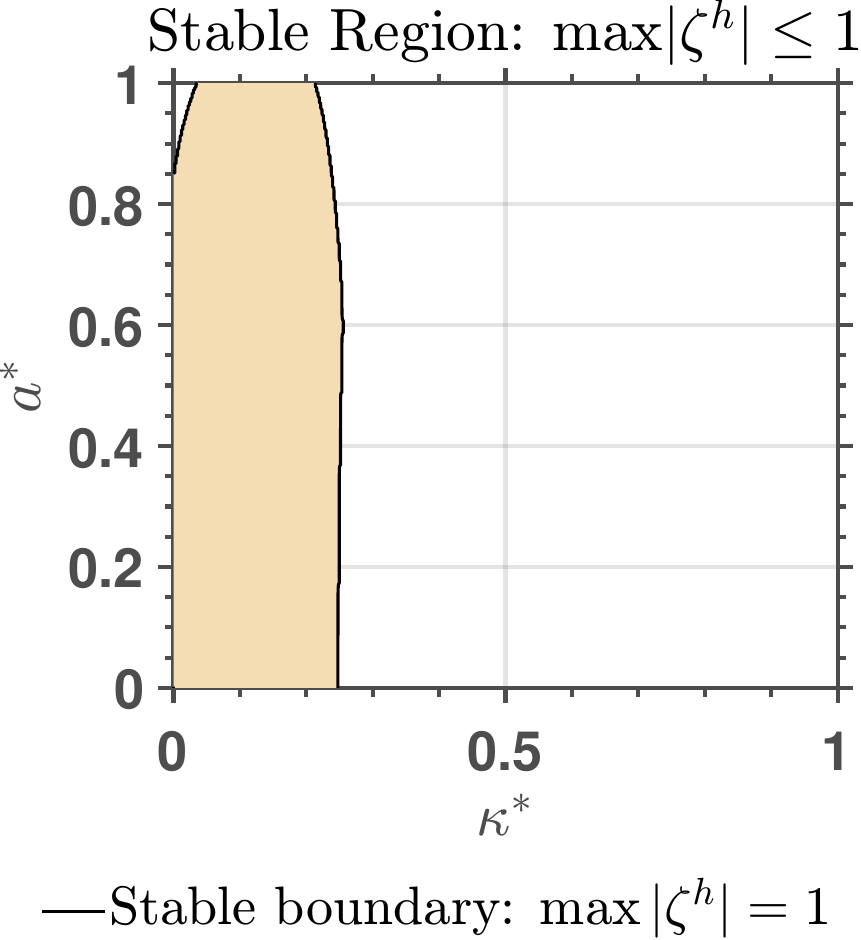} &
\includegraphics[width=0.22\linewidth, trim=0 30 0 32, clip]{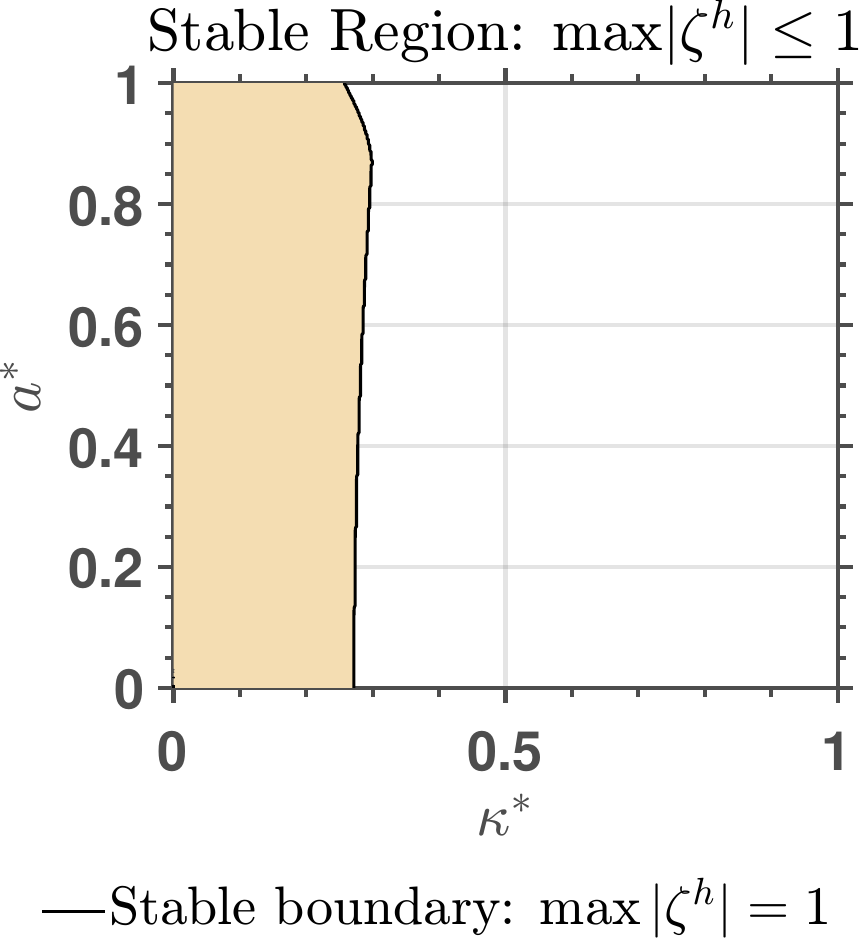} \\
& RK$(1,1)$-VMS & RK$(2,2)$-VMS & RK$(3,3)$-VMS & RK$(4,4)$-VMS \\[0.5em]
\raisebox{1\height}{\rotatebox{90}{$\tau^*=0.5$}} &
\includegraphics[width=0.22\linewidth, trim=0 30 0 32, clip]{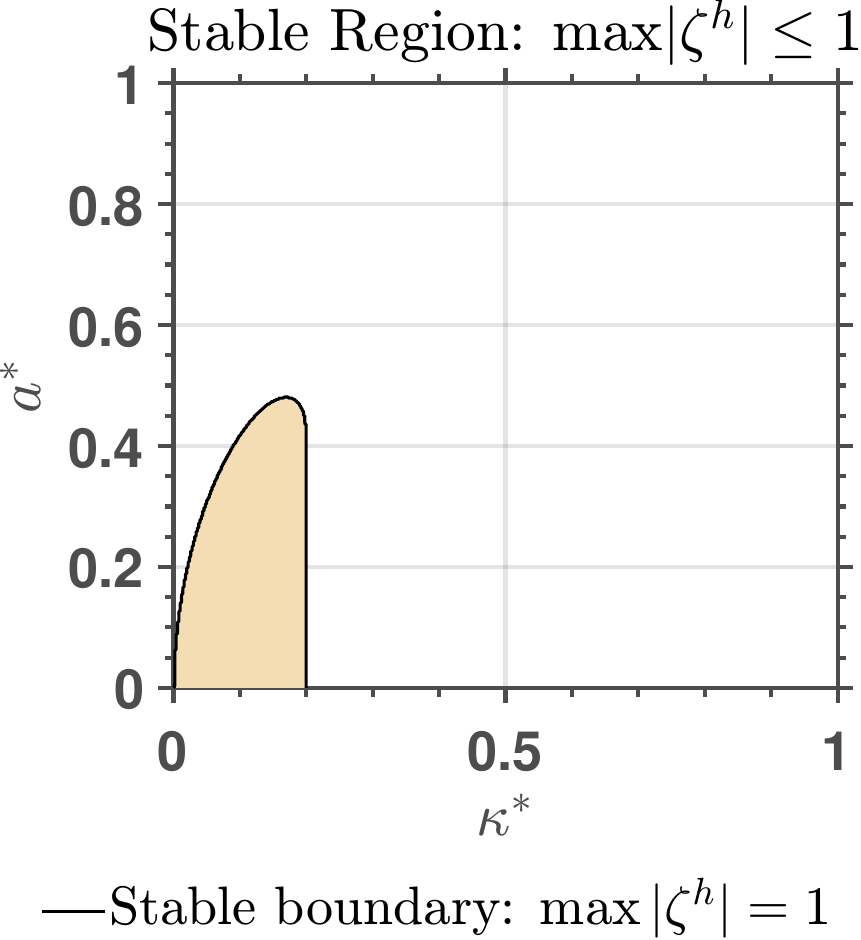} &
\includegraphics[width=0.22\linewidth, trim=0 30 0 32, clip]{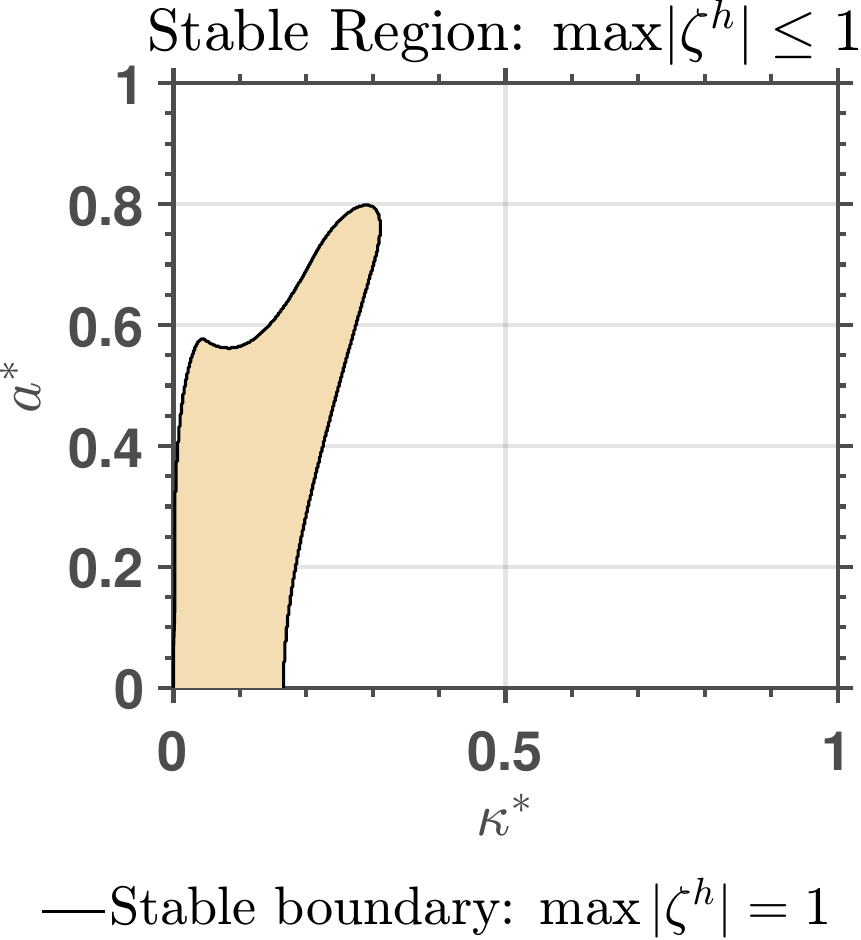} &
\includegraphics[width=0.22\linewidth, trim=0 30 0 32, clip]{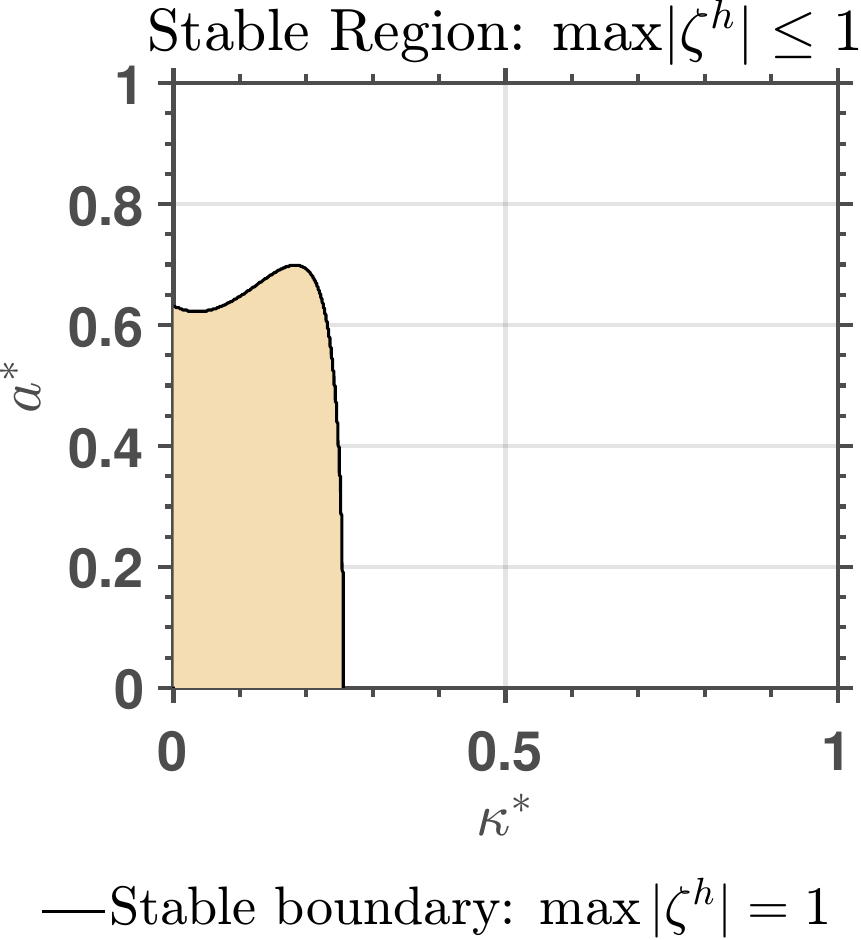} &
\includegraphics[width=0.22\linewidth, trim=0 30 0 32, clip]{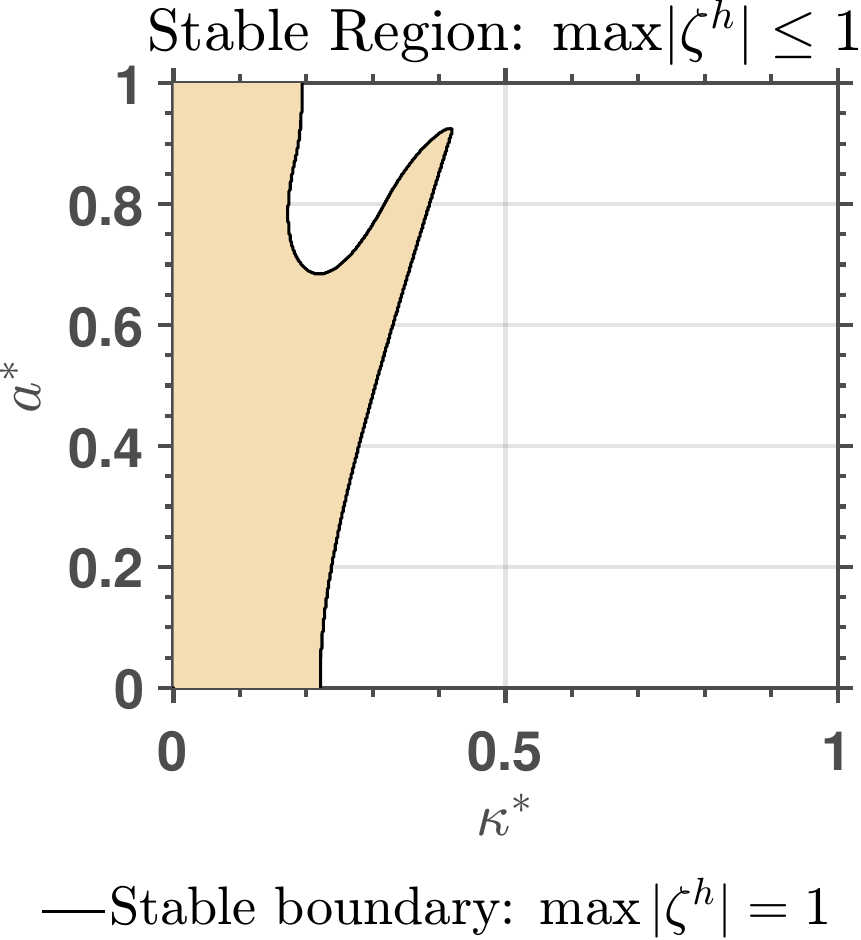} \\
& RK$(1,1)$-VMS & RK$(2,2)$-VMS & RK$(3,3)$-VMS & RK$(4,4)$-VMS \\[0.5em]
\raisebox{1\height}{\rotatebox{90}{$\tau^*=0.9$}} &
\includegraphics[width=0.22\linewidth, trim=0 30 0 32, clip]{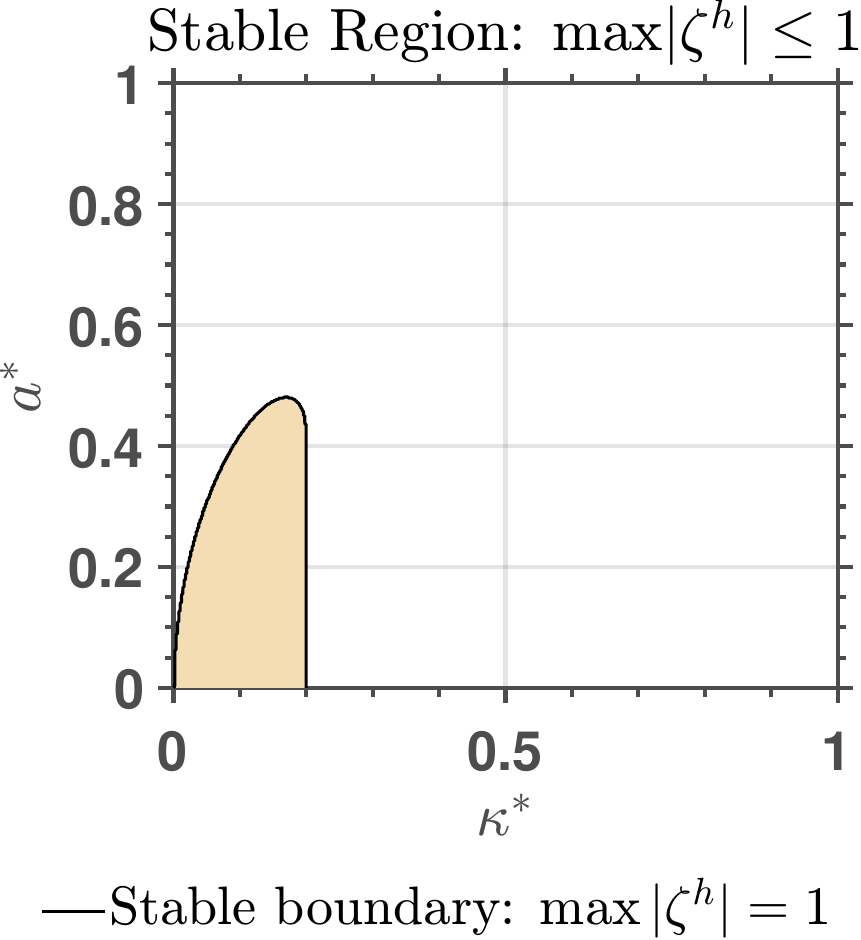} &
\includegraphics[width=0.22\linewidth, trim=0 30 0 32, clip]{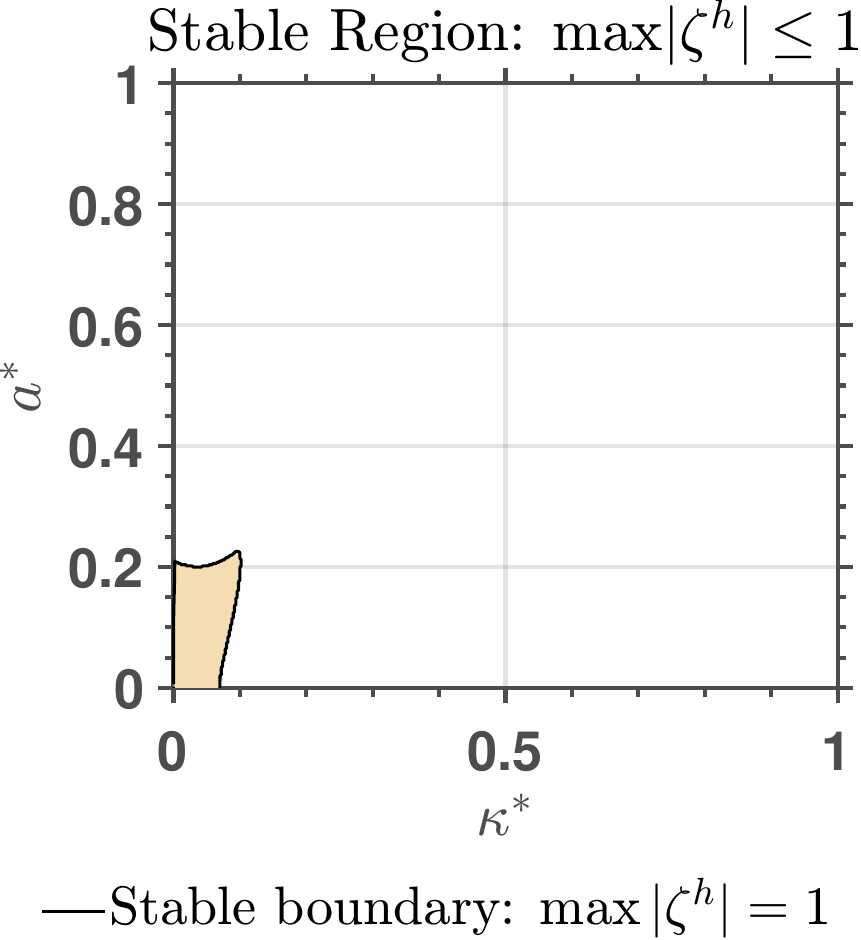} &
\includegraphics[width=0.22\linewidth, trim=0 30 0 32, clip]{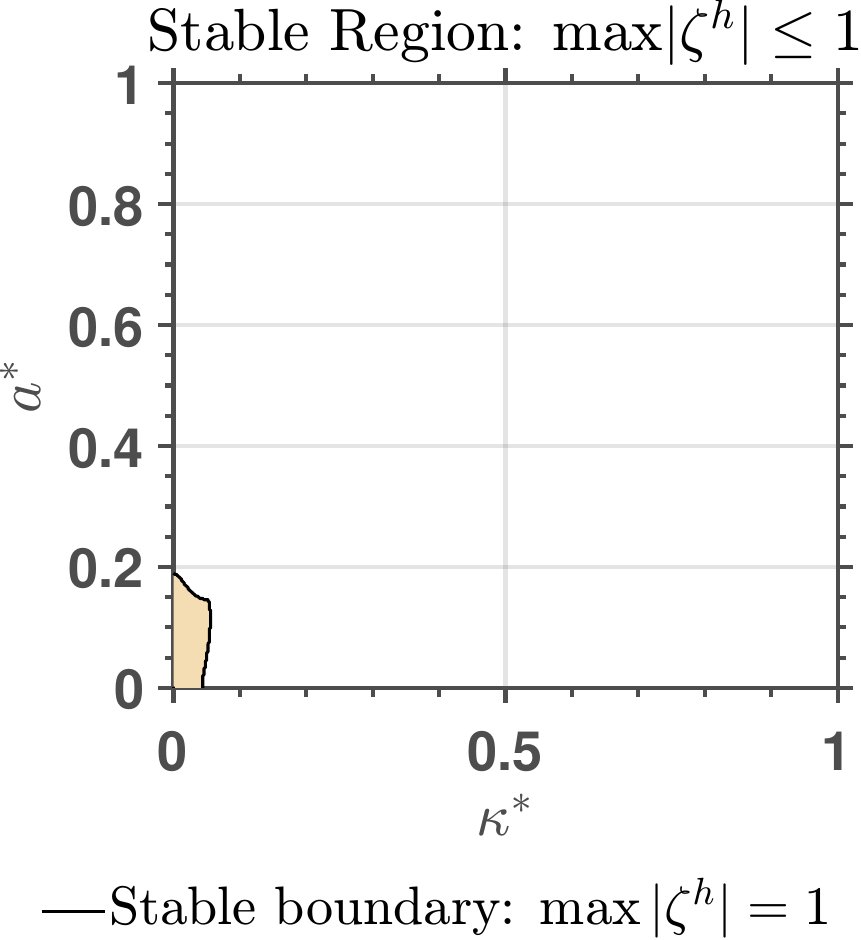} &
\includegraphics[width=0.22\linewidth, trim=0 30 0 32, clip]{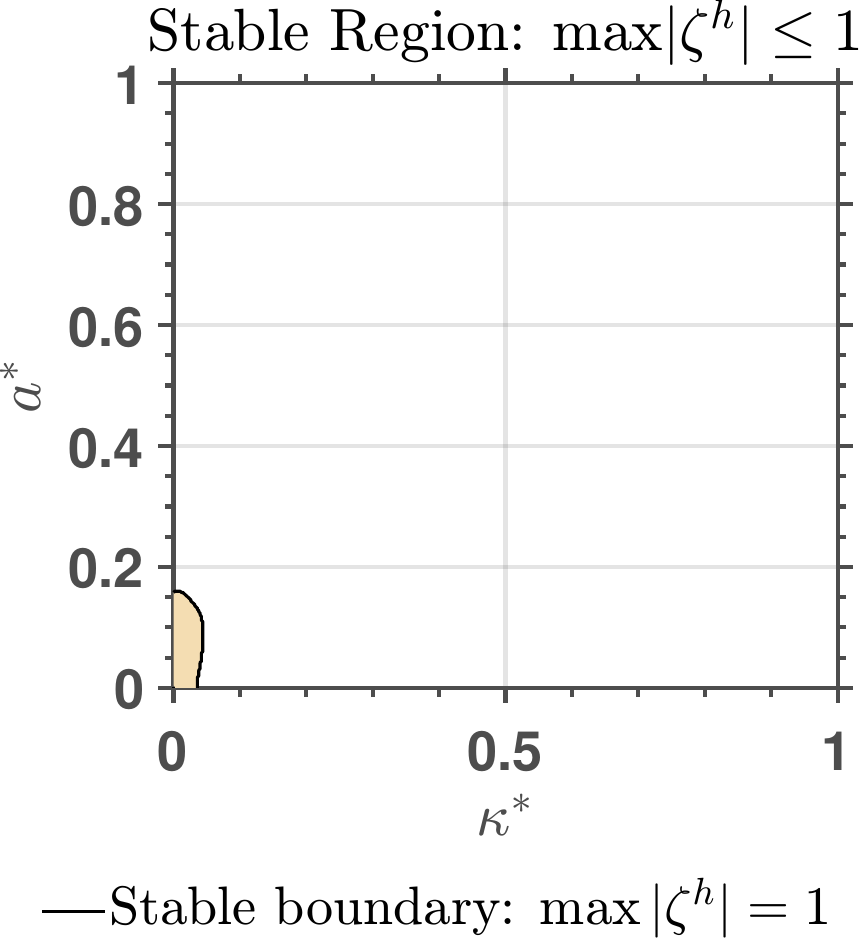} \\
& RK$(1,1)$-VMS & RK$(2,2)$-VMS & RK$(3,3)$-VMS & RK$(4,4)$-VMS \\[0.5em]
& 
\includegraphics[width=0.22\linewidth, trim=0 30 0 32, clip]{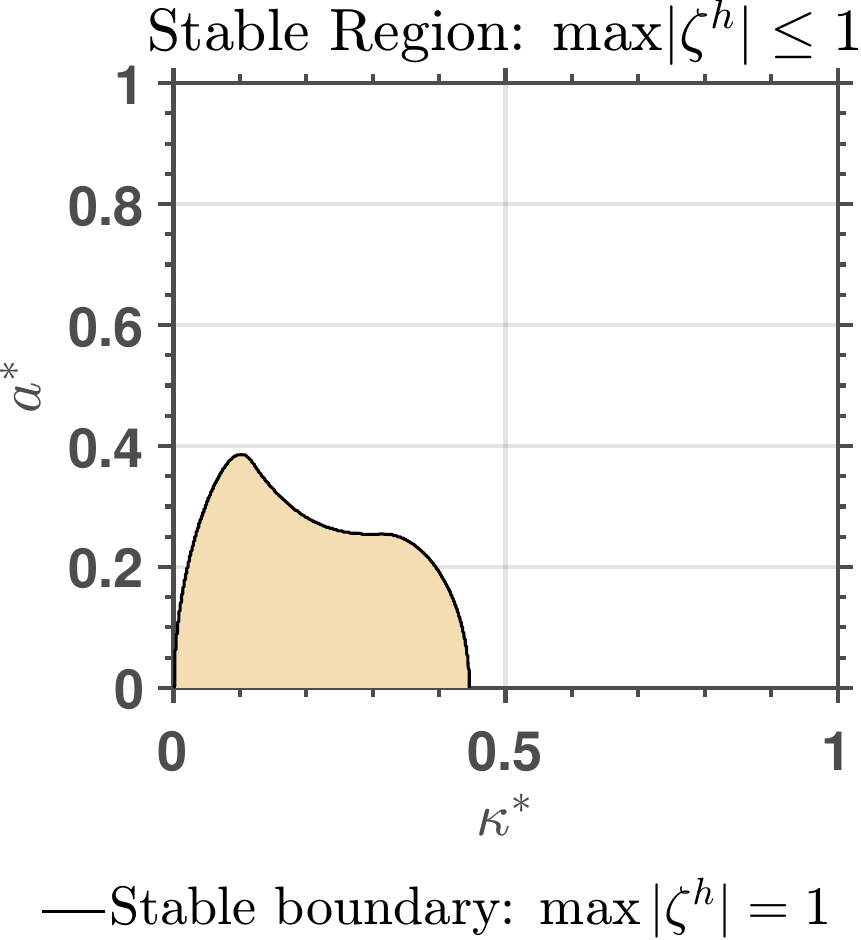} &
\includegraphics[width=0.22\linewidth, trim=0 30 0 32, clip]{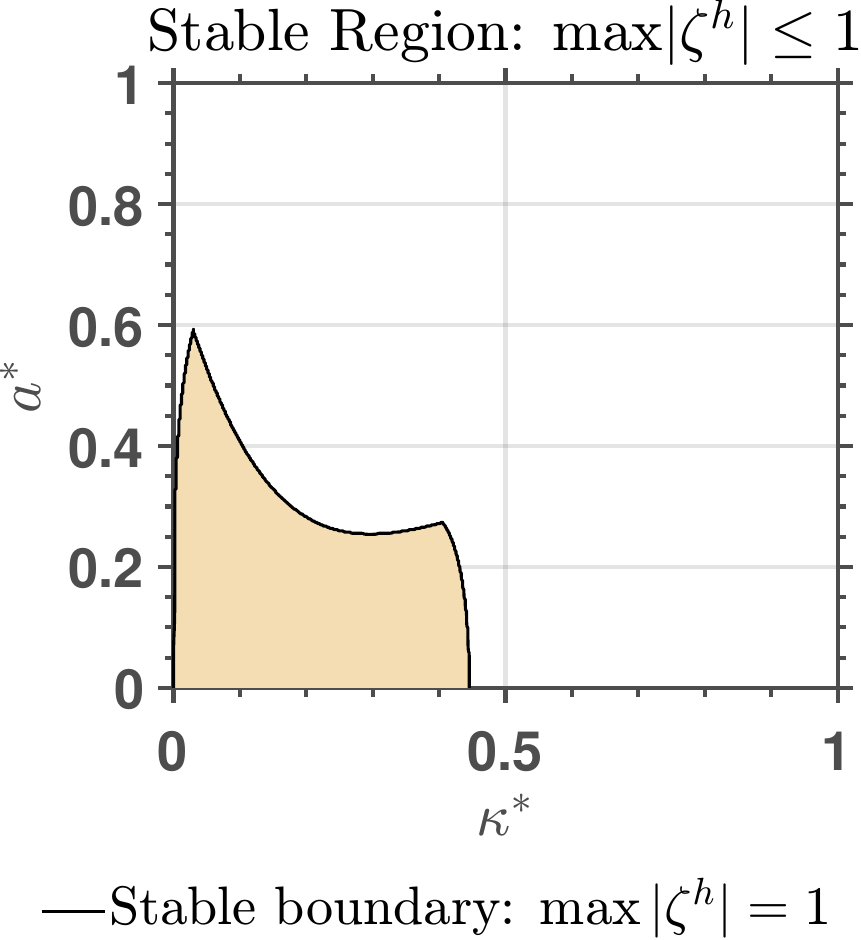} &
\includegraphics[width=0.22\linewidth, trim=0 30 0 32, clip]{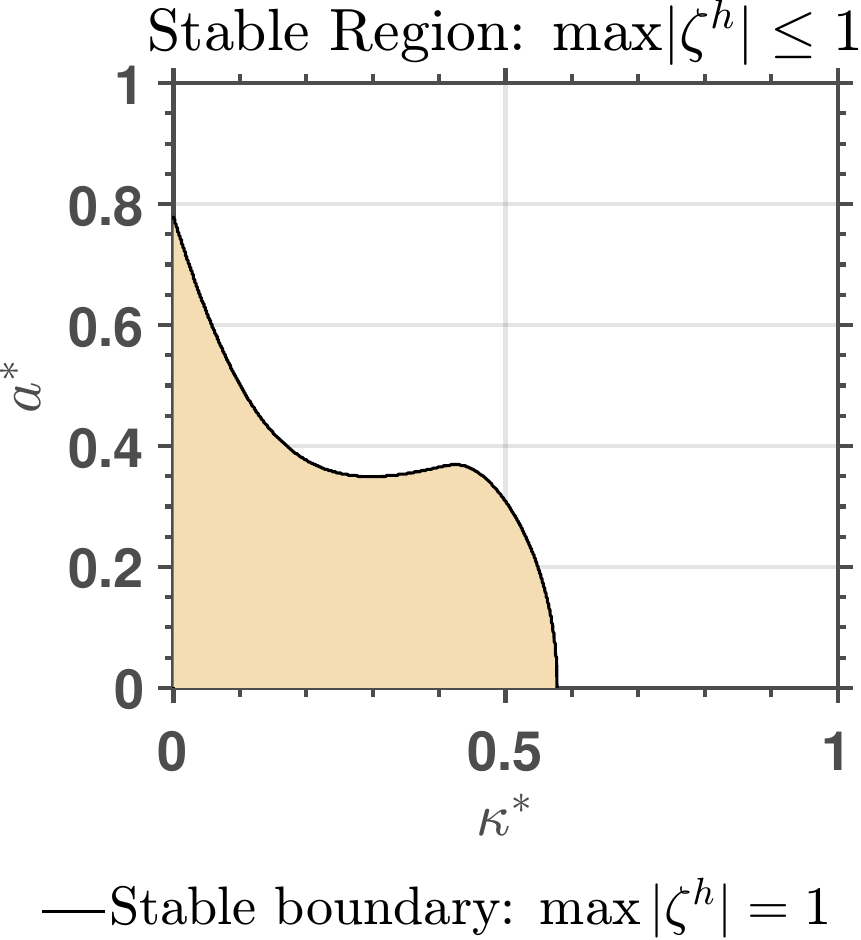} &
\includegraphics[width=0.22\linewidth, trim=0 30 0 32, clip]{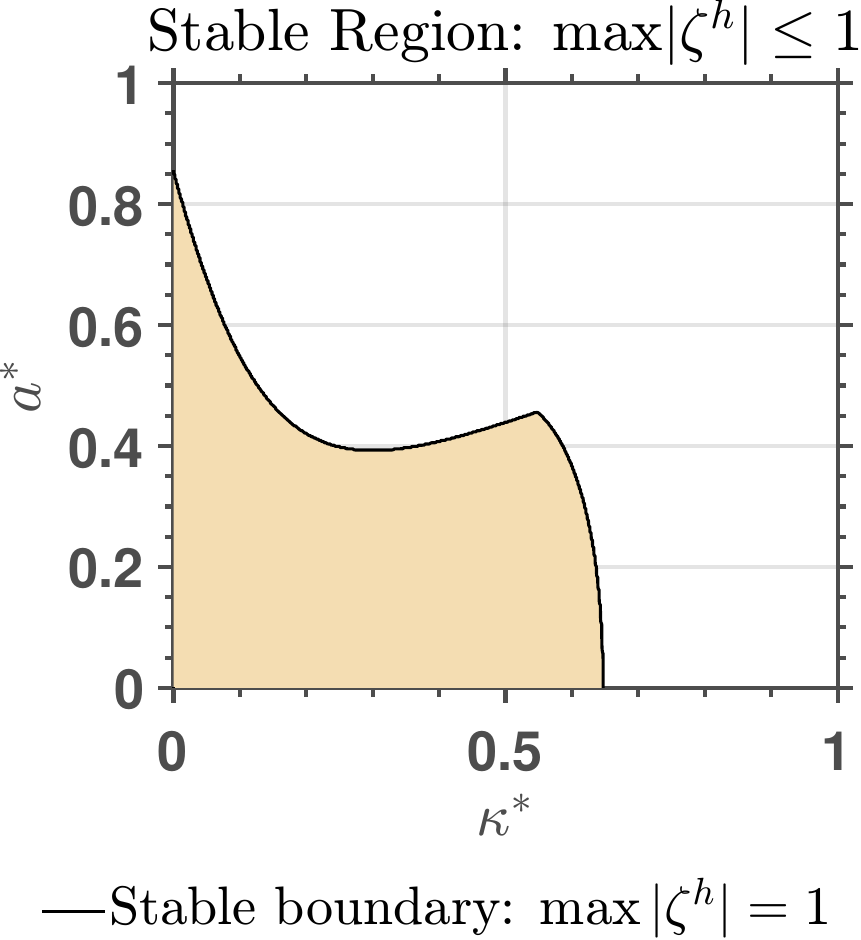} \\
& VMS-RK$(1,1)$ & VMS-RK$(2,2)$ & VMS-RK$(3,3)$ & VMS-RK$(4,4)$ \\
\end{tabular}
\caption{Stability regions for RK$(s,p)$-VMS and VMS-RK$(s,p)$ schemes.}
\label{fig:stability_region}
\end{figure}

With the amplification factors established, we now analyze the stability, which requires that the modulus of the amplification factor satisfies $|\zeta^h| \leq 1$ for all nondimensional wavenumbers $K^* \in [0,\pi]$. For the considered problem, the amplification factors also depend on $a^*$ and $\kappa^*$. Figure \ref{fig:stability_region} depicts the stability domain for $a^*\in [0,1]$ and $\kappa^* \in [0,1]$, where $|\zeta^h| \leq 1$ for all wavenumbers $K^*$. The black contour indicates the boundary of these domains. For the RK$(s,p)$-VMS scheme, stability regions are plotted for $\tau^* = 0.1$, $0.5$, and $0.9$, respectively.

As shown in Figure \ref{fig:stability_region}, the stability depends on both the order of the RK scheme and the choice of the parameter $\tau^*$. In the case of RK$(1,1)$-VMS, the fine-scale model does not contribute to the final discrete formulation, due to the use of the forward Euler method. Consequently, the stability remains unaffected by $\tau^*$. However, for higher-order schemes, increasing $\tau^*$ from $0.1$ to $0.9$ progressively reduces the stability domain. This observation is consistent with the stability analysis made in Section \ref{sec:VMS formulation}. Setting $\tau^*=1$ leads to a complete loss of stability. Except for the extreme case $\tau^*=0.9$, increasing the order of the RK methods results in a gradual expansion of the stability region. Figure \ref{fig:stability_region} also reveals that the reversal of the spatiotemporal discretization induces marked differences in the stability behavior. For a given value of $a^*$ within the stability region, the RK$(s,p)$-VMS schemes maintain stability at smaller values of $\kappa^*$, whereas for a fixed $\kappa^*$ in the stability domain, they allow larger values of $a^*$. This contrast is particularly evident when comparing RK$(4,4)$-VMS with VMS-RK$(4,4)$, where the stability regions exhibit different shapes on the $a^*$-$\kappa^*$ plane. Overall, the stability region of the RK$(s,p)$-VMS schemes extends further along the $a^*$ direction, indicating enhanced robustness in convection-dominated regimes, while the VMS-RK$(s,p)$ schemes are more amenable to diffusion-dominated regimes.

\subsubsection{Dispersion and dissipation}
With the amplification factor, it follows that $\nu^h = -\xi^h + \iota \omega^h = \ln(\zeta^h)/\Delta t$. The algorithmic damping and frequency can be obtained by $\xi^h = -\mathrm{Re}(\ln(\zeta^h)/\Delta t)$ and $\omega^h = \mathrm{Im}(\ln(\zeta^h)/\Delta t)$. The algorithmic ratios $\xi^h/\xi$ and $\omega^h / \omega$ are functions of $K^*$, $a^*$, and $\kappa^*$, and they measure the amount of numerical damping and dispersion, respectively. The accuracy of the discrete schemes can be assessed by expanding the expressions of $\xi^{h}$ and $\omega^{h}$ as power series of $\Delta t$ and $\Delta x$, where the leading-order terms reveal the asymptotic behavior of the methods. The temporal and spatial accuracy of the two schemes is summarized in Tables \ref{tab:accuracy_rk-vms} and \ref{tab:accuracy_vms-rk}. 

It can be observed that, for all schemes considered, the constant terms in the series expansions of $\xi^{h}$ and $\omega^{h}$ coincide with the exact expressions $\xi = \kappa K^{2}$ and $\omega = -aK$, confirming the consistency of the schemes. Both RK$(s,p)$-VMS and VMS-RK$(s,p)$ schemes attain the expected temporal accuracy of order $p$. However, the two schemes exhibit distinct spatial behaviors. For the RK$(s,p)$-VMS schemes, the leading spatial error terms of $\xi^{h}$ and $\omega^{h}$ are $\mathcal{O}(\Delta x^{4})$ and $\mathcal{O}(\Delta x^{6})$, respectively.  We notice that, for higher-order schemes ($p=2,3,4$), additional mixed contributions appear, such as $\Delta x^{2}\Delta t$, $\Delta x^{2}\Delta t^{2}$, and $\Delta x^{4}\Delta t$, which are also observed in the Fourier analyses of implicit stabilized formulations \cite{Dettmer2003}. For the VMS-RK$(s,p)$ schemes, there are additional mixed terms involving the parameter $\tau_{\diamond}$ (e.g., $\Delta x^{2}\tau_{\diamond}$, $\Delta x^{2}\tau_{\diamond}^{2}$, and $\Delta x^{4}\tau_{\diamond}$). As a result, their spatial accuracy depends on $\tau_{\diamond}$. In specific, when $\Delta t \ll \Delta x$, the schemes recover fourth- and sixth-order accuracy in $\xi^{h}$ and $\omega^{h}$, respectively; when $\Delta x \ll \Delta t$, the accuracy reduces to third and fourth order in advection-dominated conditions, while in diffusion-dominated regimes, the asymptotic fourth- and sixth-order behavior is restored. This dependence on whether the problem is advection- or diffusion-dominated has also been reported in previous analyses of implicit stabilized formulations \cite{Shakib1991a}.

\begin{table}[htbp]
	\centering
	\tabcolsep=0.13cm
	\renewcommand{\arraystretch}{1.3}
	\caption{Accuracy orders of RK$(s,p)$-VMS schemes, showing the lowest-order terms in $\Delta x$, $\Delta t$, and their mixed contributions.}
	\label{tab:accuracy_rk-vms}
	\begin{tabular}{@{}P{2.5cm} P{4.5cm}  P{4.75cm} @{}}
		\toprule
		Scheme & $\xi^{h}-\xi$ & $\omega^{h}-\omega$ \\
		\midrule
		RK$(1,1)$-VMS & $\mathcal{O}(\Delta t, \Delta x^{4})$ & $\mathcal{O}(\Delta t, \Delta x^{6})$ \\
		RK$(2,2)$-VMS & $\mathcal{O}(\Delta t^{2}, \Delta x^{4}, \Delta x^{2}\Delta t)$ & $\mathcal{O}(\Delta t^{2}, \Delta x^{6}, \Delta x^{4}\Delta t)$ \\
		RK$(3,3)$-VMS & $\mathcal{O}(\Delta t^{3},  \Delta x^{4},  \Delta x^{2} \Delta t)$ & $\mathcal{O}(\Delta t^{3},  \Delta x^{6},  \Delta x^{2} \Delta t^{2}, \Delta x^{4}\Delta t)$\\
		RK$(4,4)$-VMS & $\mathcal{O}(\Delta t^{4},  \Delta x^{4},  \Delta x^{2} \Delta t)$ & $\mathcal{O}(\Delta t^{4},  \Delta x^{6},  \Delta x^{2} \Delta t^{2}, \Delta x^{4}\Delta t)$\\
		\bottomrule
	\end{tabular}
\end{table}

\begin{table}[htbp]
	\centering
	\tabcolsep=0.13cm
	\renewcommand{\arraystretch}{1.3}
	\caption{Accuracy orders of VMS-RK$(s,p)$ schemes, showing the lowest-order terms in $\Delta x$, $\Delta t$, $\tau_{\diamond}$, and their mixed contributions.}
	\label{tab:accuracy_vms-rk}
	\begin{tabular}{@{}P{2.5cm} P{4.5cm}  P{4.5cm} @{}}
		\toprule
		Scheme & $\xi^{h}-\xi$ & $\omega^{h}-\omega$ \\
		\midrule
		VMS-RK$(1,1)$ & $\mathcal{O}(\Delta t, \Delta x^{4}, \Delta x^{2}\tau_{\diamond})$ & $\mathcal{O}(\Delta t,  \Delta x^{6}, \Delta x^{2}\tau_{\diamond}^{2}, \Delta x^{4}\tau_{\diamond})$ \\
		VMS-RK$(2,2)$ & $\mathcal{O}(\Delta t^{2}, \Delta x^{4}, \Delta x^{2}\tau_{\diamond})$ & $\mathcal{O}(\Delta t^{2}, \Delta x^{6}, \Delta x^{2}\tau_{\diamond}^{2}, \Delta x^{4}\tau_{\diamond})$ \\
		VMS-RK$(3,3)$ & $\mathcal{O}(\Delta t^{3}, \Delta x^{4}, \Delta x^{2}\tau_{\diamond})$ &  $\mathcal{O}(\Delta t^{3}, \Delta x^{6}, \Delta x^{2}\tau_{\diamond}^{2}, \Delta x^{4}\tau_{\diamond})$ \\
		VMS-RK$(4,4)$ & $\mathcal{O}(\Delta t^{4}, \Delta x^{4}, \Delta x^{2}\tau_{\diamond})$ &  $\mathcal{O}(\Delta t^{4}, \Delta x^{6}, \Delta x^{2}\tau_{\diamond}^{2}, \Delta x^{4}\tau_{\diamond})$ \\
		\bottomrule
	\end{tabular}
\end{table}

\begin{figure}[htbp]
\centering
\begin{tabular}{cc}
\includegraphics[width=0.3\linewidth, trim=0 30 0 0, clip]{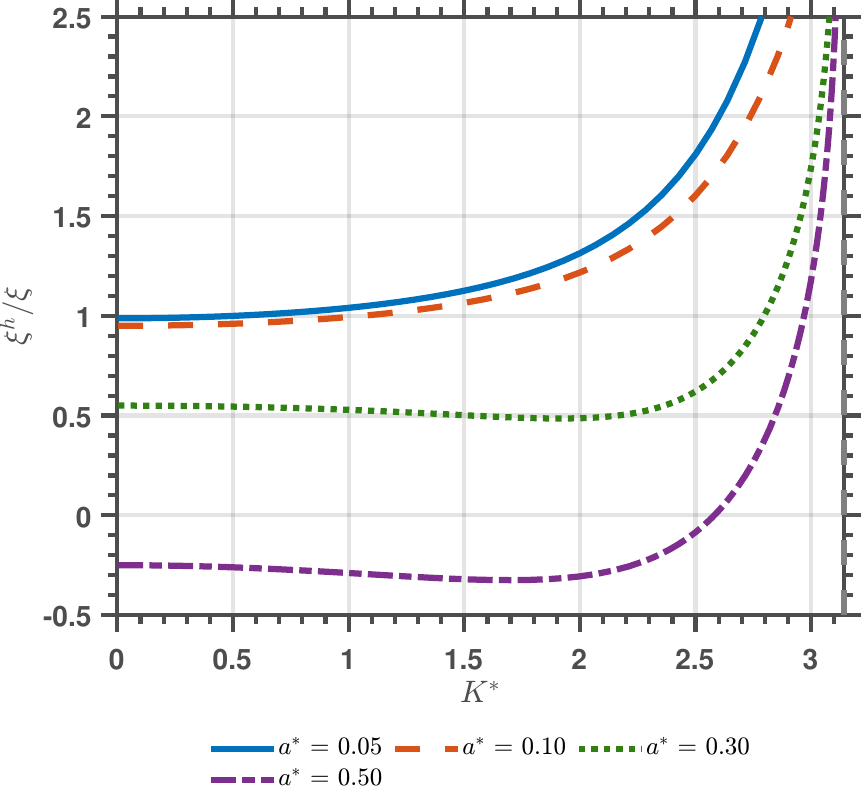} &
\includegraphics[width=0.3\linewidth, trim=0 30 0 0, clip]{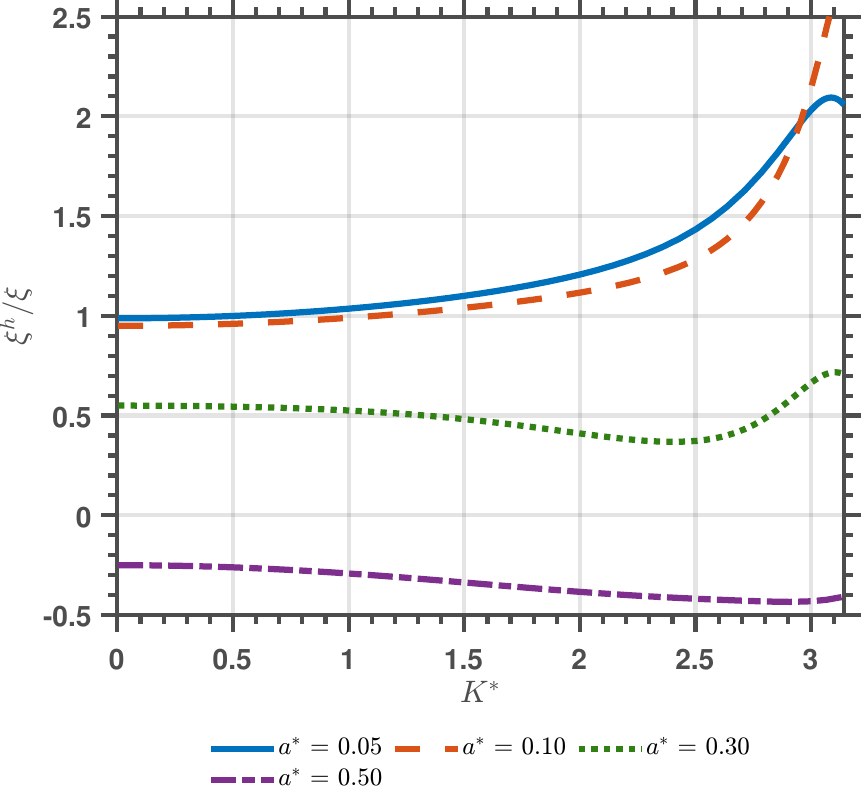} \\
RK$(1,1)$-VMS & VMS-RK$(1,1)$  \\[0.5em]
\includegraphics[width=0.3\linewidth, trim=0 30 0 0, clip]{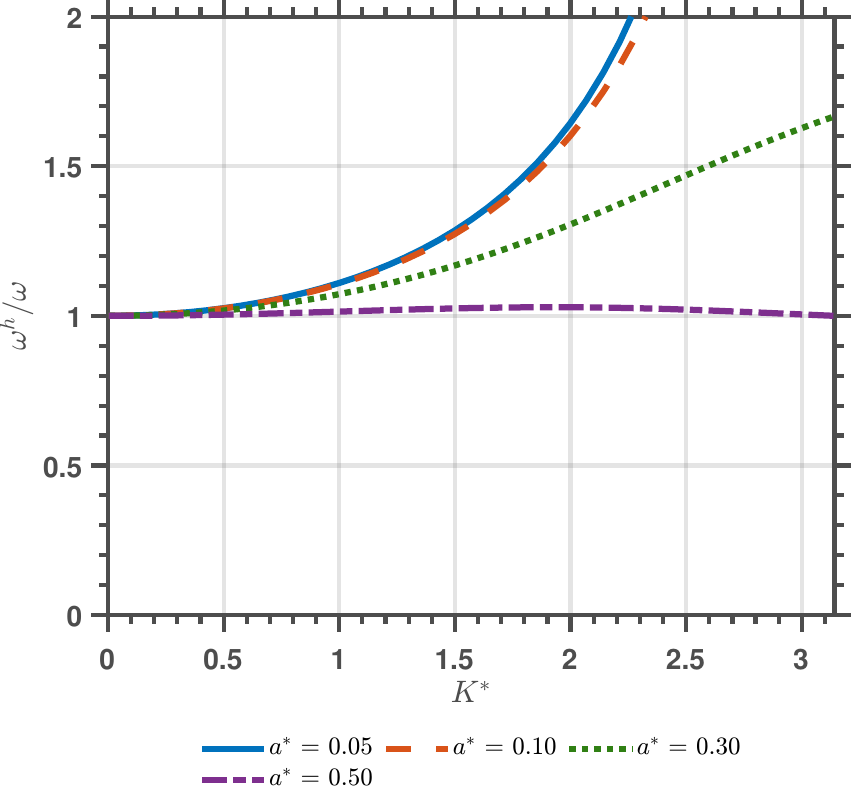} &
\includegraphics[width=0.3\linewidth, trim=0 30 0 0, clip]{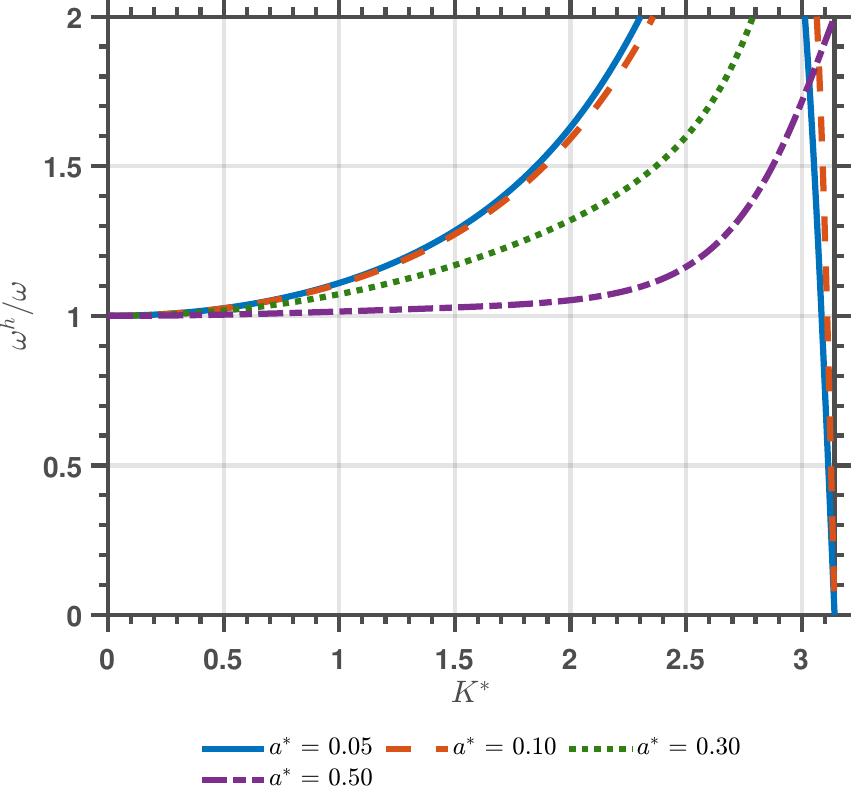}  \\
RK$(1,1)$-VMS & VMS-RK$(1,1)$  \\[0.5em]
\end{tabular}
\includegraphics[width=0.5\linewidth, trim=0 0 0 0, clip]{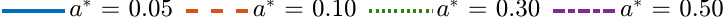}  \\[-0.5em]
\caption{Algorithmic damping (top) and frequency ratios (bottom) for schemes with $s=p=1$. }
\label{fig:comparison of vms_RK1 and RK1_vms}
\end{figure}

\begin{figure}[htbp]
\centering
\begin{tabular}{cc}
\includegraphics[width=0.3\linewidth, trim=0 30 0 0, clip]{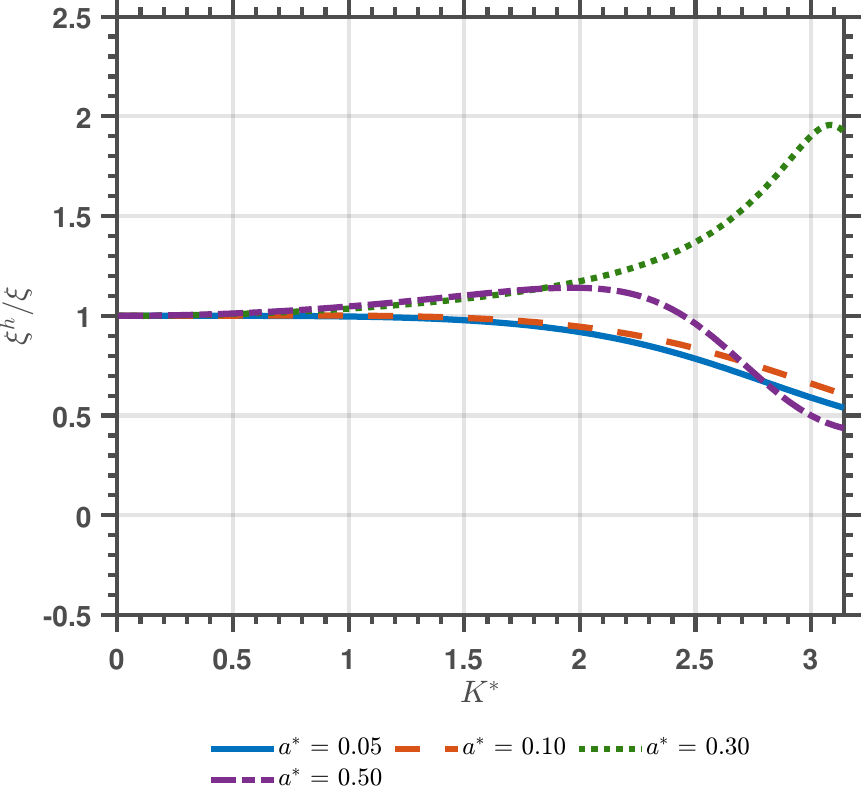} &
\includegraphics[width=0.3\linewidth, trim=0 30 0 0, clip]{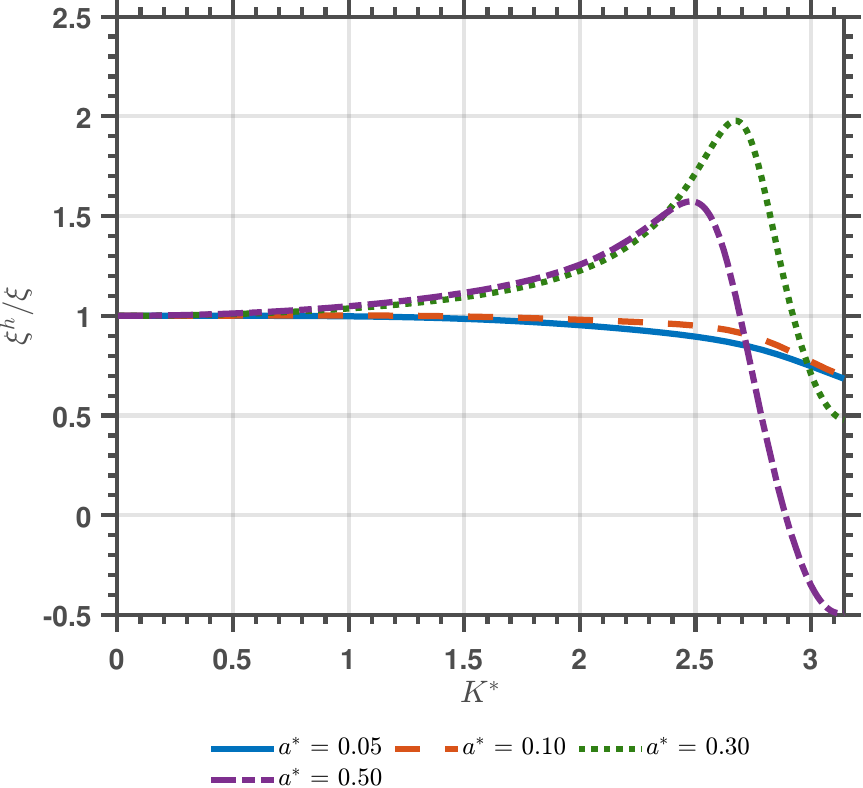} \\
RK$(2,2)$-VMS & VMS-RK$(2,2)$  \\[0.5em]
\includegraphics[width=0.3\linewidth, trim=0 30 0 0, clip]{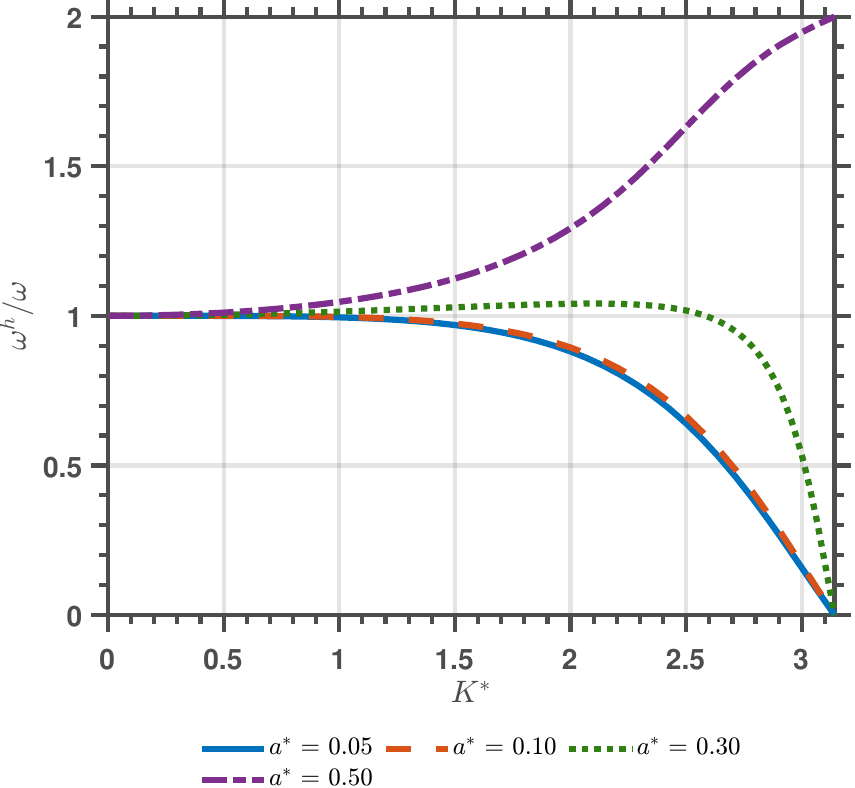} &
\includegraphics[width=0.3\linewidth, trim=0 30 0 0, clip]{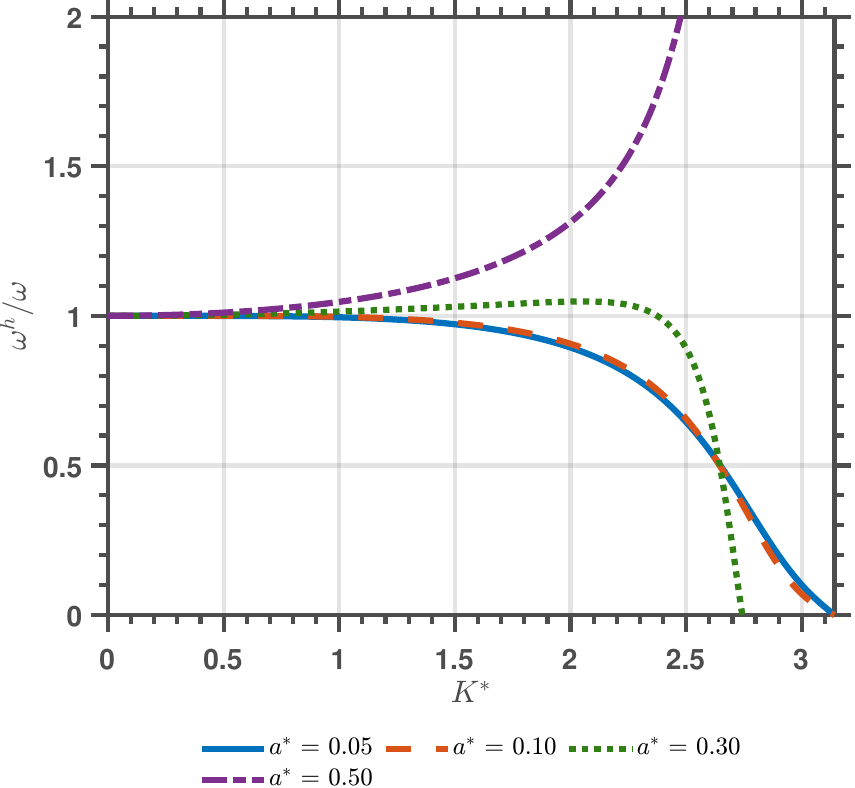}  \\
RK$(2,2)$-VMS & VMS-RK$(2,2)$ \\[0.5em]
\end{tabular}
\includegraphics[width=0.5\linewidth, trim=0 0 0 0, clip]{./numerical_dissipation_and_dispersion_legend.pdf}  \\[-0.5em]
\caption{Algorithmic damping (top) and frequency ratios (bottom) for schemes with $s=p=2$. }
\label{fig:comparison of vms_RK2 and RK2_vms}
\end{figure}

\begin{figure}[htbp]
\centering
\begin{tabular}{cc}
\includegraphics[width=0.3\linewidth, trim=0 30 0 0, clip]{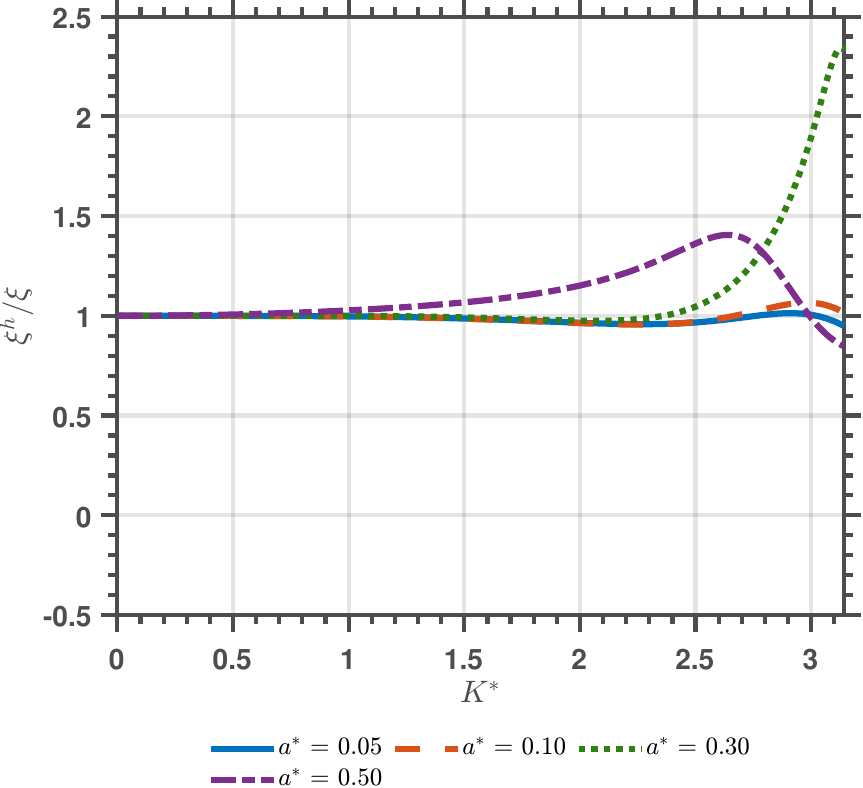} &
\includegraphics[width=0.3\linewidth, trim=0 30 0 0, clip]{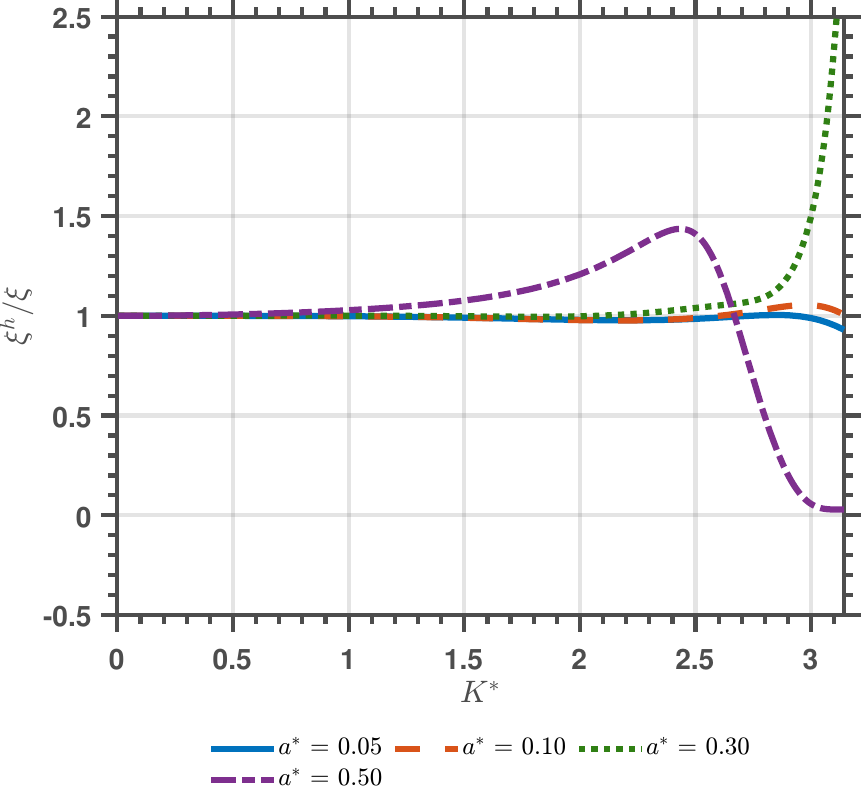} \\
RK$(3,3)$-VMS & VMS-RK$(3,3)$  \\[0.5em]
\includegraphics[width=0.3\linewidth, trim=0 30 0 0, clip]{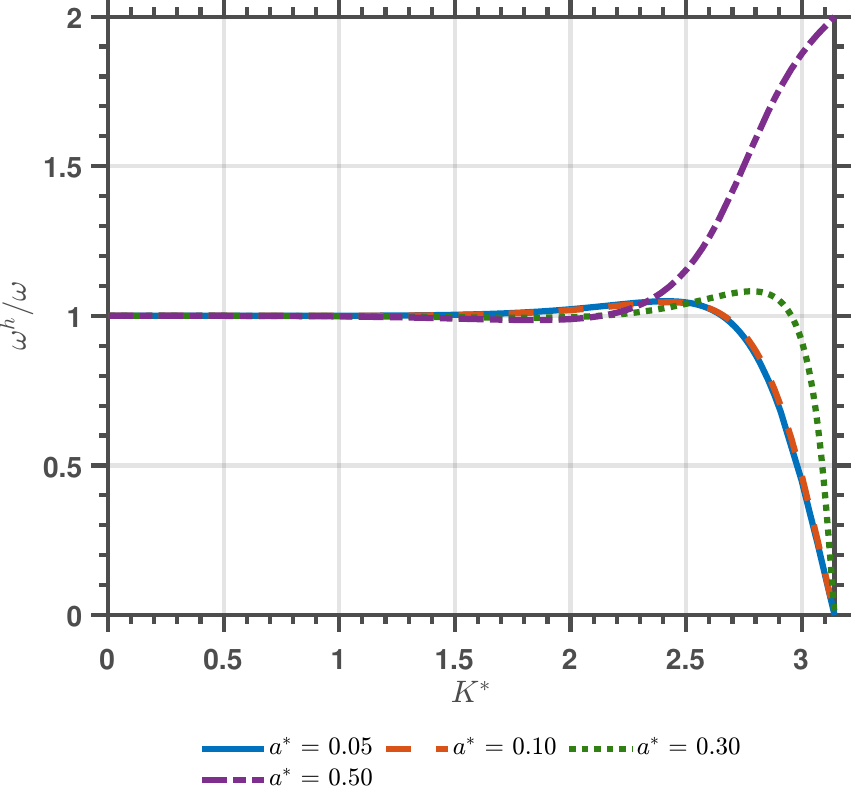} &
\includegraphics[width=0.3\linewidth, trim=0 30 0 0, clip]{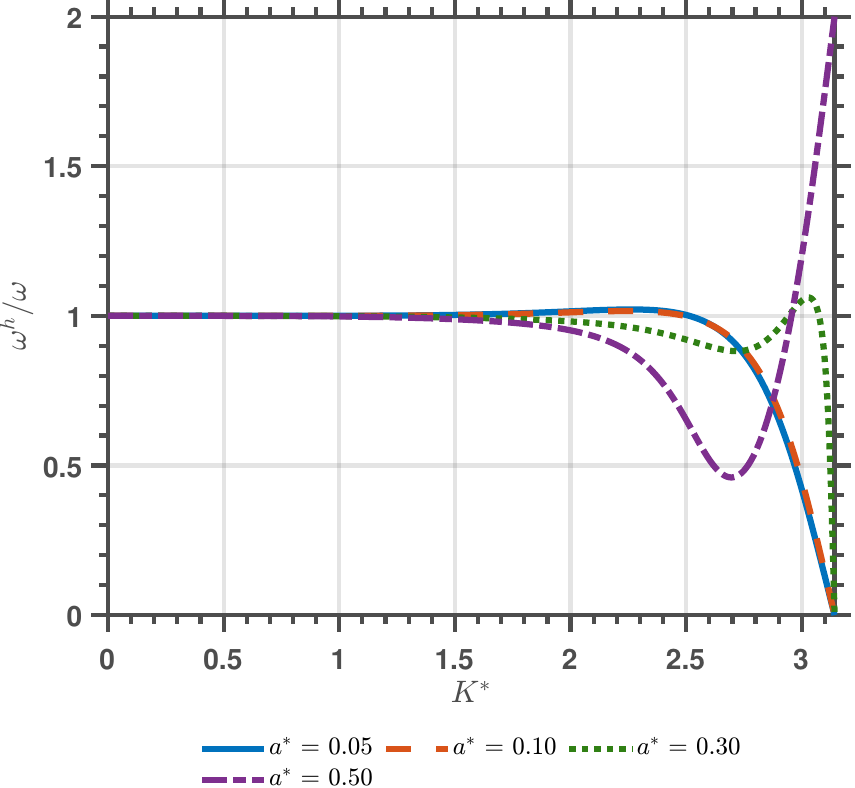}  \\
RK$(3,3)$-VMS & VMS-RK$(3,3)$ \\[0.5em]
\end{tabular}
\includegraphics[width=0.5\linewidth, trim=0 0 0 0, clip]{./numerical_dissipation_and_dispersion_legend.pdf}  \\[-0.5em]
\caption{Algorithmic damping (top) and frequency ratios (bottom) for schemes with $s=p=3$. }
\label{fig:comparison of vms_RK3 and RK3_vms}
\end{figure}

\begin{figure}[htbp]
\centering
\begin{tabular}{cc}
\includegraphics[width=0.3\linewidth, trim=0 30 0 0, clip]{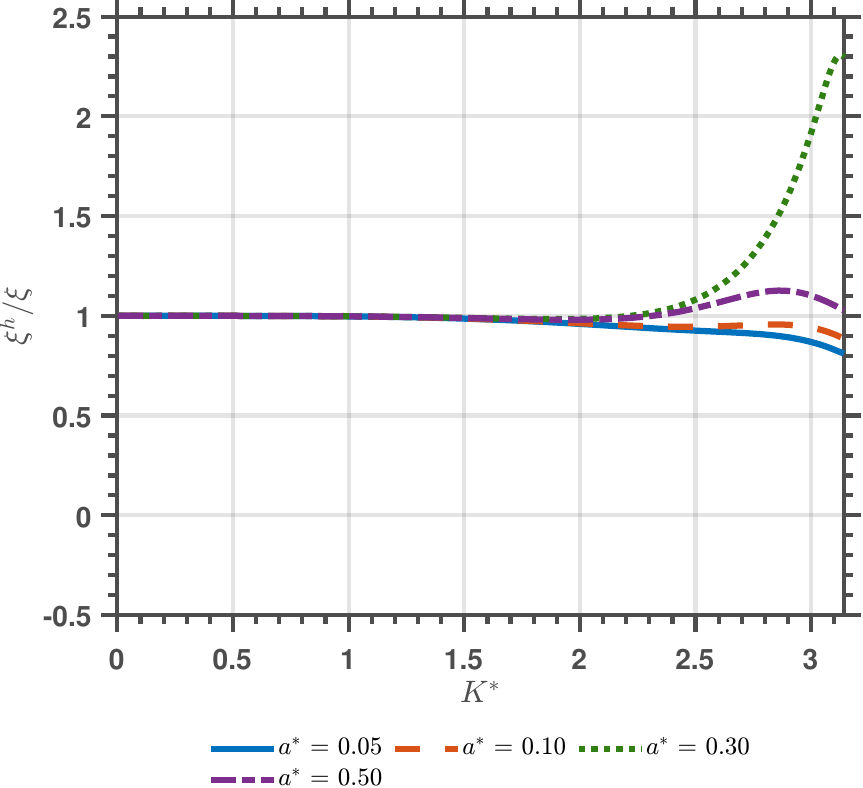} &
\includegraphics[width=0.3\linewidth, trim=0 30 0 0, clip]{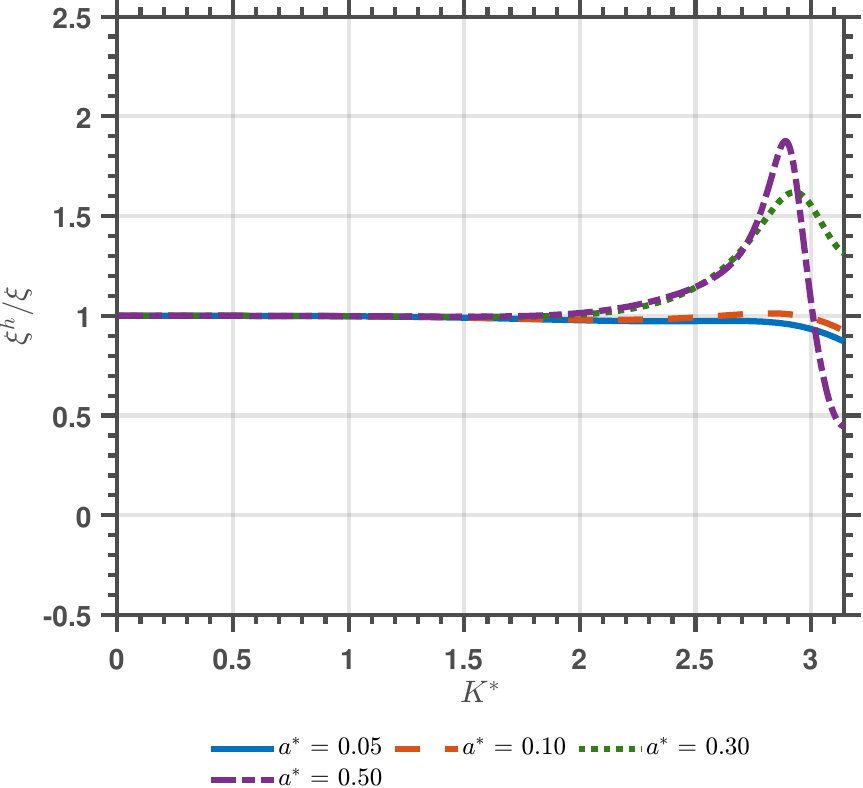} \\
RK$(4,4)$-VMS & VMS-RK$(4,4)$ \\[0.5em]
\includegraphics[width=0.3\linewidth, trim=0 30 0 0, clip]{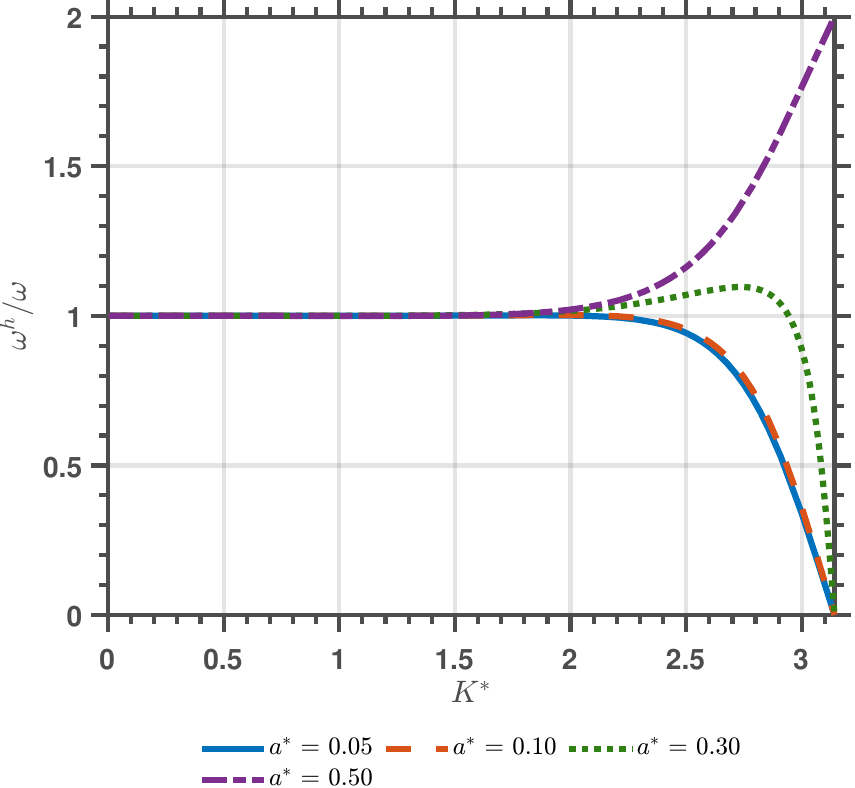} &
\includegraphics[width=0.3\linewidth, trim=0 30 0 0, clip]{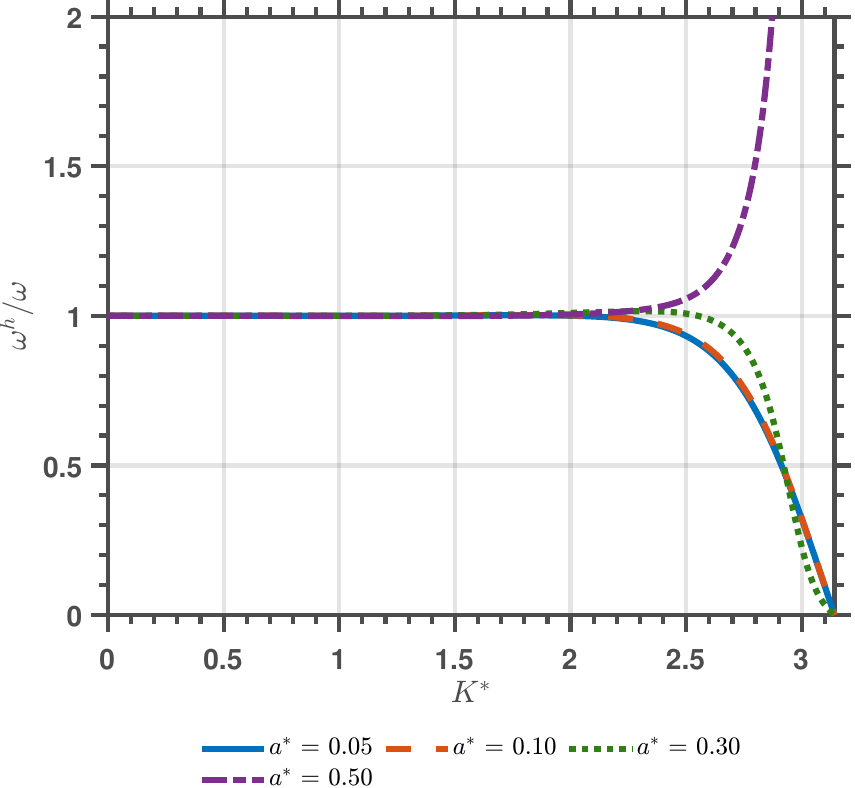}  \\
RK$(4,4)$-VMS & VMS-RK$(4,4)$ \\[0.5em]
\end{tabular}
\includegraphics[width=0.5\linewidth, trim=0 0 0 0, clip]{./numerical_dissipation_and_dispersion_legend.pdf}  \\[-0.5em]
\caption{Algorithmic damping (top) and frequency ratios (bottom) for schemes with $s=p=4$. }
\label{fig:comparison of vms_RK4 and RK4_vms}
\end{figure}

The dissipation and dispersion properties of the RK$(s,p)$-VMS and VMS-RK$(s,p)$ schemes are assessed, with the parameter set to $\tau^{*}=1/2$ for the RK$(s,p)$-VMS scheme. The nondimensional diffusivity is fixed as $\kappa^* = 0.1$, a relatively small value that reflects the strongly convective nature of turbulent flows. The nondimensional advective speed is varied with values $a^* = 0.05$, $0.1$, $0.3,$ and $0.5$ to examine a broad range of Courant numbers. As indicated by Figures \ref{fig:stability_region}, most schemes remain stable for the considered values of $\kappa^*$ and $a^*$, with the exceptions being schemes of $p=1$ and the VMS-RK$(2,2)$ scheme, which lose stability at $a^*=0.5$. This can be confirmed from Figures \ref{fig:comparison of vms_RK1 and RK1_vms} and \ref{fig:comparison of vms_RK2 and RK2_vms}, where negative numerical dissipation $\xi^h$ is observed. As shown in Figures \ref{fig:comparison of vms_RK1 and RK1_vms}-\ref{fig:comparison of vms_RK4 and RK4_vms}, the two schemes with different orders exhibit nearly identical behavior in the low-wavenumber regime, while differences arise at high wavenumbers. For $p=1$, the VMS-RK$(1,1)$ scheme exhibits slightly reduced numerical dissipation compared with RK$(1,1)$-VMS, but the latter shows more accurate dispersion behavior at large values of $a^*$. For higher-order schemes ($p=2,3,4$), the two schemes yield nearly indistinguishable results for small values of $a^*$, indicating that the influence of the order of the space-time discretization diminishes when the advection is relatively weak. When the advective speed is large, the RK$(s,p)$-VMS schemes generally exhibit smaller dissipation and dispersion errors than the corresponding VMS-RK$(s,p)$ schemes. These results demonstrate that the RK$(s,p)$-VMS schemes offer improved spectral accuracy and robustness in convection-dominated regimes, making them well suited for flow problems.

\section{Implementation}
The VMS formulations \eqref{eq:herk_vms_coarse_closed} and \eqref{eq:herk_vms_pres_coarse_closed} determine the velocity and pressure fields at each time step. In the following, we derive the corresponding matrix systems and propose an iterative solver design by leveraging their structural properties.

\subsection{Consistent tangent matrix}
We consider the equation \eqref{eq:herk_vms_coarse_closed}, which represents the momentum and continuity equations at the $i$-th RK stage. We invoke equal-order interpolation and use $N_A$ to denote the basis function for the velocity and pressure fields. Recall that $\bm{e}_m$ stands for the $m$-th Cartesian basis vector. Choosing the test functions as $\lbrace N_A \bm{e}_m, 0 \rbrace$ and $\lbrace \bm 0, N_A \rbrace$, respectively, in \eqref{eq:herk_vms_coarse_closed}, we obtain the residual vectors as 
\begin{align*}
	\boldsymbol{\mathrm{R}}_i^{\mathrm{M}} &= \left[\mathrm{R}^{\mathrm{M}}_{i;A,m}\right], &&
	\mathrm{R}^{\mathrm{M}}_{i;A,m} := 	B^{\mathrm{VMS}}_{i}\left(\left\{N_A \bm{e}_m, 0\right\},\left\{\bar{\bm{u}}_i,\bar{p}_{n(i)} \right\}\right)-L^{\mathrm{VMS}}_{i}\left(\left\{N_A \bm{e}_m, 0\right\}\right), \nonumber \displaybreak[2] \\
	\boldsymbol{\mathrm{R}}_i^{\mathrm{C}} &= \left[\mathrm{R}^{\mathrm{C}}_{i;A}\right], &&
	\mathrm{R}^{\mathrm{C}}_{i;A} := B^{\mathrm{VMS}}_{i}\left(\left\{\bm{0}, N_A\right\},\left\{\bar{\bm{u}}_i,\bar{p}_{n(i)} \right\}\right)-L^{\mathrm{VMS}}_{i}\left(\left\{\bm{0}, N_A\right\}\right).
\end{align*}
Here, the superscripts $\mathrm M$ and $\mathrm C$ indicate that the vectors are related to the momentum and continuity equations, respectively. Solving the algebraic equations $\boldsymbol{\mathrm{R}}_i^{\mathrm{M}}=0$ and $\boldsymbol{\mathrm{R}}_i^{\mathrm{C}}=0$ leads to the velocity and pressure fields at the $i$-th stage. Due to their linearity, these equations can be solved in a single pass of a predictor-corrector algorithm. To describe the solution procedure, we denote $\mathsf{U}_i$ and $\mathsf P_i$ as the vectors of nodal or control point degrees of freedom corresponding to the velocity and pressure fields at the $i$-th stage, respectively. The consistent tangent matrix takes the form
\begin{align*}
	\boldsymbol{\mathrm K}_{i} :=
	\begin{bmatrix}
		\boldsymbol{\mathrm A}_{i} & \boldsymbol{\mathrm B}_{i} \\[0.3mm]
		\boldsymbol{\mathrm C}_{i} & \boldsymbol{\mathrm D}_{i}
	\end{bmatrix},
\end{align*}
with block components defined as
\begin{align}
\label{eq:Ki_block_matrix}
\boldsymbol{\mathrm{A}}_{i} =\frac{\partial \boldsymbol{\mathrm{R}}_i^{\mathrm{M}} \big( \mathsf{U}_i, \mathsf{P}_i\big)}{\partial \mathsf{U}_i}, \quad
\boldsymbol{\mathrm{B}}_{i} =\frac{\partial \boldsymbol{\mathrm{R}}_i^{\mathrm{M}} \big( \mathsf{U}_i, \mathsf{P}_i \big)}{\partial \mathsf{P}_i}, \quad
\boldsymbol{\mathrm{C}}_{i} =\frac{\partial \boldsymbol{\mathrm{R}}_i^{\mathrm{C}} \big( \mathsf{U}_i, \mathsf{P}_i \big)}{\partial \mathsf{U}_i}, \quad
\boldsymbol{\mathrm{D}}_{i} =\frac{\partial \boldsymbol{\mathrm{R}}_i^{\mathrm{C}} \big( \mathsf{U}_i, \mathsf{P}_i \big)}{\partial \mathsf{P}_i}.
\end{align}
The explicit formulas for the above block matrices are given as
\begin{align}
	\label{eq:Ai_block_matrix}
	\boldsymbol{\mathrm{A}}_{i} &= \left[\mathrm{A}^{mn}_{i;AB}\right], &&
	\mathrm{A}^{mn}_{i;AB} = \left(1-\tau\right)\frac{\rho}{\Delta t_n}\big(N_A, N_B \big) \delta_{mn}, \displaybreak[2] \\
	\label{eq:Bi_block_matrix}
	\boldsymbol{\mathrm{B}}_{i} &= \left[\mathrm{B}^{m}_{i;AB}\right], &&
	\mathrm{B}^{m}_{i;AB} = -{\alpha}_{ii}\big(N_{A,m}, N_B\big)  -{\alpha}_{ii}\tau\big(N_{A}, N_{B,m}\big), \displaybreak[2] \\
	\label{eq:Ci_block_matrix}
	\boldsymbol{\mathrm{C}}_{i} &= \left[\mathrm{C}^{m}_{i;AB}\right], &&
	\mathrm{C}^{m}_{i;AB} = \alpha_{ii} \big(N_A, N_{B,m} \big) + \alpha_{ii} \tau\big(N_{A,m}, N_B \big), \displaybreak[2] \\
	\label{eq:Di_block_matrix}
	\boldsymbol{\mathrm{D}}_{i} &= \left[\mathrm{D}_{i;AB}\right], &&
	\mathrm{D}_{i;AB} = {\alpha}^2_{ii}\tau \frac{\Delta t_n}{\rho} \sum_{l=1}^{3}\big( N_{A,l}, N_{B,l}\big),
\end{align}
where $\delta_{mn}$ denotes the Kronecker delta, and our choice for the parameter $\tau$ is $1/2$, as was given in \eqref{eq:tau_equal_1_2}. By inspecting the expressions, we notice that $\boldsymbol{\mathrm{A}}_{i}$ is independent of the RK stage $i$, and we therefore may simply denote it as $\boldsymbol{\mathrm{A}}$, and $\boldsymbol{\mathrm{A}}$ is in fact a scaled mass matrix on the velocity space. We also have $\boldsymbol{\mathrm{C}}_i = - \boldsymbol{\mathrm{B}}_{i}^T$. In the meantime, $\boldsymbol{\mathrm{B}}_{i}$ and $\boldsymbol{\mathrm{D}}_{i}$ are linearly scaled by the RK coefficient $\alpha_{ii}$ and the time-step size and can be factored as $\boldsymbol{\mathrm{B}}_{i} = \alpha_{ii} \boldsymbol{\mathrm{B}}$ and $\boldsymbol{\mathrm{D}}_{i} = \alpha^2_{ii} \Delta t_n \boldsymbol{\mathrm{D}}$, where the constant matrices $\boldsymbol{\mathrm{B}}$ and $\boldsymbol{\mathrm{D}}$ have components
\begin{align}
\mathrm{B}^{m}_{AB} = -\big(N_{A,m}, N_B\big)  - \tau\big(N_{A}, N_{B,m}\big) \quad \mbox{and} \quad
\mathrm{D}_{AB} = \frac{\tau}{\rho} \sum_{l=1}^{3}\big( N_{A,l}, N_{B,l} \big).
\end{align}
This analysis implies that only the matrices $\boldsymbol{\mathrm{A}}$, $\boldsymbol{\mathrm{B}}$, and $\boldsymbol{\mathrm{D}}$ need to be assembled once at the beginning of the computation, after which the corresponding block matrices of $\boldsymbol{\mathrm K}_{i}$ at each RK stage can be directly constructed using the corresponding RK coefficient and the time-step size. This not only reduces the computational cost of matrix assembly but also facilitates the development of flexible and efficient linear solver strategies.

To solve the equations $\boldsymbol{\mathrm{R}}_i^{\mathrm{M}}= \bm 0$ and $\boldsymbol{\mathrm{R}}_i^{\mathrm{C}}=\bm 0$, an initial guess for $\{\mathsf{U}_i, \mathsf{P}_i\}$ is introduced as the predicted solution, typically taken from the solution of the previous step or stage. Based on the initial guess, the residual vectors on the right-hand side are assembled and are denoted by a tilde. The incrementals are determined from the following linear system, 
\begin{align}
	\label{eq:Ki_corrector_eqn}
	\begin{bmatrix}
		\boldsymbol{\mathrm A} & \alpha_{ii} \boldsymbol{\mathrm B} \\[0.3mm]
		\alpha_{ii} \boldsymbol{\mathrm B}^{T} & -\alpha^2_{ii}\Delta t_n \boldsymbol{\mathrm D}
	\end{bmatrix}
	\begin{bmatrix}
		\Delta \mathsf{U}_i \\[0.3mm]
		\Delta \mathsf{P}_i
	\end{bmatrix}
	=\begin{bmatrix}
		-\tilde{\boldsymbol{\mathrm{R}}}_i^{\mathrm{M}} \\[0.3mm]
		\tilde{\boldsymbol{\mathrm{R}}}_i^{\mathrm{C}}
	\end{bmatrix}.
\end{align}
Here, we reorganized the equation associated with the continuity equation to make the matrix symmetric in \eqref{eq:Ki_corrector_eqn}. The solution is then updated by applying the corrections $\mathsf{U}_i \leftarrow \mathsf{U}_i + \Delta \mathsf{U}_i$ and $\mathsf{P}_i \leftarrow \mathsf{P}_i + \Delta \mathsf{P}_i$, which yields the velocity and pressure fields that satisfy $\boldsymbol{\mathrm{R}}_i^{\mathrm{M}} = \bm 0$ and $\boldsymbol{\mathrm{R}}_i^{\mathrm{C}} = \bm 0$. The linear system associated with the VMS formulation \eqref{eq:herk_vms_pres_coarse_closed}, which determines $p_{n+1}$, exhibits a structure analogous to that of \eqref{eq:Ki_corrector_eqn}. Notably, by setting $\Delta t_n = 1$ and $\alpha_{ii} = 1$ in the definitions \eqref{eq:Ai_block_matrix}-\eqref{eq:Di_block_matrix}, one directly recovers the associated block matrices for that problem.

\subsection{Preconditioned iterative solver}
In the following, we focus on the linear system \eqref{eq:Ki_corrector_eqn} and develop an iterative solution procedure for it. The matrix in that system is symmetric and definite owing to the coercivity of the underlying bilinear form, as shown in \eqref{eq:Bvms_stability}. Therefore, the linear system  \eqref{eq:Ki_corrector_eqn} can be effectively addressed by the conjugate gradient (CG) method, which is implemented in terms of an efficient three-term recurrence strategy. To accelerate the convergence of the Krylov iteration, an effective preconditioner can be designed, motivated by the block structure of $\boldsymbol{\mathrm K}_{i}$. A block factorization can be performed as
\begin{align*}
	\boldsymbol{\mathrm K}_i =
	\begin{bmatrix}
		\boldsymbol{\mathrm I} & \boldsymbol{\mathrm O} \\[0.3em]
		\alpha_{ii} \boldsymbol{\mathrm B}^{T} \boldsymbol{\mathrm A}^{-1} & \boldsymbol{\mathrm I}
	\end{bmatrix}
	\begin{bmatrix}
		\boldsymbol{\mathrm A} & \boldsymbol{\mathrm O} \\[0.3em]
		\boldsymbol{\mathrm O} & \alpha^2_{ii} \boldsymbol{\mathrm S}
	\end{bmatrix}
	\begin{bmatrix}
		\boldsymbol{\mathrm I} & \alpha_{ii}\boldsymbol{\mathrm A}^{-1} \boldsymbol{\mathrm B} \\[0.3em]
		\boldsymbol{\mathrm O} & \boldsymbol{\mathrm I}
	\end{bmatrix}, \quad \mbox{with} \quad \boldsymbol{\mathrm S} := - \boldsymbol{\mathrm B}^{T} \boldsymbol{\mathrm A}^{-1} \boldsymbol{\mathrm B}-\Delta t_n \boldsymbol{\mathrm D}.
\end{align*}
Let $\hat{\boldsymbol{\mathrm S}} := - \boldsymbol{\mathrm B}^{T} ( \mathrm{diag} \boldsymbol{\mathrm A}) ^{-1} \boldsymbol{\mathrm B}-\Delta t_n \boldsymbol{\mathrm D}$ be a symmetric sparse approximation of $\boldsymbol{\mathrm S}$, we introduce a block preconditioner defined as
 \begin{align*}
	\boldsymbol{\mathrm P}_i =
	\begin{bmatrix}
		\boldsymbol{\mathrm I} & \boldsymbol{\mathrm O} \\[0.3em]
		\alpha_{ii} \boldsymbol{\mathrm B}^{T} \boldsymbol{\mathrm A}^{-1} & \boldsymbol{\mathrm I}
	\end{bmatrix}
	\begin{bmatrix}
		\boldsymbol{\mathrm A} & \boldsymbol{\mathrm O} \\[0.3em]
		\boldsymbol{\mathrm O} & \alpha^2_{ii} \hat{\boldsymbol{\mathrm S}}
	\end{bmatrix}
	\begin{bmatrix}
		\boldsymbol{\mathrm I} & \alpha_{ii}\boldsymbol{\mathrm A}^{-1} \boldsymbol{\mathrm B} \\[0.3em]
		\boldsymbol{\mathrm O} & \boldsymbol{\mathrm I}
	\end{bmatrix}.
\end{align*}
This preconditioner is built directly based on the full factorization of $\boldsymbol{\mathrm K}_{i}$. Alternatively, one may simply use the block diagonal part $\mathrm{diag}[\boldsymbol{\mathrm A}, \alpha^2_{ii} \hat{\boldsymbol{\mathrm S}}]$, which is an effective candidate for the stabilized formulation for the Stokes equation \cite{Wathen1993}. We also mention that $\hat{\boldsymbol{\mathrm S}}$ provides a straightforward sparse approximation for $\boldsymbol{\mathrm S}$, though other more sophisticated approximations may yield improved performance. Readers are referred to \cite{Benzi2005} for a thorough discussion on this topic. Each Krylov iteration requires the action of $\boldsymbol{\mathrm K}_{i}\boldsymbol{\mathrm P}_{i}^{-1}$ on a vector $\bm z$, considering a right preconditioner. This operation can be realized by solving a linear problem $\boldsymbol{\mathrm P}_{i} \bm y = \bm z$ and then performing a matrix-vector multiplication on $\bm y$ with $\boldsymbol{\mathrm K}_{i}$. The action of the preconditioner $\boldsymbol{\mathrm P}_i$ can be summarized in the following procedures.
\begin{enumerate}
\item Solve for an intermediate velocity $\hat{\bm{y}}^{\bm v}$ from the equation $\boldsymbol{\mathrm A} \hat{\bm{y}}^{\bm v} = \bm{z}^{\bm v}$.
\item Update the continuity residual by $\bm{z}^{p} \gets \bm{z}^{p} - \alpha_{ii}\boldsymbol{\mathrm B}^{T}\hat{\bm{y}}^{\bm v}$.
\item Solve for ${\bm{y}}^{p}$ from the equation $\hat{\boldsymbol{\mathrm {S}}} {\bm{y}}^{p} = \bm{z}^{p}$
and set $\bm{y}^{p} \gets \alpha^{-2}_{ii} \bm y^p$.
\item Update the momentum residual by $\bm{z}^{\bm v} \gets \bm{z}^{\bm v} - \alpha_{ii}\boldsymbol{\mathrm B} {\bm{y}}^{p}$.
\item Solve for ${\bm{y}}^{\bm v}$ from the equation $\boldsymbol{\mathrm A} {\bm{y}}^{\bm v} = \bm{z}^{\bm v}$.
\end{enumerate}
In this procedure, the two linear solves involving the matrix $\boldsymbol{\mathrm A}$ can be performed with the CG method or simply the Richardson iterative method preconditioned by a lumped mass matrix. Using the mass-lumping approach is effective for low-order elements. Yet, we observed performance degradation with higher-order elements. This behavior is somewhat expected, as mass lumping is generally less effective for higher-order spline elements in explicit dynamics \cite{Cottrell2006}. Therefore, in this study, we employ an additive Schwarz preconditioner paired with incomplete LU factorization for the matrix $\boldsymbol{\mathrm A}$. Meanwhile, recent advances in explicit dynamics with higher-order spline elements may offer new insights for improving the solution strategy for this matrix problem \cite{Nguyen2023,Voet2026}. In Step 3 of the above procedure, one needs to solve a linear system involving $\hat{\boldsymbol{\mathrm {S}}}$, which can likewise be solved with the CG method. If the time-step size remains constant, this matrix remains unchanged; otherwise, it can be updated conveniently from the block matrices. This matrix is analogous to a pressure Laplacian and can be preconditioned by multigrid or domain decomposition methods. In this study, we employ the same additive Schwarz with incomplete LU factorization for $\hat{\boldsymbol{\mathrm {S}}}$. Because highly accurate solutions of the subproblems associated with $\boldsymbol{\mathrm A}$ and $\hat{\boldsymbol{\mathrm {S}}}$ are not required in the action of $\boldsymbol{\mathrm P}_{i}$, the convergence criteria for these problems can be relaxed.  As a result, the algebraic definition of $\boldsymbol{\mathrm P}_{i}$ varies across iterations, and we consequently have to invoke the flexible CG \cite{Notay2000} for the outer iteration associated with $\boldsymbol{\mathrm K}_i$. Finally, we note that the proposed solver technology leverages the block matrices $\boldsymbol{\mathrm{A}}$, $\boldsymbol{\mathrm{B}}$, and $\boldsymbol{\mathrm{D}}$ and allows reusing them for the definition of the matrix $\boldsymbol{\mathrm K}_i$ and the preconditioner $\boldsymbol{\mathrm P}_i$ across time steps and RK stages. The block structure enables the construction of effective preconditioners, yet there is still ample room for further improvement of the solver design, which will be pursued in future work.

\section{Numerical examples}
In what follows, all examples are presented in dimensionless form, eliminating the need to assign physical units while maintaining the rigor of the numerical verification. We invoke the forward Euler, Heun, strong stability preserving, and classical fourth-order RK schemes \cite{Butcher2016,Gottlieb2001} and denote them as the HERK(1,1), HERK(2,2), HERK(3,3), and HERK(4,4) schemes, respectively. Their corresponding Butcher tableaux are provided in Appendix \ref{sec:butcher-tableaux-explicit-RK-in-examples}. We adopt the parameter $\tau$ as $1/2$ in all HERK$(s,p)$-VMS schemes.

\subsection{Manufactured solution}		     
To verify the spatial and temporal accuracy of the proposed scheme, we consider a manufactured solution given by 
\begin{align*}
\bm{v} = \frac{1}{2}\left(1-\cos(20\pi t) \right)
\begin{bmatrix} 
-\sin(\pi x)\cos(\pi y)\\
\cos(\pi x)\sin(\pi y)\\
0
\end{bmatrix} 
\quad \mbox{and} \quad
p = p_0 + \frac{1}{8}\left(1-\cos(20\pi t) \right) \left( \cos(2\pi x) + \cos(2\pi y) \right),
\end{align*}
defined on the domain $[0.25, 2.25] \times [0.25, 2.25] \times [-0.1, 0.1]$, where $p_0=1$. This domain is intentionally chosen to avoid zero normal velocity on the boundary. In the present tests, the fluid density and dynamic viscosity are set to $1.0$ and $0.01$, respectively. A traction boundary condition is applied on the surface at $z = 0.1$, while time-dependent Dirichlet boundary conditions are prescribed on the remaining five boundaries. The simulation is advanced in time with the final time $T = 0.05$, at which both velocity and pressure reach their peak magnitudes. To eliminate the influence of linear solver errors, the relative tolerance for solving the linear systems is set to $1.0 \times 10^{-13}$.

\begin{figure}[htbp]
\centering	
\begin{tabular}{cc}
\includegraphics[width=0.3\linewidth, trim=0 20 0 20, clip]{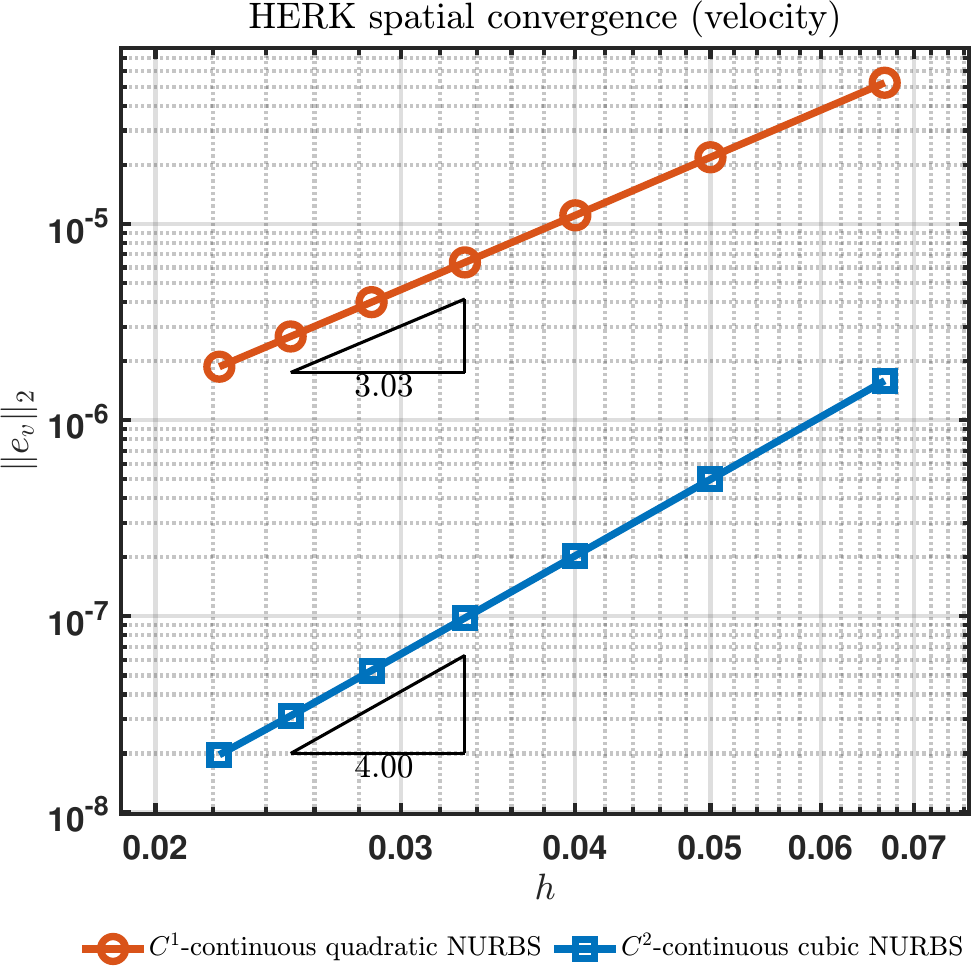} &
\includegraphics[width=0.3\linewidth, trim=0 20 0 20, clip]{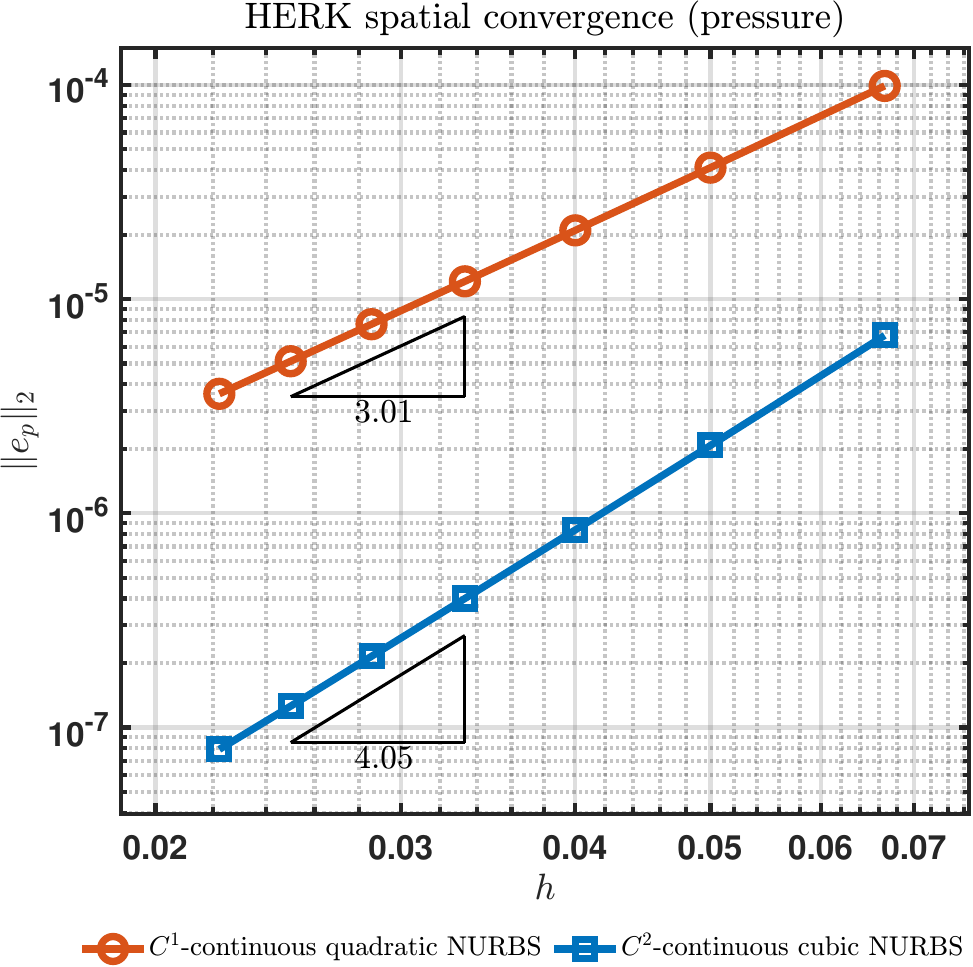} \\
(a) & (b)  \\[0.5em]
\includegraphics[width=0.3\linewidth, trim=0 20 0 20, clip]{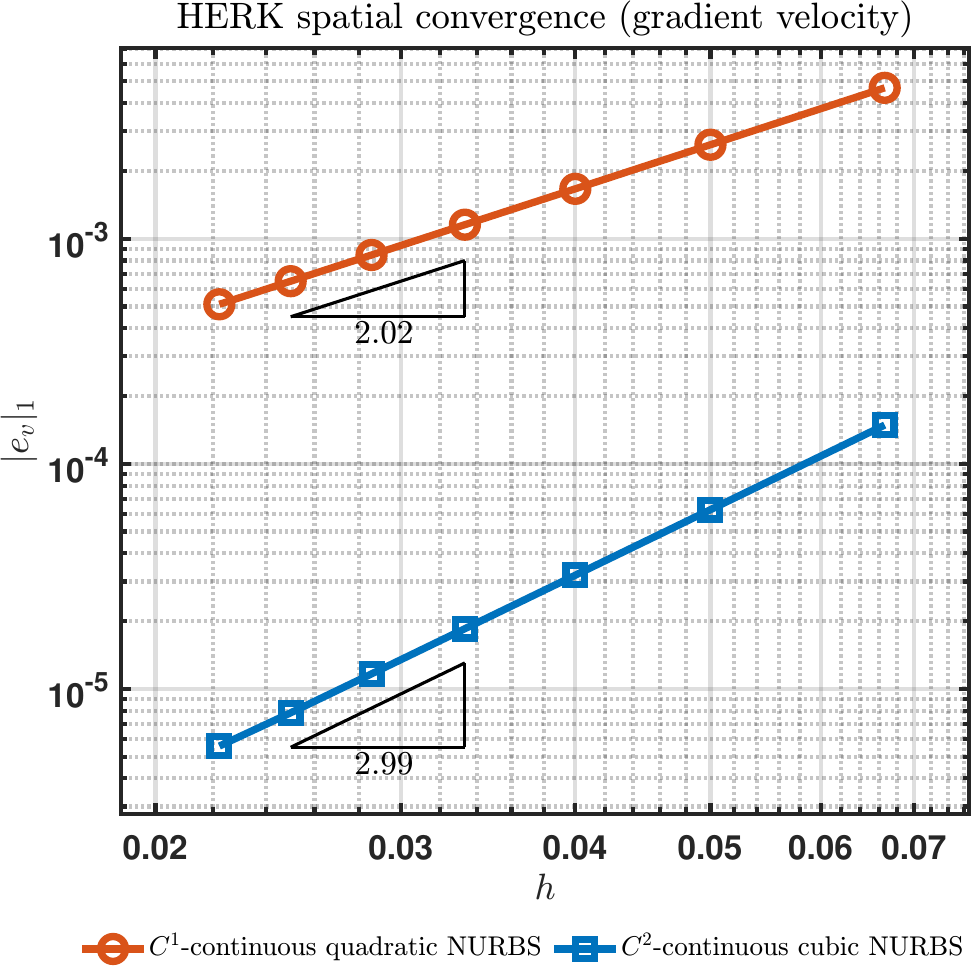} &
\includegraphics[width=0.3\linewidth, trim=0 20 0 20, clip]{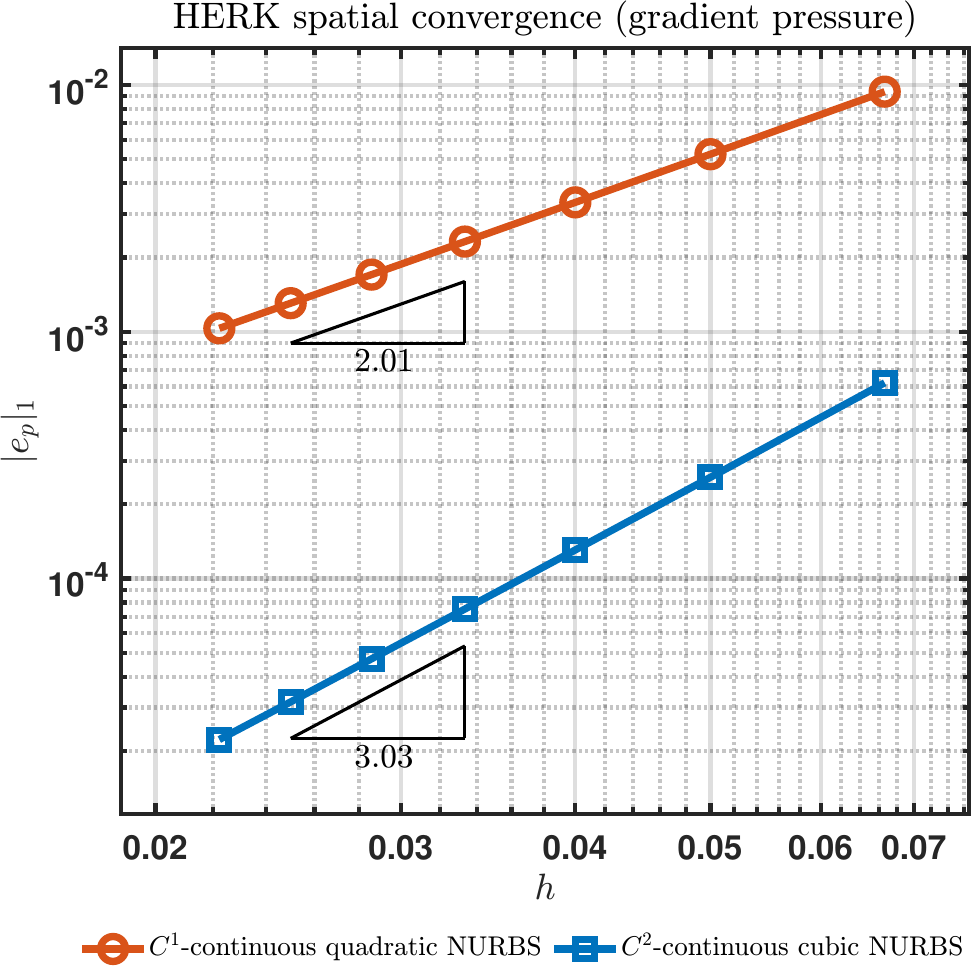}  \\
(c) & (d)  \\[0.5em]
\end{tabular}
\includegraphics[width=0.55\linewidth, trim=0 1.5 0 0, clip]{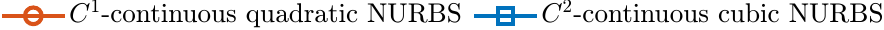}  \\[-0.5em]
\caption{Spatial error convergence rates for $C^1$-continuous quadratic and $C^2$-continuous cubic NURBS.}
\label{fig:spatial_convergence}
\end{figure}

\begin{figure}[htbp]
\centering
\begin{tabular}{ccc}
\includegraphics[width=0.3\linewidth, trim=0 22 0 20, clip]{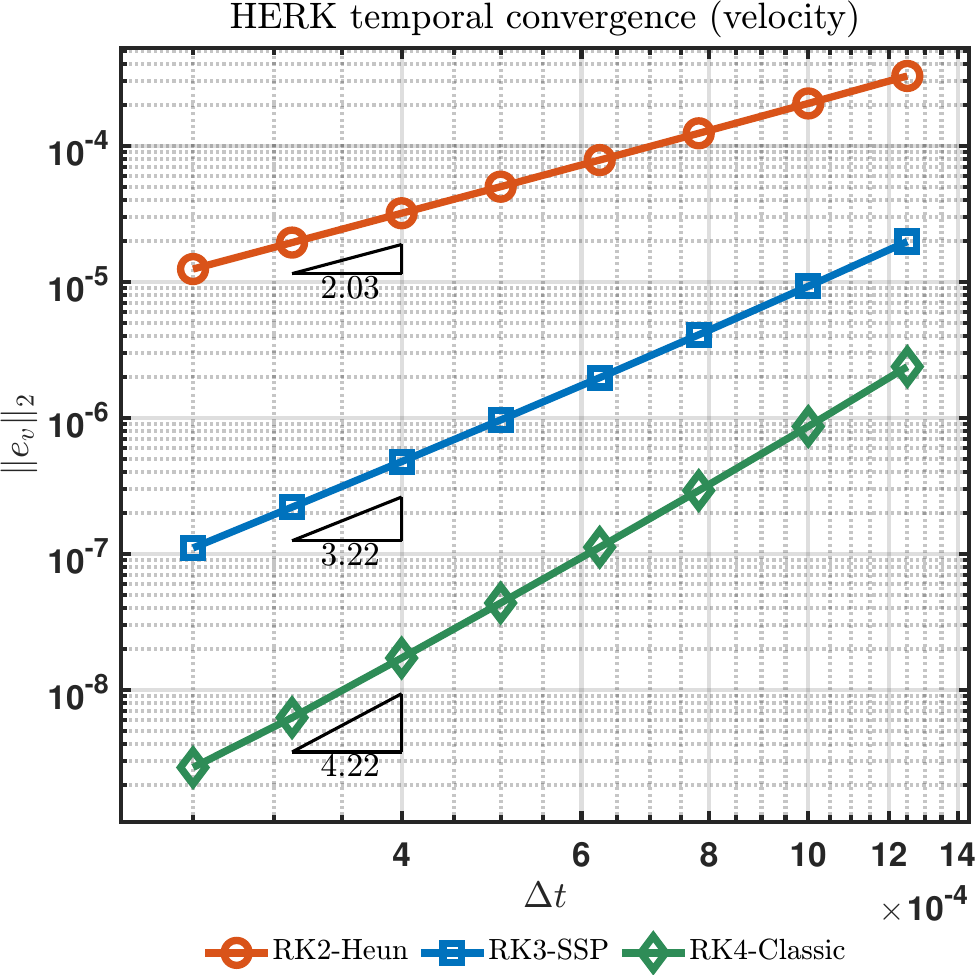} &
\includegraphics[width=0.3\linewidth, trim=0 22 0 20, clip]{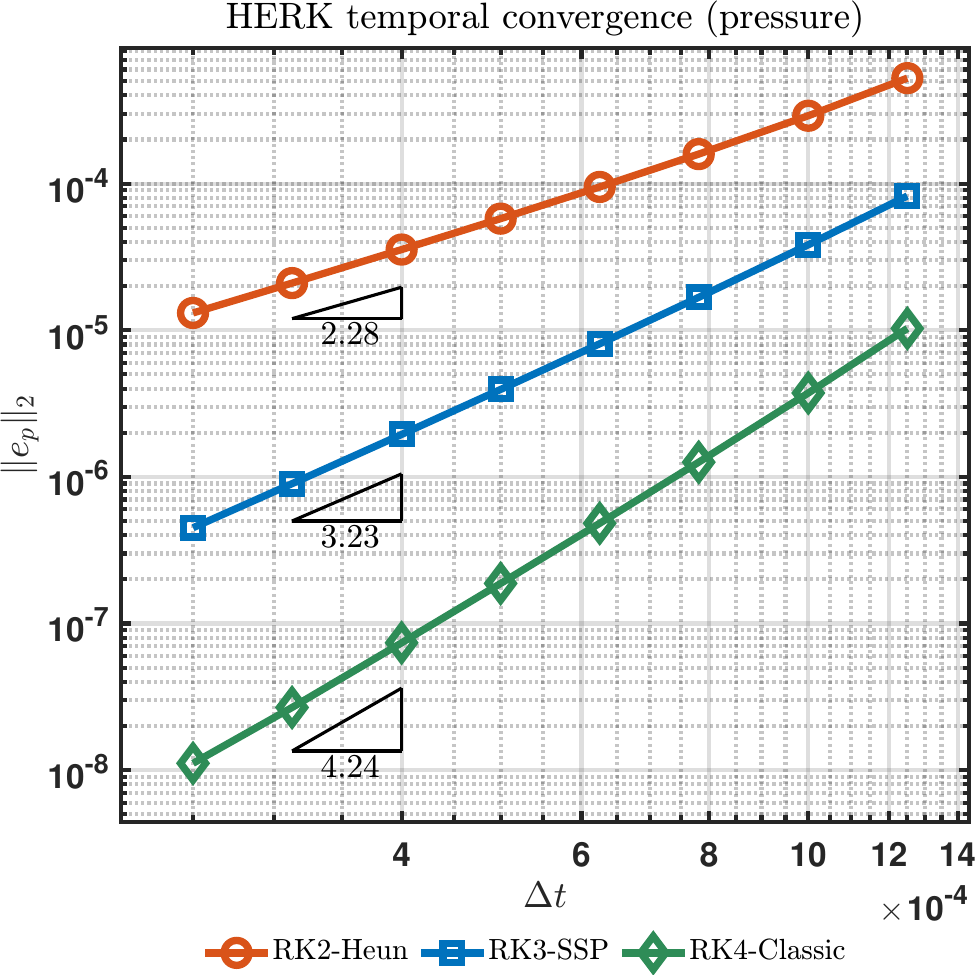} &
\includegraphics[width=0.3\linewidth, trim=0 22 0 20, clip]{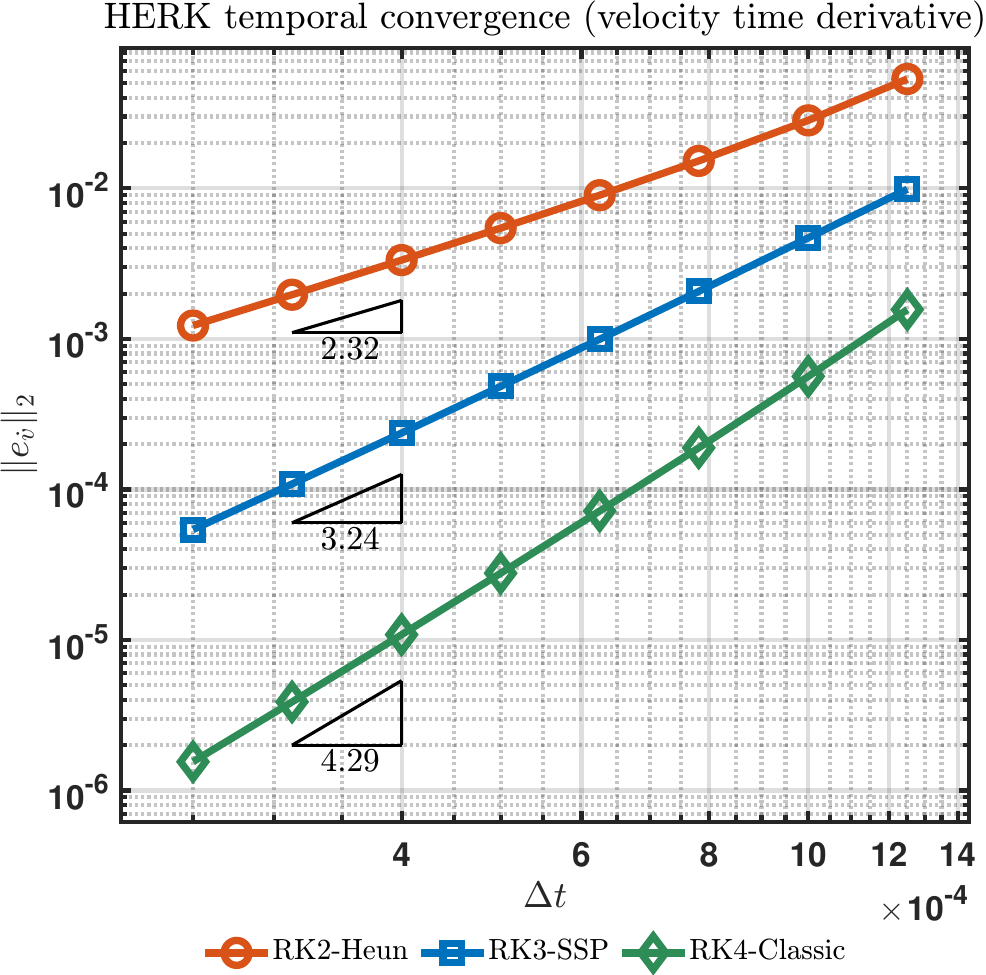} \\
(a) & (b) & (c) \\
\end{tabular}
\includegraphics[width=0.45\linewidth, trim=0 2 0 0, clip]{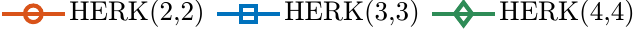} \\[-0.5em]
\caption{Temporal error convergence rates for HERK(2,2), HERK(3,3), and HERK(4,4).}
\label{fig:temporal_convergence}
\end{figure}

\paragraph{Spatial accuracy}
We first examine the spatial accuracy. To ensure that the temporal error is negligible, we employ HERK(4,4) with a sufficiently small fixed time-step size of $\Delta t_n = 2.5 \times 10^{-4}$. Both $C^1$-continuous quadratic and $C^2$-continuous cubic NURBS elements are considered, using equal-order interpolation for the velocity and pressure. A sequence of mesh refinement is performed, and the spatial accuracy is quantified using the $L^2$-norm and $H^1$-seminorm of the velocity and pressure errors, with results presented in Figure \ref{fig:spatial_convergence}. All convergence rates reported in the figures are obtained by least-squares fitting of the error data. Specifically, quadratic (cubic) elements yield third-order (fourth-order) convergence in the $L^2$-norm and second-order (third-order) convergence in the $H^1$-seminorm, respectively, for both velocity and pressure. The observed rates are higher than those from the error analysis for the Darcy equations, and such superconvergent behavior has been reported in previous numerical studies for the Darcy equations \cite{Masud2002,Badia2010}. The observed optimal accuracy here supports the effectiveness of the proposed numerical formulation.

\paragraph{Temporal accuracy}
To evaluate the temporal accuracy, we vary the time-step size while employing a sufficiently refined spatial discretization consisting of $80 \times 80 \times 4$ quartic NURBS elements with $C^3$-continuity. Simulations are performed using a sequence of time-step sizes from $2.5 \times 10^{-4}$ to $1.25 \times 10^{-3}$. Under this setting, the spatial discretization error remains sufficiently small in comparison with the temporal error. Three RK schemes are tested, including the HERK(2,2), HERK(3,3), and HERK(4,4) methods. To assess temporal accuracy, we compute the $L^2$-norm of the velocity and pressure errors. Since the time derivative of the velocity, $\dot{\bm{v}}$, is computed during the pressure calculation, we also evaluate the $L^2$-norm of its error, denoted by $\|\bm{e}_{\dot{\bm v}}\|_{2}$. Figure \ref{fig:temporal_convergence} presents the temporal convergence, and all convergence curves are annotated with slopes obtained by linear regression of all data points in the logarithmic scale. As expected, all schemes achieve their respective optimal orders of convergence.

\subsection{Taylor-Green vortex at Re=1600}
The Taylor-Green vortex (TGV) at Re=1600 is employed to assess the proposed numerical schemes. In this problem, the initially laminar vortices stretch and form vortex sheets, which subsequently roll up, reconnect, and trigger a transition to turbulence. This eventually induces rapid kinetic energy dissipation \cite{Pereira2021,Najafiyazdi2023}. Owing to these characteristics, the TGV has become a canonical benchmark for assessing numerical methods and turbulence models in their capability in capturing the energy spectrum and dissipation rate \cite{Bull2015,Tsoutsanis2021,Tsoutsanis2025}. Within turbulence modeling, classical eddy viscosity approaches, such as the Smagorinsky, WALE, and Vreman models, typically yield comparable results in TGV simulations, while VMS outperforms conventional eddy-viscosity models by providing more accurate dissipation behavior and capturing smaller vortex structure \cite{Evans2018}. Accordingly, the residual-based VMS framework provides an appropriate basis for further assessment of the proposed schemes.


The computational domain is a triply periodic cube $\Omega = [-\pi, \pi]^3$, and the flow field is initialized as
\begin{align*}
\bm{v} =
\begin{bmatrix} 
		\sin(x)\cos(y)\cos(z)\\
		-\cos(x)\sin(y)\cos(z)\\
		0
\end{bmatrix} \quad \mbox{and} \quad
	p = p_0 + \frac{1}{16}\left(\cos(2x) + \cos(2y) \right)\left(\cos(2z) + 2\right),
\end{align*}
with $p_0$ set to $1$ \cite{Bull2015, Najafiyazdi2023}. Table \ref{tab:tgv_3d_setting} summarizes the numerical settings adopted for all test cases. All computations are carried out on structured meshes using $C^1$-continuous quadratic NURBS elements. The notation ``VMS-Gen-$\alpha$($\rho_\infty$)" refers to the vertical method of lines, where the spatial discretization is first performed with the VMS formulation, followed by time integration using the generalized-$\alpha$ method with the specified spectral radius of the amplification matrix at the highest mode $\rho_\infty$. The stabilization parameters in the VMS-Gen-$\alpha$($\rho_\infty$) formulation are chosen consistently with those in \cite{Bazilevs2007}. In contrast, the notation ``HERK$(s,p)$-VMS" refers to the horizontal method of lines, in which the HERK$(s,p)$ scheme is used for time integration first, and the VMS formulation is subsequently applied to the resulting steady problem.

\begin{table}[htbp]
	\centering
	\caption{Summary of simulation cases and numerical settings.}
	\label{tab:tgv_3d_setting}
	\setlength{\tabcolsep}{22pt} 
	\begin{tabular}{cccc}
	\toprule
	Case & Discretization scheme & Number of elements & $\Delta t$ \\
	\midrule
	1 & VMS-Gen-$\alpha$(0.5) & $64^3$  & $5.0\times10^{-3}$ \\
	2 & VMS-Gen-$\alpha$(0.5) & $96^3$  & $5.0\times10^{-3}$ \\ 
	3 & VMS-Gen-$\alpha$(0.5) & $128^3$ & $5.0\times10^{-3}$ \\ 
	4 & VMS-Gen-$\alpha$(0.5) & $128^3$ & $1.0\times10^{-3}$ \\ 
	5 & VMS-Gen-$\alpha$(0.5) & $128^3$ & $1.0\times10^{-2}$ \\
	6 & VMS-Gen-$\alpha$(0.5) & $128^3$ & $2.5\times10^{-2}$ \\
	7 & VMS-Gen-$\alpha$(0.5) & $128^3$ & $5.0\times10^{-2}$ \\
	8 & VMS-Gen-$\alpha$(1.0) & $128^3$ & $1.0\times10^{-2}$ \\
	9 & HERK$(1,1)$-VMS & $128^3$ & $1.0\times10^{-2}$  \\
	10 & HERK$(2,2)$-VMS & $128^3$ & $1.0\times10^{-2}$  \\
	11 & HERK$(3,3)$-VMS & $64^3$  & $5.0\times10^{-3}$  \\
	12 & HERK$(3,3)$-VMS & $96^3$  & $5.0\times10^{-3}$  \\
	13 & HERK$(3,3)$-VMS & $128^3$ & $1.0\times10^{-3}$  \\
	14 & HERK$(3,3)$-VMS & $128^3$ & $5.0\times10^{-3}$  \\
	15 & HERK$(3,3)$-VMS & $128^3$ & $1.0\times10^{-2}$  \\
	16 & HERK$(4,4)$-VMS & $128^3$ & $1.0\times10^{-2}$  \\  
	\bottomrule
\end{tabular}
\end{table}

\begin{figure}[htbp]
\centering
\begin{tabular}{cc}
VMS-Gen-$\alpha$(0.5) & HERK$(3,3)$-VMS \\[0.5em]
\includegraphics[width=0.4\linewidth, trim=0 60 0 0, clip]{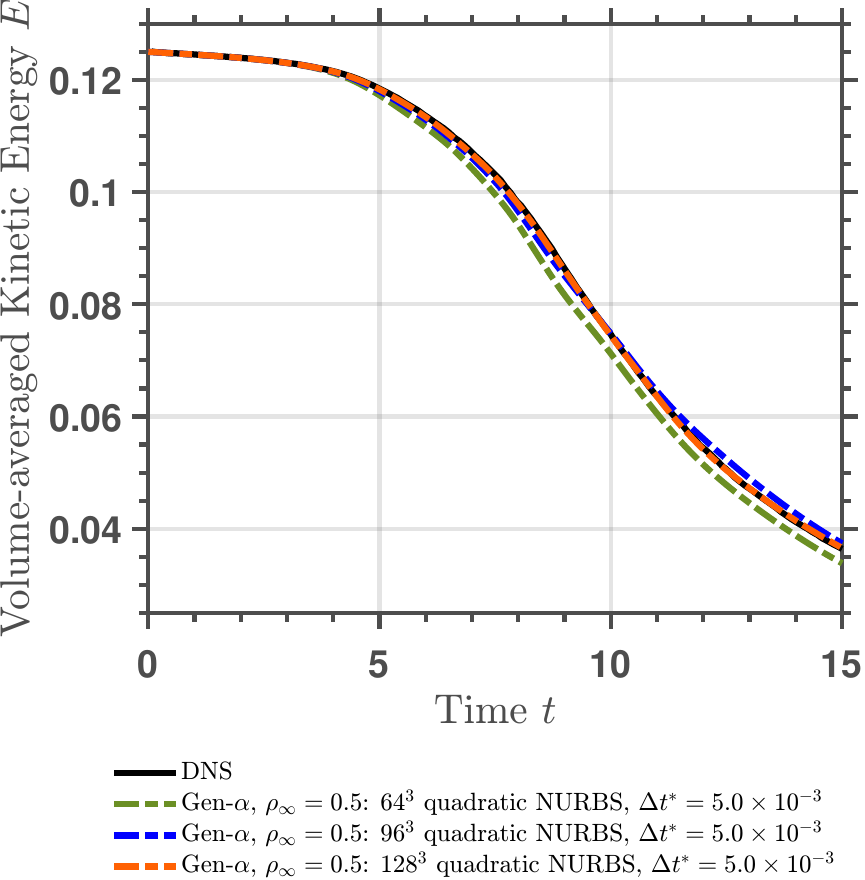} &
\includegraphics[width=0.4\linewidth, trim=0 60 0 0, clip]{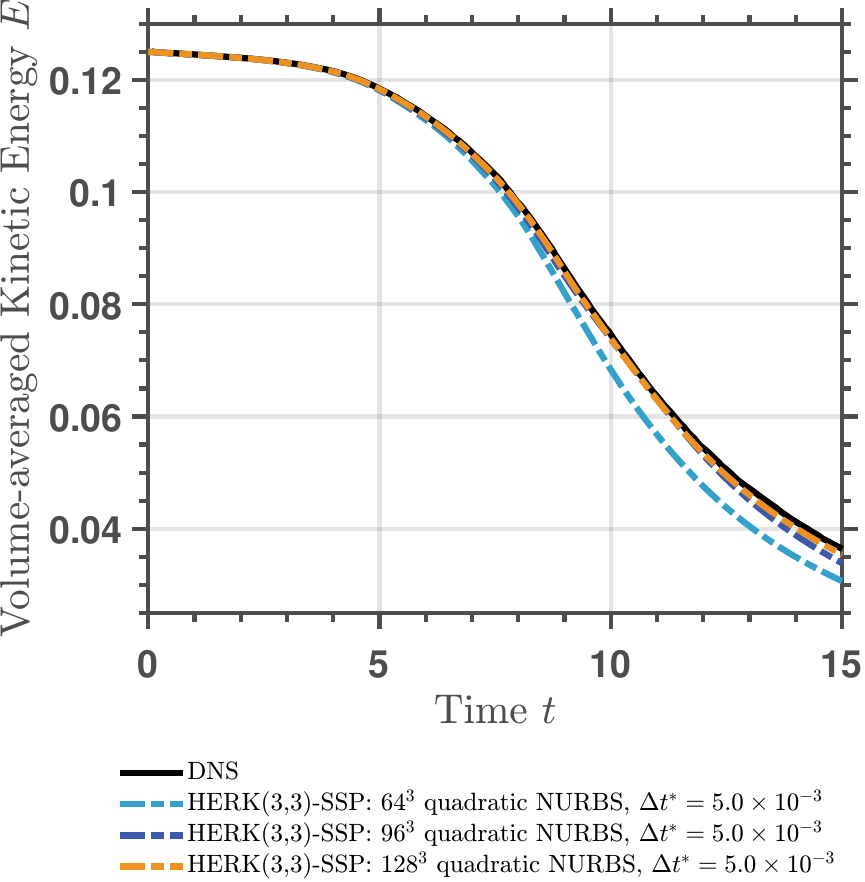} \\
\includegraphics[width=0.42\linewidth, trim=0 60 0 0, clip]{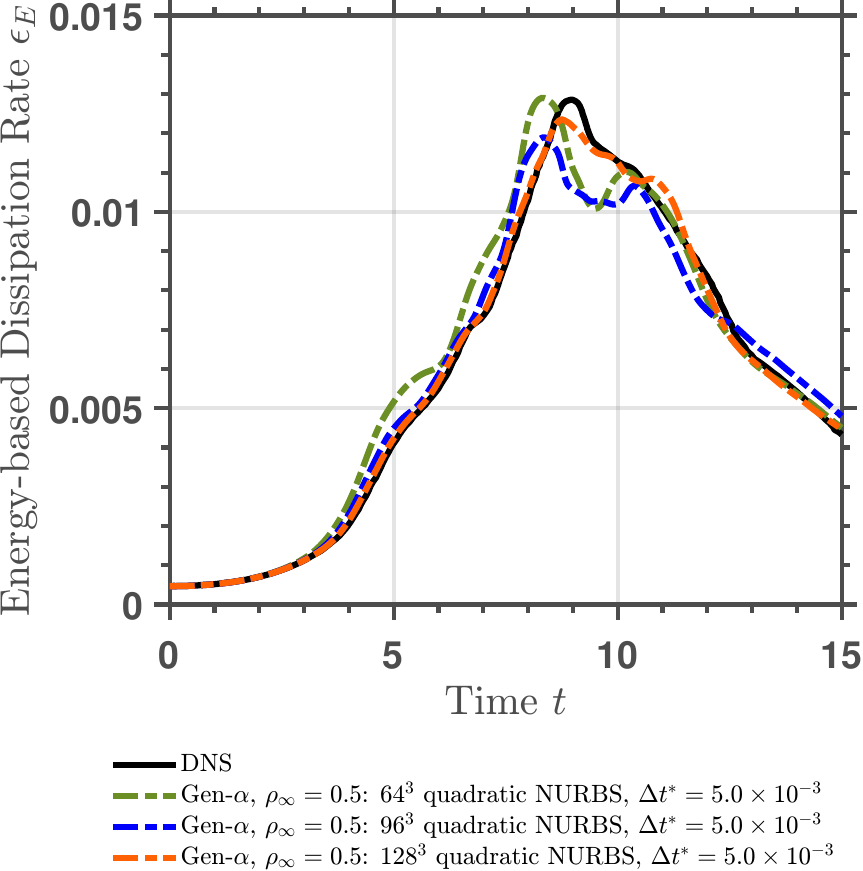} &
\includegraphics[width=0.42\linewidth, trim=0 60 0 0, clip]{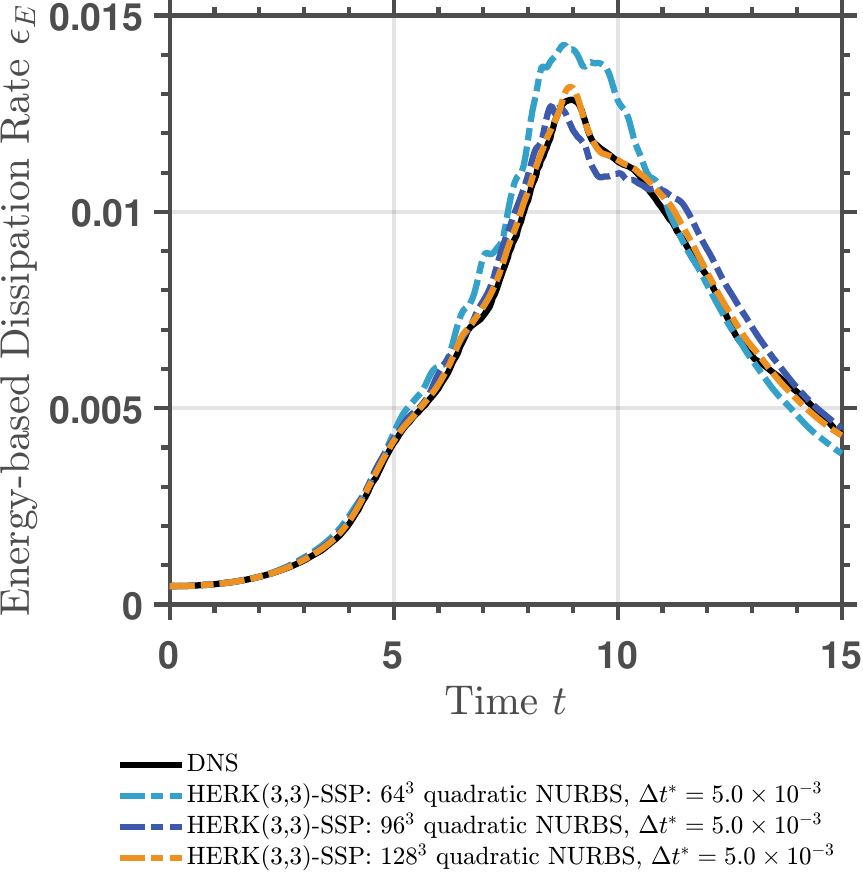} \\
\end{tabular}
\includegraphics[width=0.8\linewidth, trim=0 1 0 0, clip]{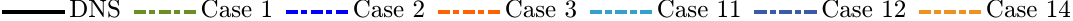}  \\[-0.5em]
\caption{Volume-averaged kinetic energy and energy-based dissipation rate comparison between two schemes on different mesh resolutions.}
\label{fig:TGV_energy_GA_vs_HERK3_different_mesh_compare}
\end{figure}

\paragraph{Time evolution of the kinetic energy}
One of the key quantities of interest is the kinetic energy dissipation rate. Following the definition given in \cite{Bull2015}, this quantity can be expressed as
\begin{align}
	\label{eq:energy_based_dissipation_def}
	\epsilon_E := -\frac{dE}{dt}, \quad \mbox{with} \quad E := \frac{1}{2(2\pi)^3} \left(\bm v, \bm v \right),
\end{align}
where $E$ denotes the volume-averaged kinetic energy. We evaluate these quantities under the settings listed in Table \ref{tab:tgv_3d_setting}, and compare them against reference direct numerical simulation (DNS) data \cite{Wang2013}. The reference simulation employed a $512^3$ mesh and a dealiased pseudo-spectral method with a low-storage, three-stage RK scheme for temporal integration. We first examine the behavior of the VMS-Gen-$\alpha$(0.5) and HERK$(3,3)$-VMS schemes at a fixed time-step size $\Delta t_n = 5.0 \times 10^{-3}$ across various mesh resolutions, with results shown in Figure \ref{fig:TGV_energy_GA_vs_HERK3_different_mesh_compare}. Under spatial mesh refinement, predictions from both schemes progressively converge toward the DNS results. Across all mesh resolutions, the VMS-Gen-$\alpha$(0.5) scheme exhibits an anomalous double-peaked structure in the dissipation rate, with the secondary peak occurring at $t \approx 11$. This behavior, also observed in \cite{Drikakis2007,Tsoutsanis2023short}, is typically attributed to the poor numerical dispersion property of the scheme. In contrast, the HERK$(3,3)$-VMS scheme does not exhibit such spurious behavior and, at a resolution of $128^3$ elements, closely matches the DNS dissipation rate.

\begin{figure}[htbp]
\centering
\begin{tabular}{cc}
VMS-Gen-$\alpha$(0.5) & HERK$(3,3)$-VMS \\[0.5em]
\includegraphics[width=0.4\linewidth, trim=0 90 0 0, clip]{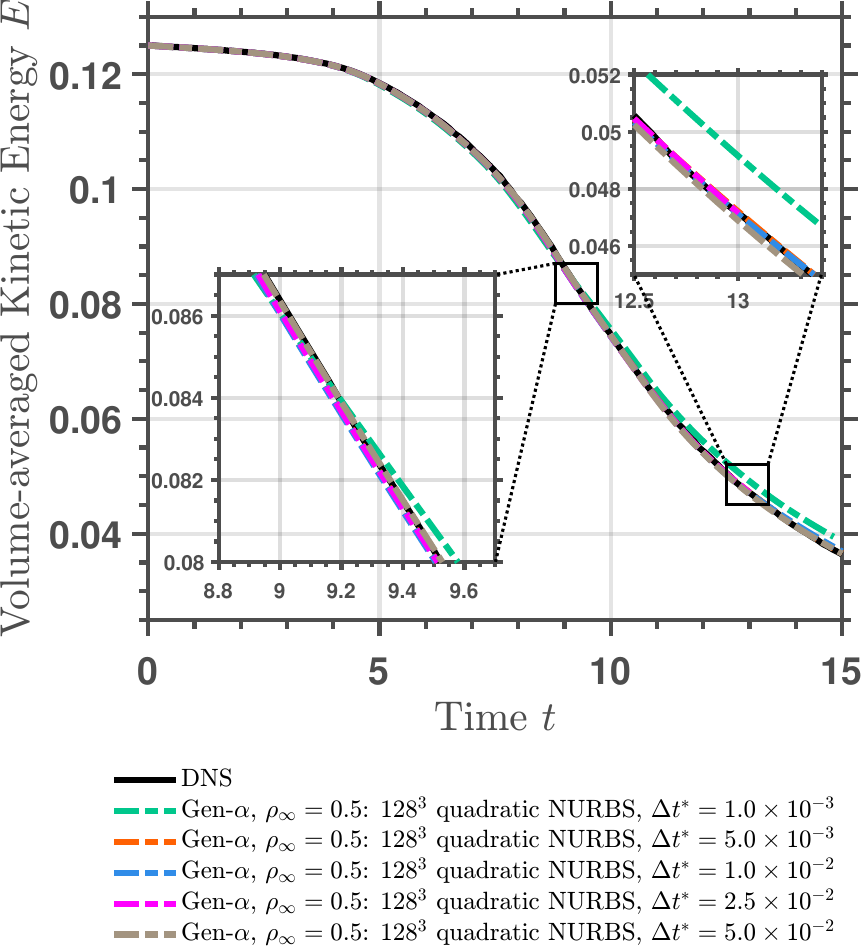} &
\includegraphics[width=0.4\linewidth, trim=0 60 0 0, clip]{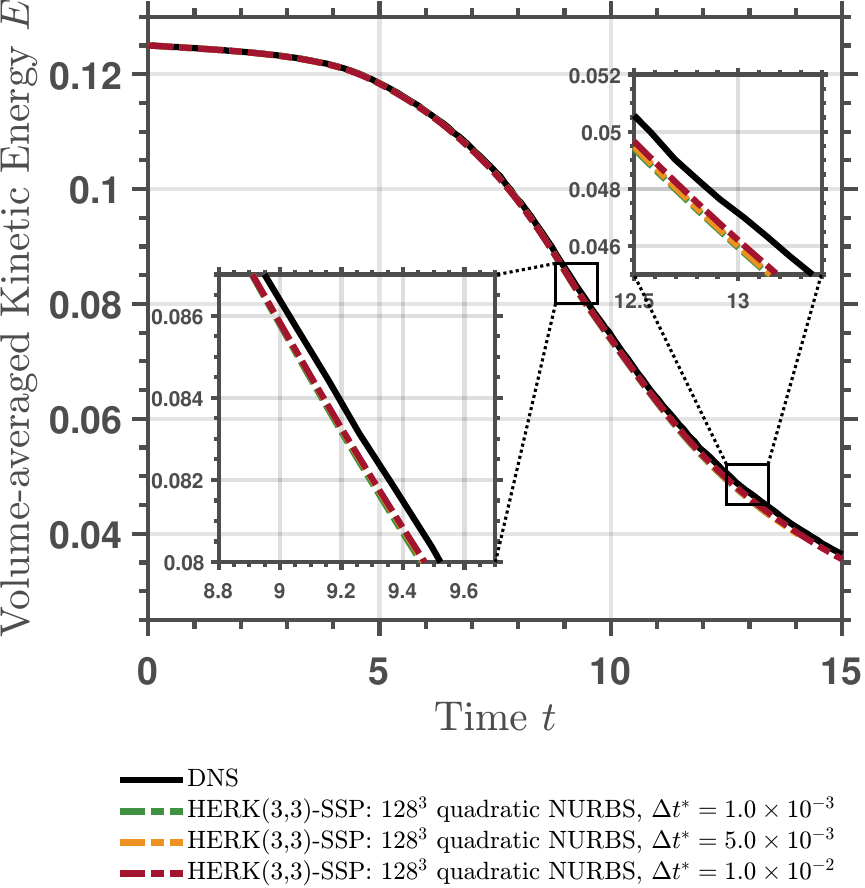} \\
\includegraphics[width=0.42\linewidth, trim=0 90 0 0, clip]{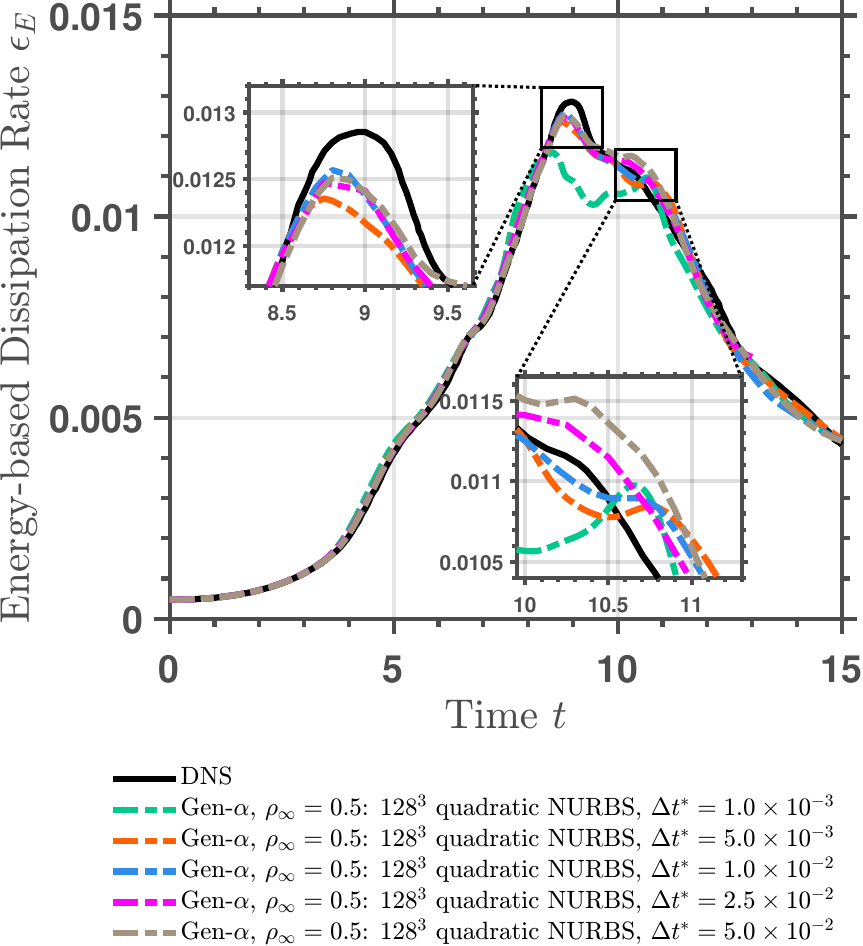} &
\includegraphics[width=0.42\linewidth, trim=0 60 0 0, clip]{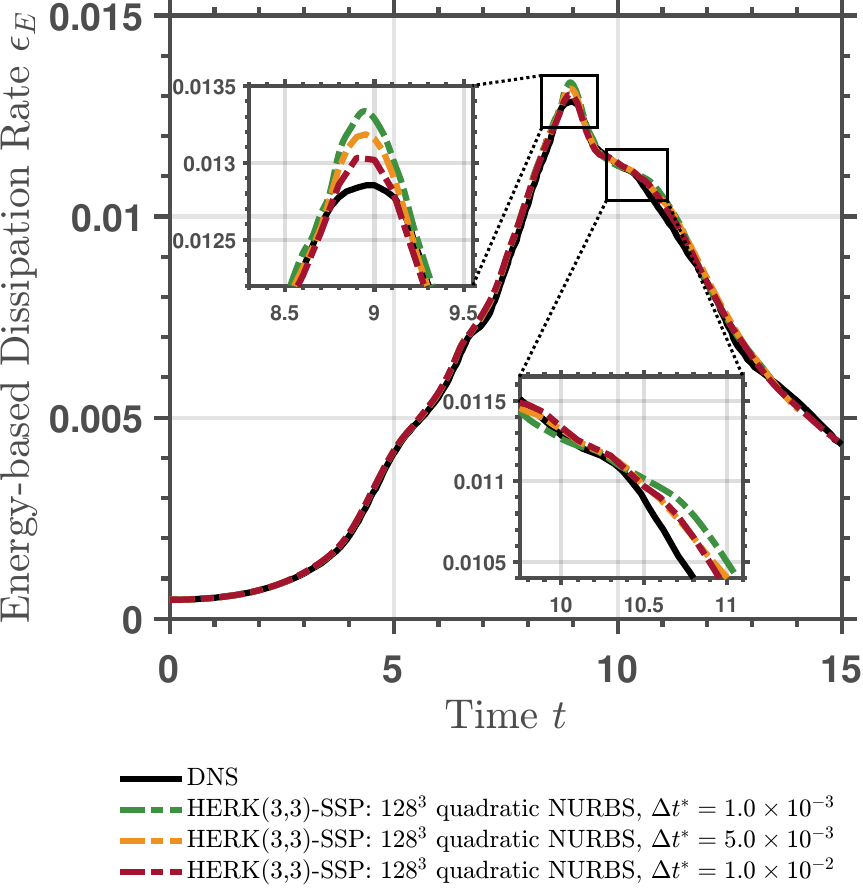} \\
\end{tabular}
\includegraphics[width=0.6\linewidth, trim=0 0 0 0, clip]{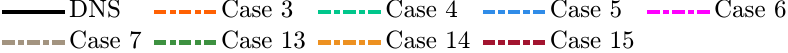}  \\[-0.5em]
\caption{Volume-averaged kinetic energy and energy-based dissipation rate comparison between two schemes on different time-step size.}
\label{fig:TGV_energy_GA_vs_HERK3_different_dt_compare}
\end{figure}

Figure \ref{fig:TGV_energy_GA_vs_HERK3_different_dt_compare} compares the kinetic energy and dissipation rate obtained using the VMS-Gen-$\alpha$(0.5) and HERK$(3,3)$-VMS schemes for different time-step sizes at a fixed spatial resolution. For the VMS-Gen-$\alpha$(0.5) scheme, reducing the time-step size from $5 \times 10^{-2}$ to $1 \times 10^{-2}$ improves the agreement with DNS results. However, a further reduction to $1 \times 10^{-3}$ leads to anomalous behavior. The obtained dissipation rate progressively deviates from the DNS solution, showing pronounced discrepancies near the dissipation peak. This phenomenon is not unexpected, as the scheme is known to suffer from the small-time-step instabilities that degrade the LES quality \cite{Bochev2007,Hsu2010}. For the HERK$(3,3)$-VMS scheme, results with time-step size greater than $1 \times 10^{-2}$ are not shown due to the stability constraint. The scheme exhibits only mild sensitivity to the time-step size. As the time-step size decreases from $1 \times 10^{-2}$ to $1 \times 10^{-3}$, a slight deviation from DNS is observed near the dissipation peak, and the overall influence of the time-step size remains minimal. In this sense, HERK$(3,3)$-VMS scheme demonstrates markedly lower sensitivity to time-step variations than the classical residual-based VMS scheme.

\begin{figure}[htbp]
\centering
\begin{tabular}{cc}
\includegraphics[width=0.4\linewidth, trim=0 110 0 0, clip]{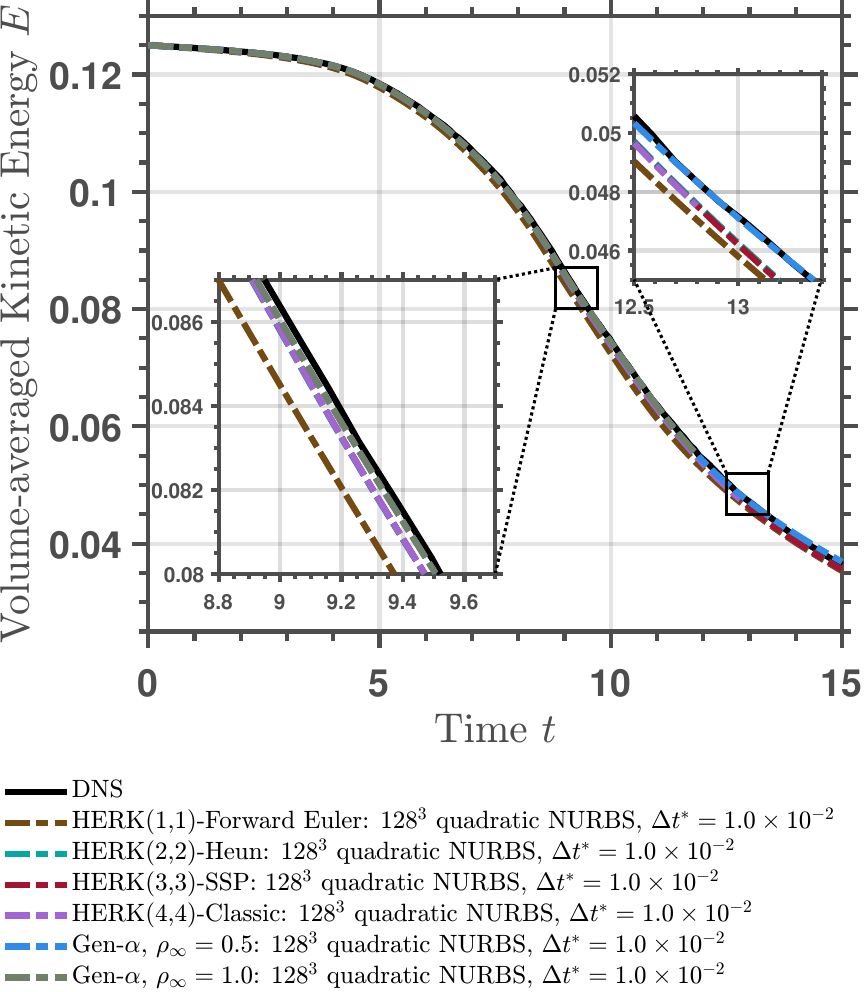} &
\includegraphics[width=0.4\linewidth, trim=0 110 0 0, clip]{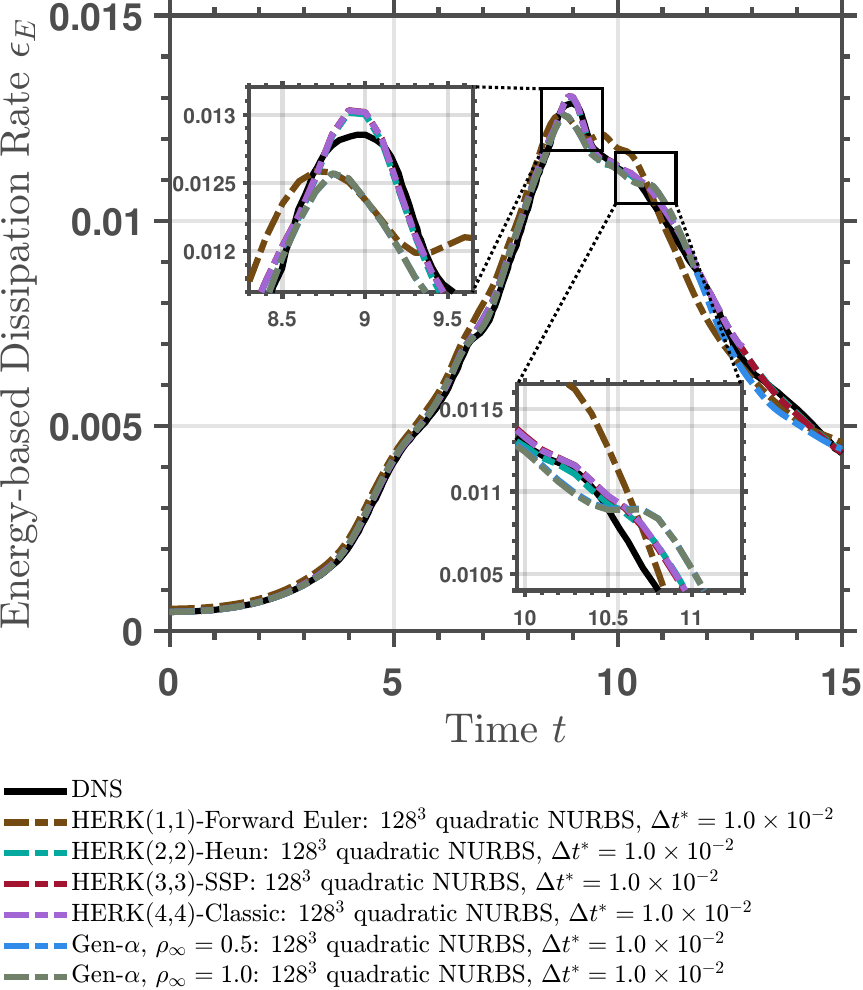} \\
\end{tabular}
\includegraphics[width=0.8\linewidth, trim=0 1 0 0, clip]{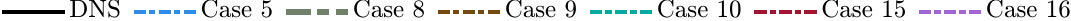}  \\[-0.5em]
\caption{Comparison of volume-averaged kinetic energy and energy dissipation rate between VMS-Gen-$\alpha$($\rho_\infty$) schemes with varying values of $\rho_\infty$ and HERK$(s,p)$-VMS schemes with different orders.}
\label{fig:TGV_energy_RK_vs_GA}
\end{figure}

\begin{figure}[htbp]
\centering
\begin{tabular}{cc}
VMS-Gen-$\alpha$(0.5) & HERK$(3,3)$-VMS \\[0.5em]
\includegraphics[width=0.45\linewidth, trim=0 95 0 0, clip]{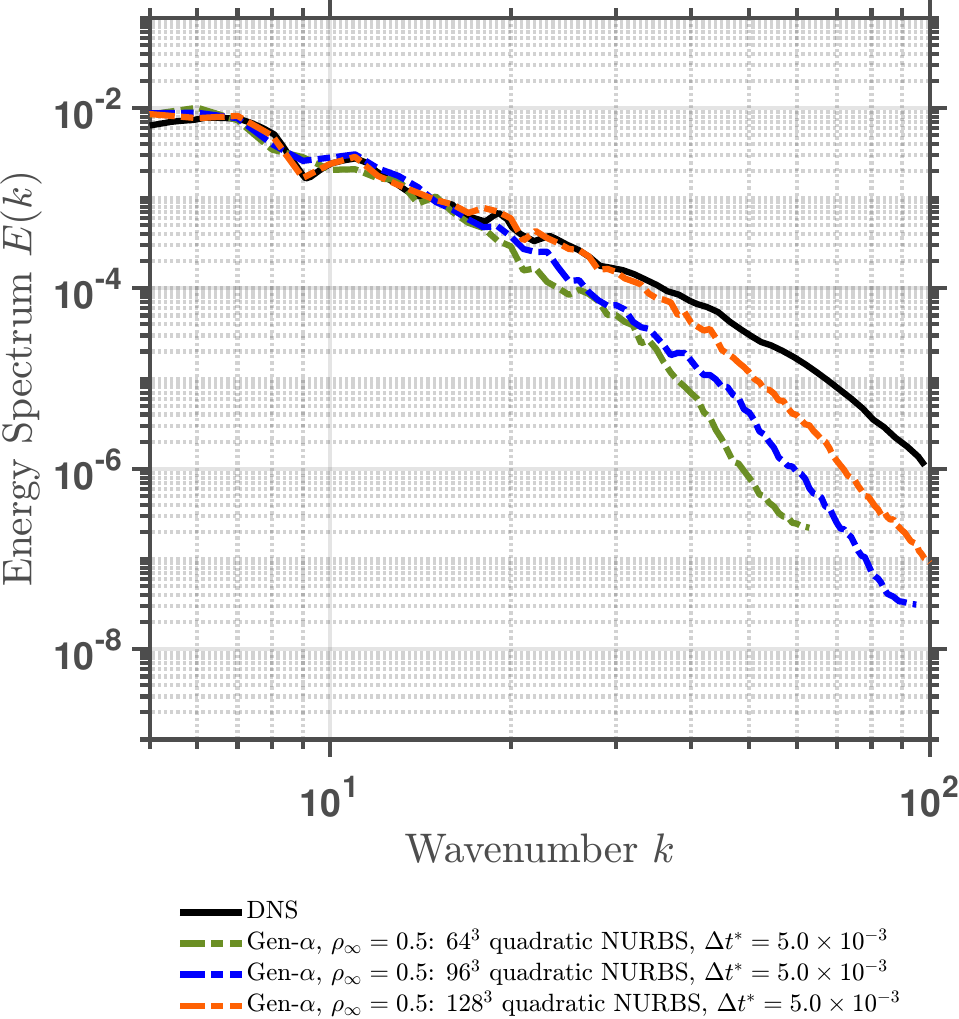} &
\includegraphics[width=0.45\linewidth, trim=0 95 0 0, clip]{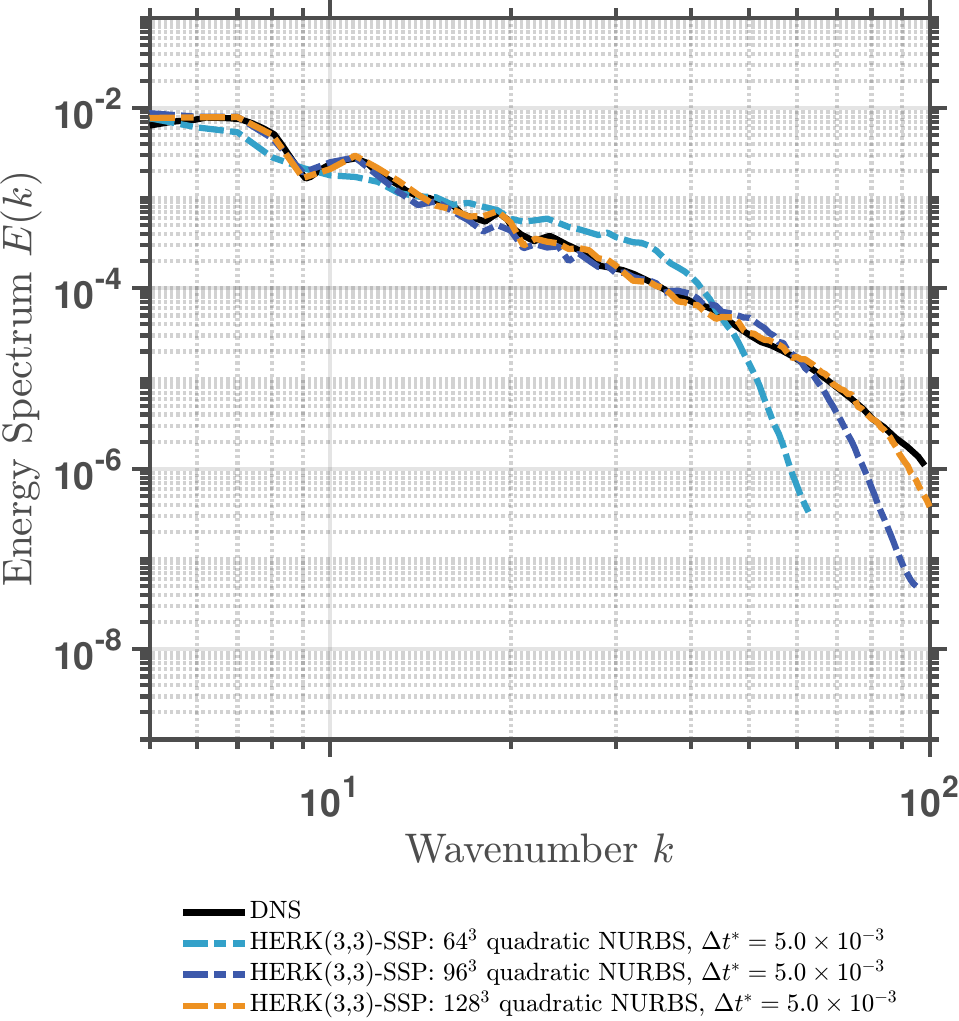}
\end{tabular}
\includegraphics[width=0.8\linewidth, trim=0 0 0 0, clip]{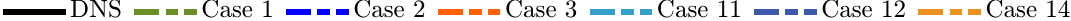}  \\[-0.5em]
\caption{Comparison of energy spectra between two schemes on different mesh resolutions.}
\label{fig:TGV_spectrum_different_mesh_compare}
\end{figure}

Figure \ref{fig:TGV_energy_RK_vs_GA} compares results obtained at identical mesh resolutions and time steps for HERK schemes of different temporal orders and for the generalized-$\alpha$ method with varying values of $\rho_\infty$. The parameter $\rho_\infty$ in the generalized-$\alpha$ method controls the level of high-frequency dissipation, with smaller values expected to enhance damping of high-frequency modes. However, the kinetic energy and dissipation rate curves for $\rho_\infty=1$ and $\rho_\infty=0.5$ are nearly indistinguishable, indicating that the TGV problem is insensitive to this parameter. Among all tested schemes, the HERK(1,1)-VMS scheme exhibits the largest deviation from DNS in the dissipation rate, due to its lowest temporal accuracy and the absence of the fine-scale contribution. In contrast, higher-order HERK schemes with $p=2,3,4$ exhibit dissipation that closely matches the DNS data and consistently outperform the generalized-$\alpha$ scheme. These results indicate that, for the TGV problem, temporal accuracy has a pronounced impact on accuracy, highlighting the superior fidelity of the higher-order HERK formulations.

\begin{figure}[htbp]
\centering
\begin{tabular}{cc}
VMS-Gen-$\alpha$(0.5) & HERK$(3,3)$-VMS \\[0.5em]
\adjustbox{valign=t, raise=6.45cm}{\includegraphics[width=0.45\linewidth, trim=0 120 0 0, clip]{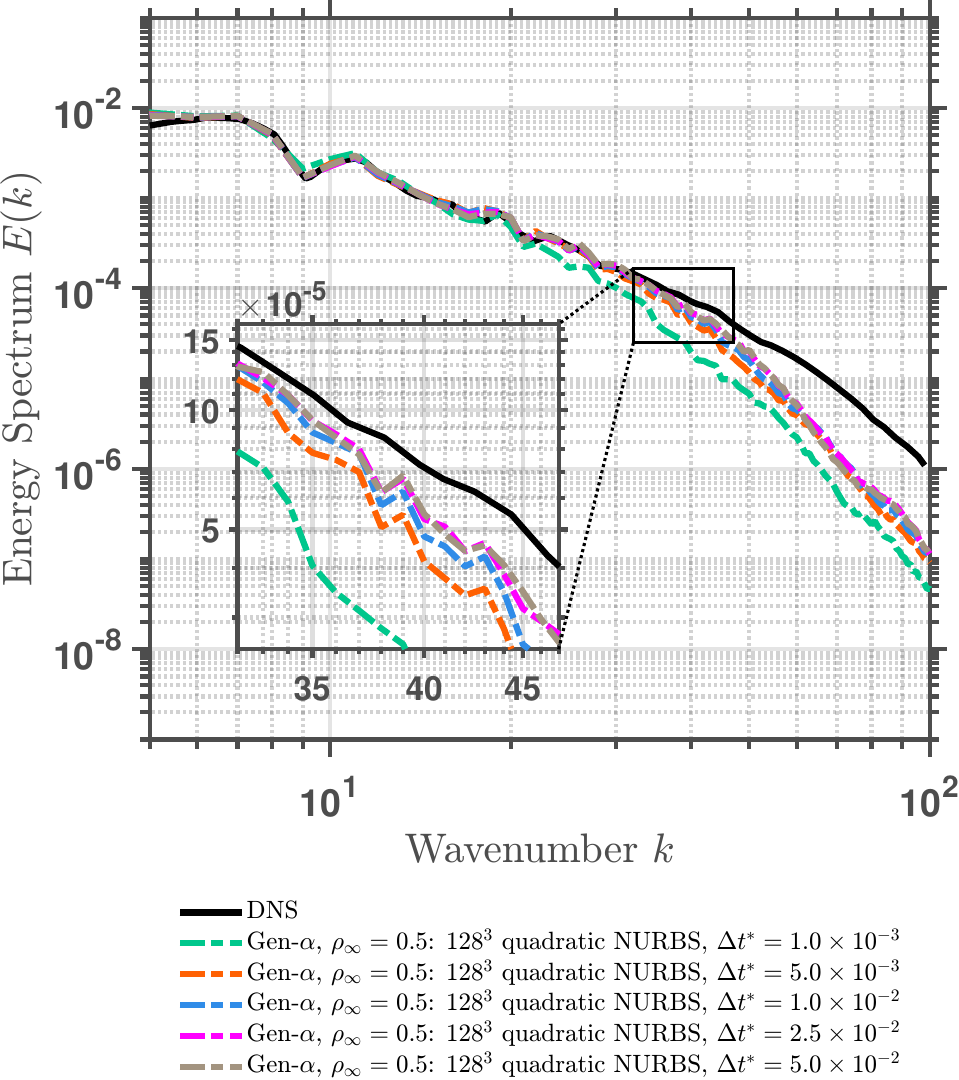}} &
\includegraphics[width=0.45\linewidth, trim=0 95 0 0, clip]{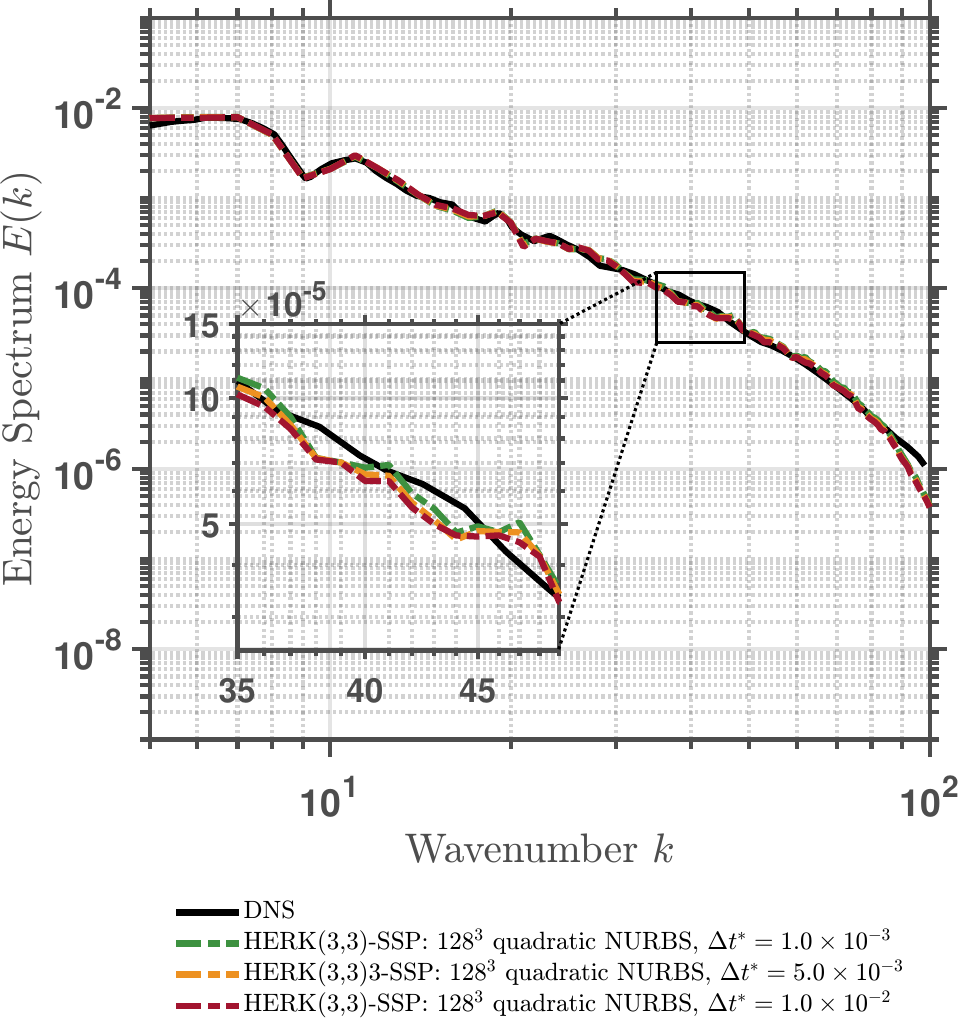}
\end{tabular}
\includegraphics[width=0.55\linewidth, trim=0 0 0 0, clip]{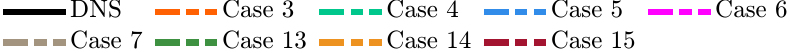}  \\[-0.5em]
\caption{Comparison of energy spectra between two schemes on different time-step sizes.}
\label{fig:TGV_spectrum_dt_compare}
\end{figure}

\begin{figure}[htbp]
\centering
\includegraphics[width=0.45\linewidth, trim=0 135 0 0, clip]{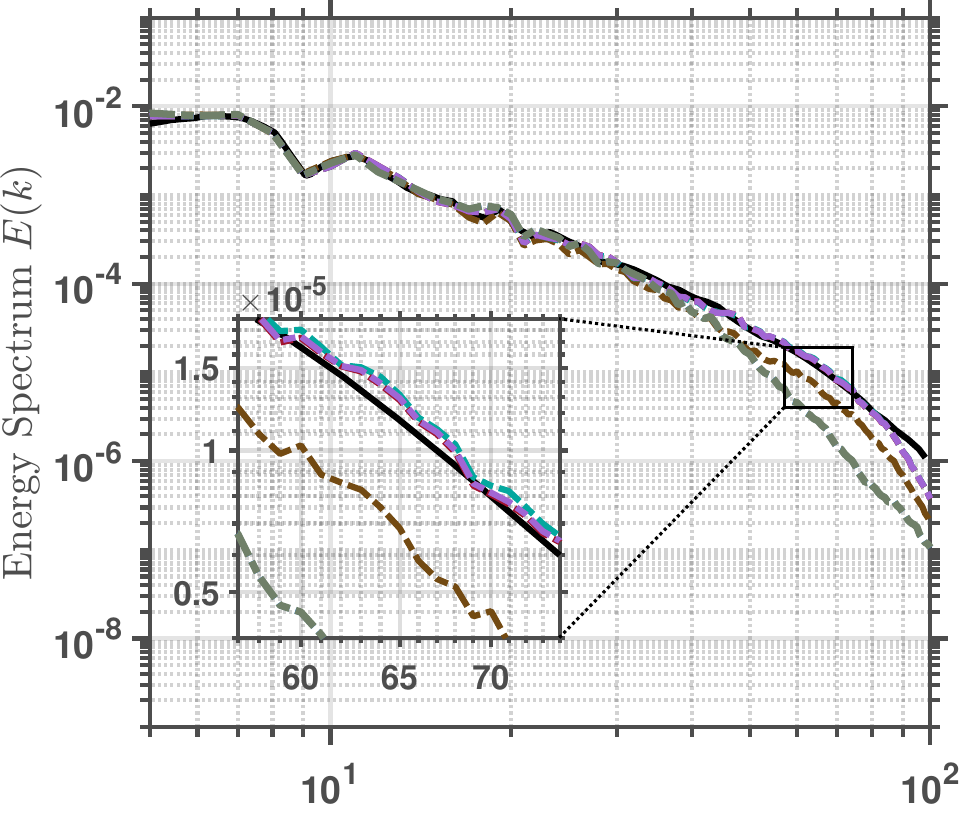}
\includegraphics[width=0.75\linewidth, trim=0 0 0 0, clip]{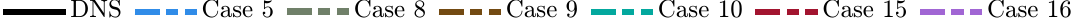}  \\[-0.5em]
\caption{Comparison of energy spectra between VMS-Gen-$\alpha$($\rho_\infty$) schemes with varying values of $\rho_\infty$ and HERK$(s,p)$-VMS schemes with different orders.}
\label{fig:TGV_spectrum_RK_vs_GA}
\end{figure}

\begin{figure}[htbp]
\centering
\begin{tabular}{cc}
	\includegraphics[width=0.33\linewidth, trim=300 20 300 50, clip]{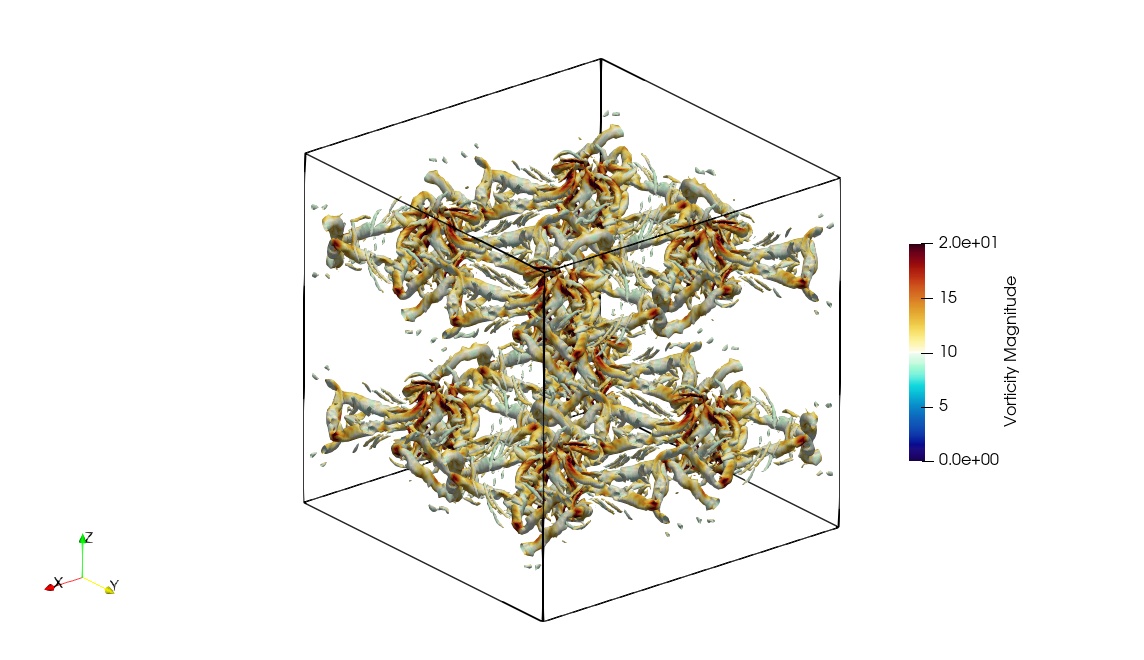} &
	\includegraphics[width=0.33\linewidth, trim=300 20 300 50, clip]{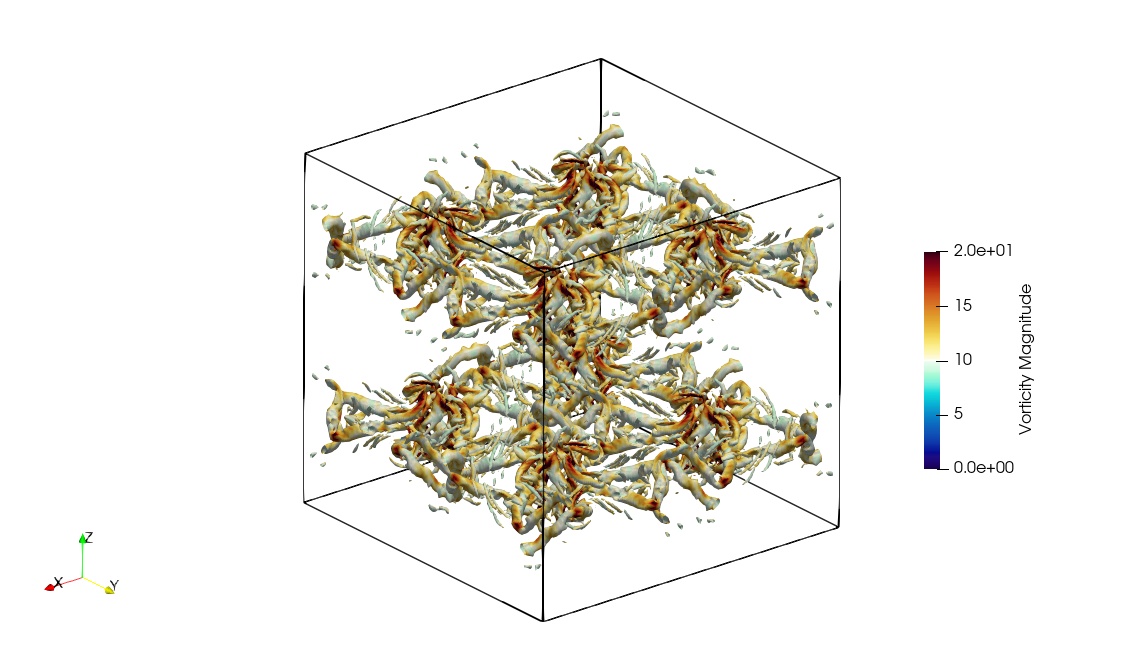} \\
	Case 5 & Case 8 \\
	\includegraphics[width=0.33\linewidth, trim=300 20 300 50, clip]{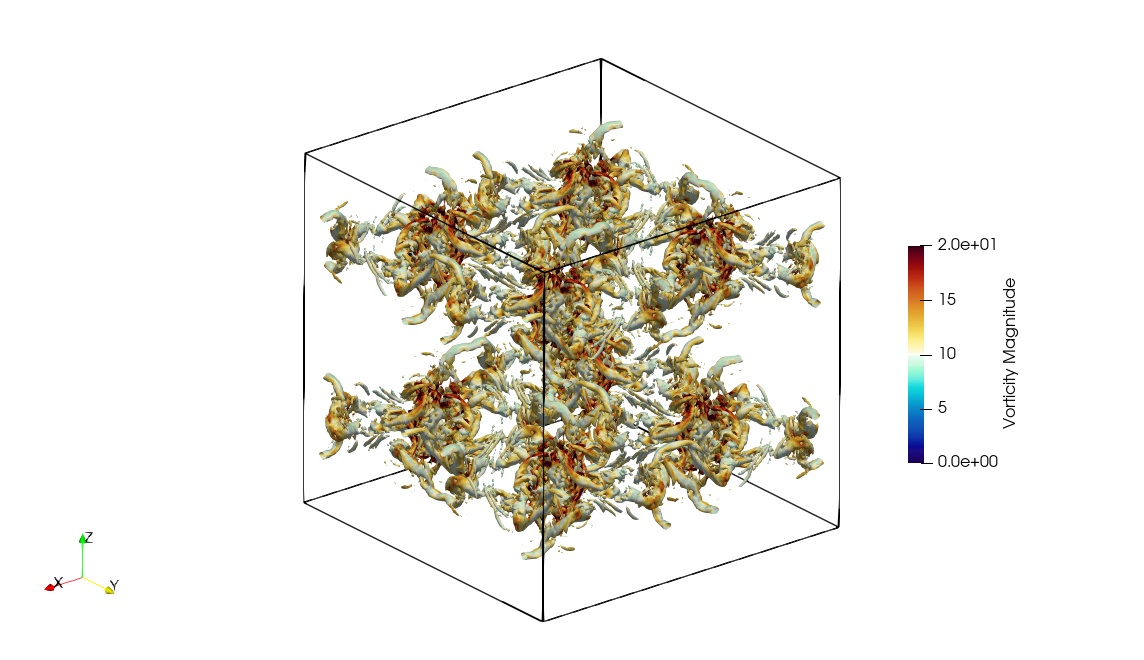} &
	\includegraphics[width=0.33\linewidth, trim=300 20 300 50, clip]{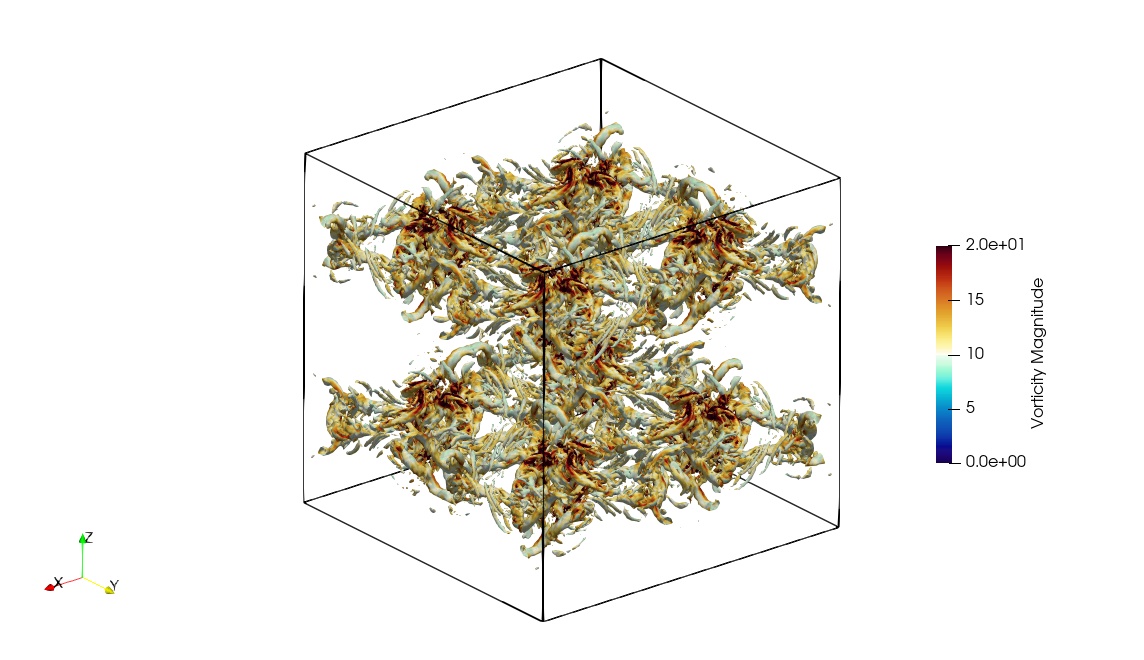} \\
	Case 9 & Case 10 \\[0.5em]
	\includegraphics[width=0.33\linewidth, trim=300 20 300 50, clip]{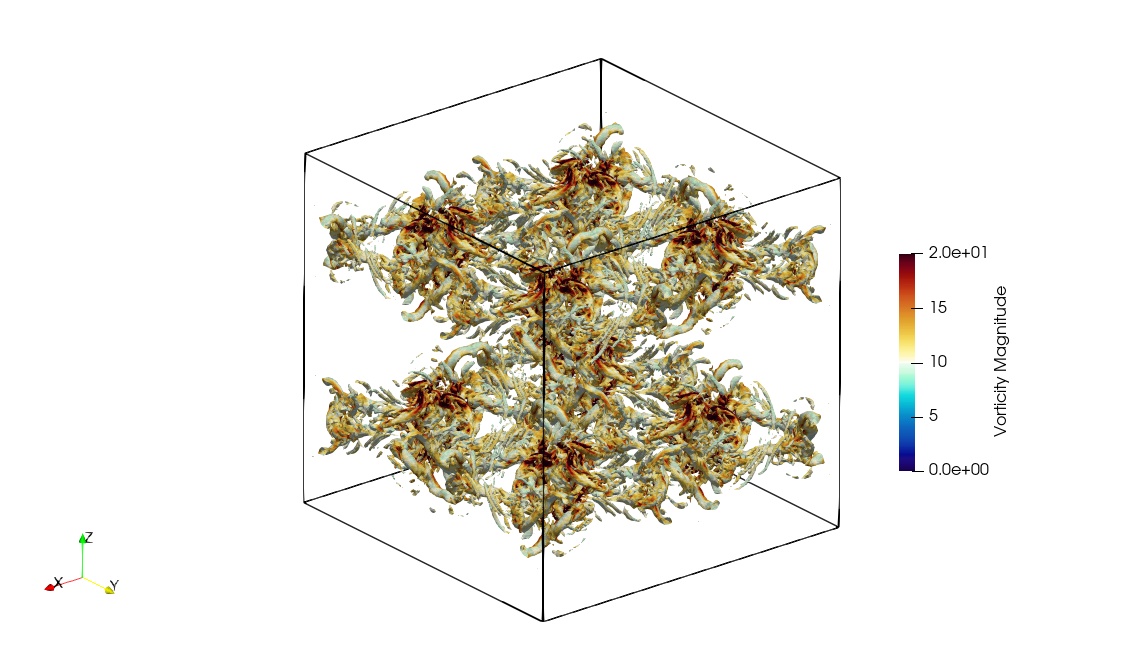} &
	\includegraphics[width=0.33\linewidth, trim=300 20 300 50, clip]{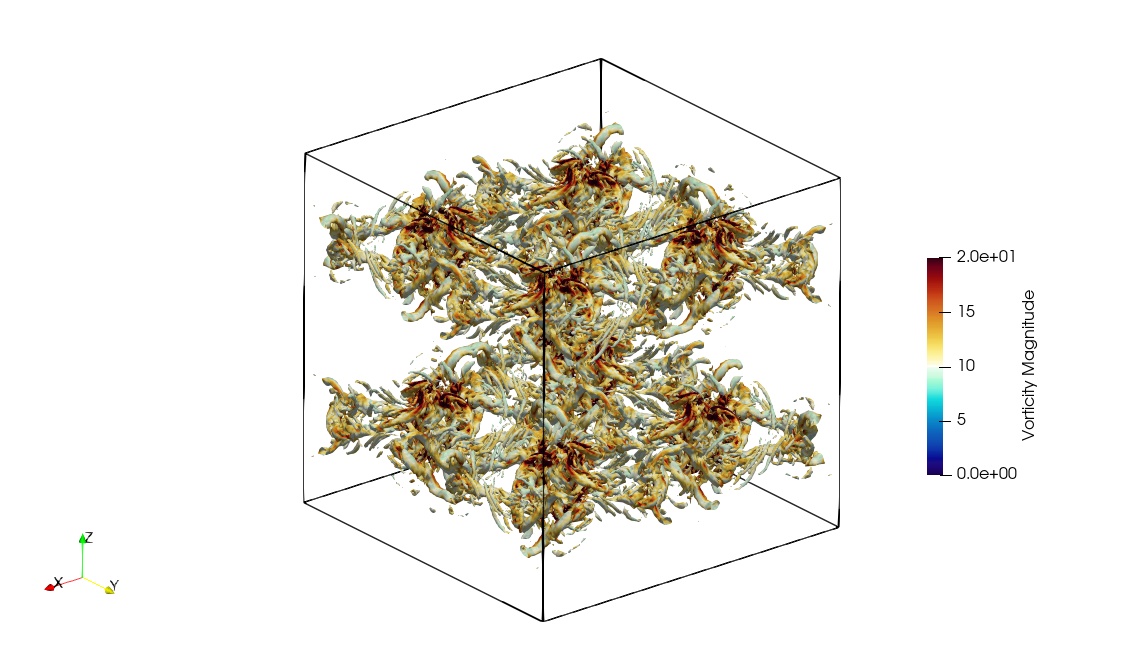} \\
	Case 15 & Case 16 \\
\end{tabular}
\vspace{0.5em}
\includegraphics[width=0.55\linewidth, trim=200 300 200 270, clip]{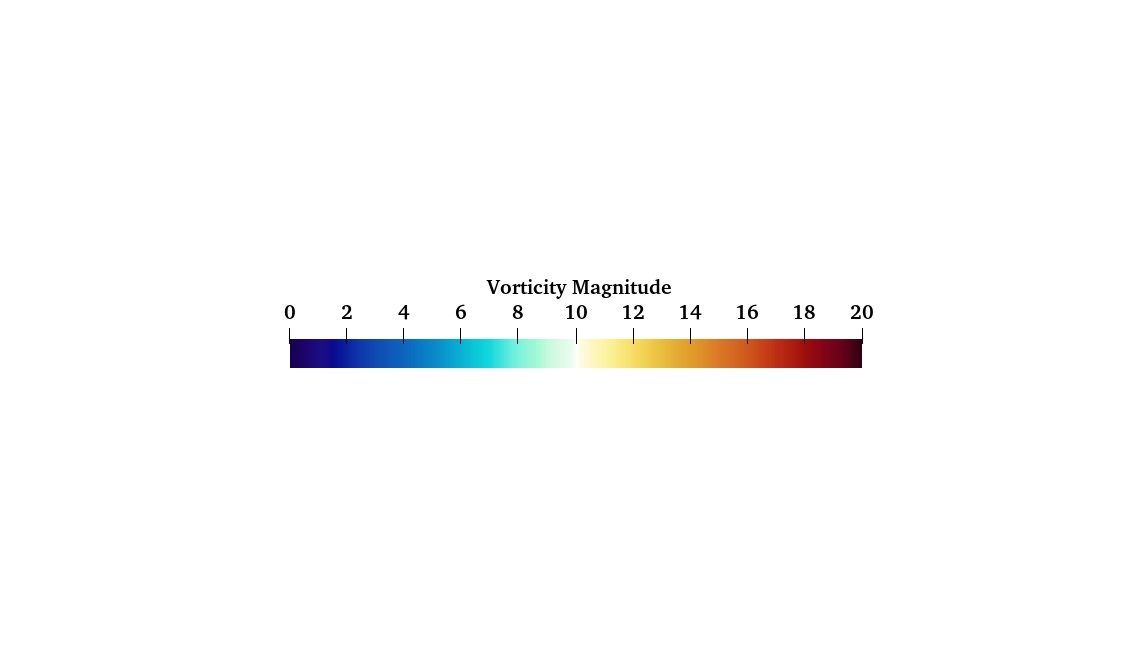} \\[-0.5em]
\caption{Isosurfaces of $Q=20$ colored by the vorticity magnitude at $t=9.0$.}
\label{fig:tgv_vortex_q_9}
\end{figure}

\begin{figure}[htbp]
	\centering
	\begin{tabular}{cc}
	\includegraphics[width=0.33\linewidth, trim=300 20 300 50, clip]{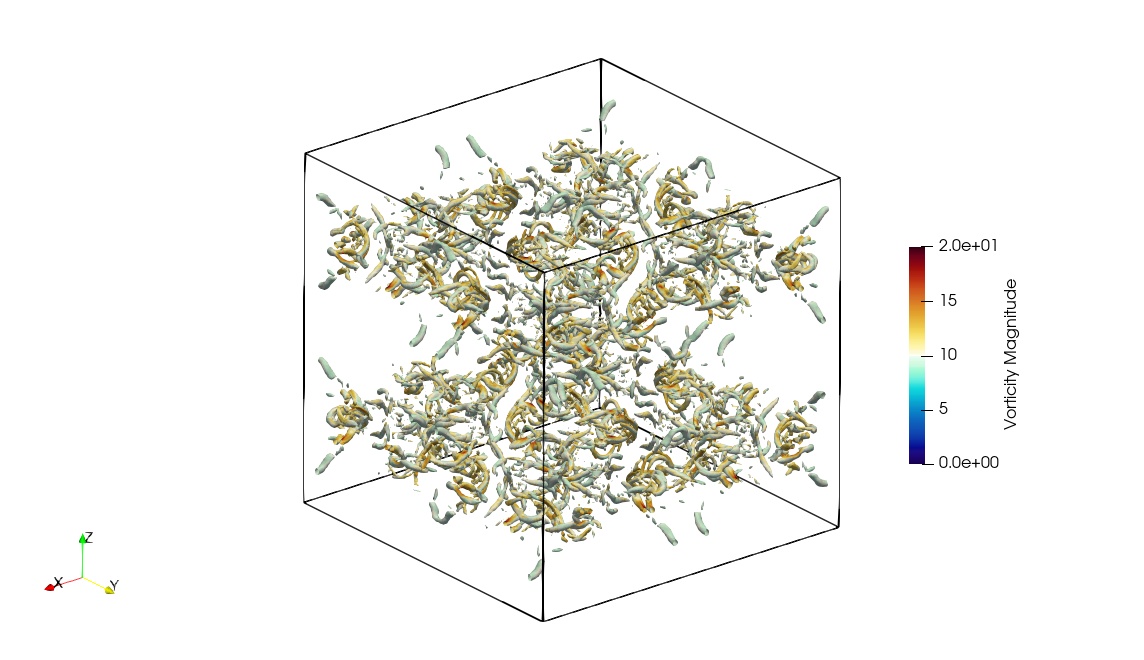} &
	\includegraphics[width=0.33\linewidth, trim=300 20 300 50, clip]{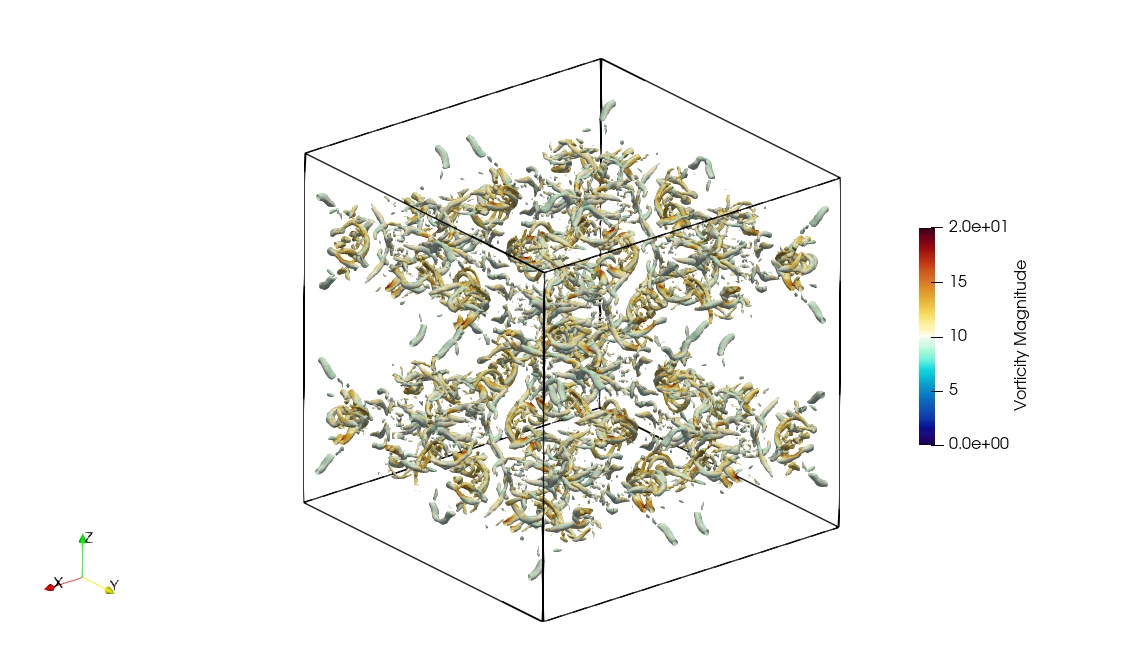} \\
	Case 5 & Case 8 \\
	\includegraphics[width=0.33\linewidth, trim=300 20 300 50, clip]{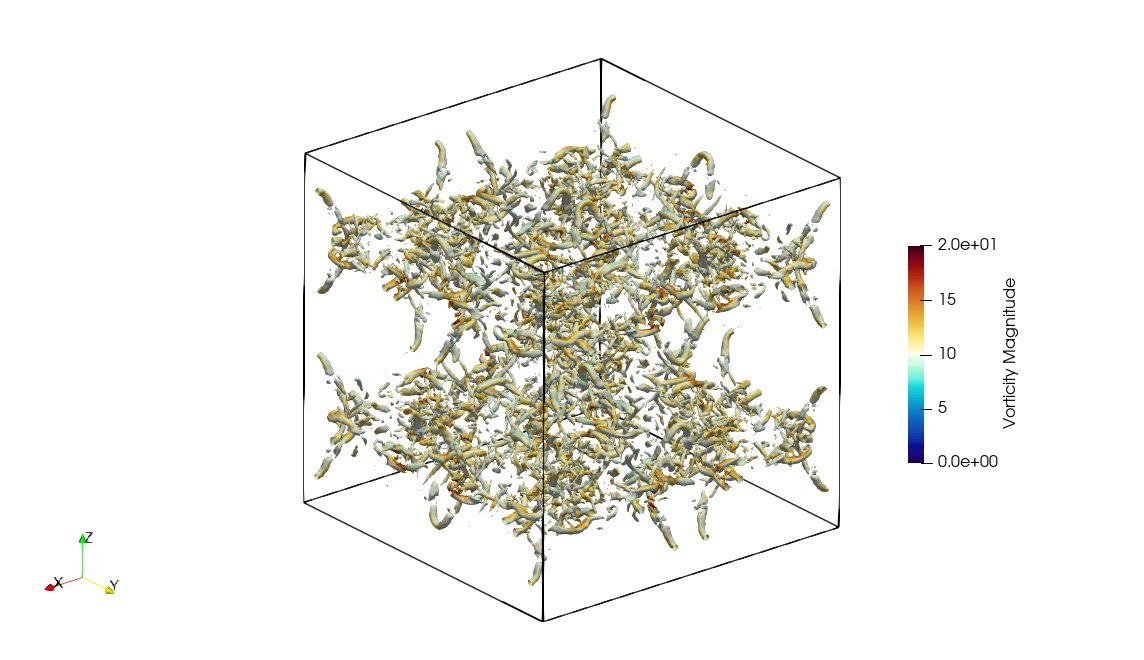} &
	\includegraphics[width=0.33\linewidth, trim=300 20 300 50, clip]{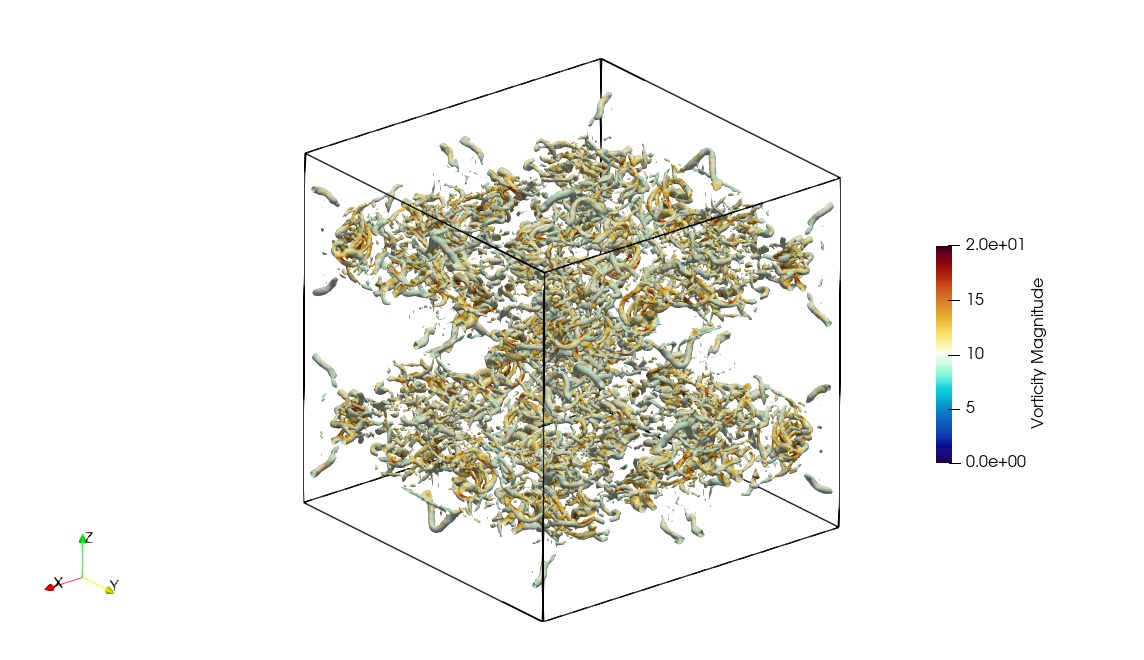} \\
	Case 9 & Case 10 \\[0.5em]
	\includegraphics[width=0.33\linewidth, trim=300 20 300 50, clip]{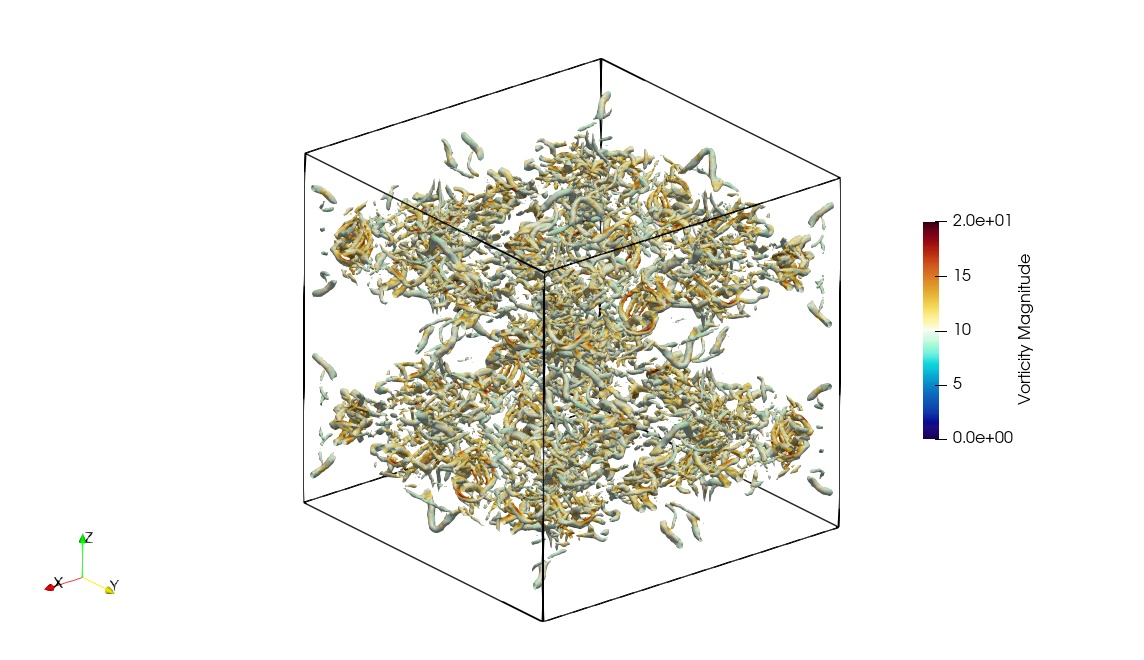} &
	\includegraphics[width=0.33\linewidth, trim=300 20 300 50, clip]{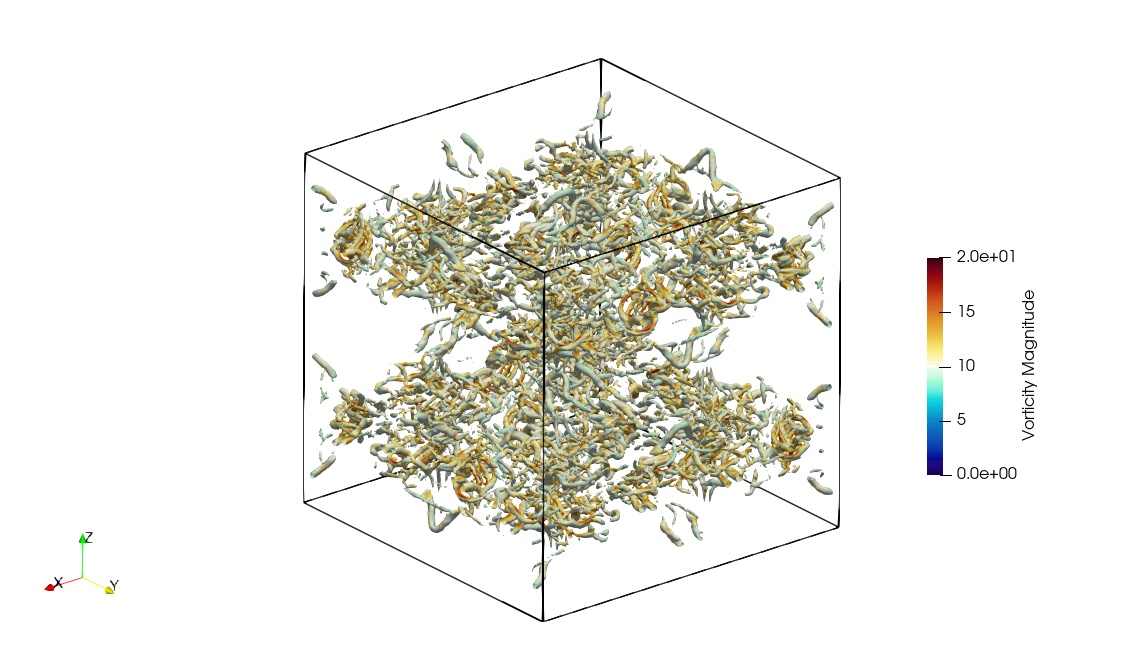} \\
	Case 15 & Case 16 \\
	\end{tabular}
\vspace{0.5em}
\includegraphics[width=0.55\linewidth, trim=200 300 200 270, clip]{./legend_q_128_3_2p_dt_0_01_new.jpeg} \\[-0.5em]
\caption{Isosurfaces of $Q=20$ colored by the vorticity magnitude at $t=12.0$.}
\label{fig:tgv_vortex_q_12}
\end{figure}

\paragraph{Kinetic energy spectra}
Particular attention is paid to the energy spectrum at the dissipation peak ($t = 9$). Figure \ref{fig:TGV_spectrum_different_mesh_compare} presents the energy spectrum obtained at different mesh resolutions, with DNS results from \cite{Wang2013} included for comparison. The results show that, with mesh refinement, the spectra of all schemes progressively converge toward the DNS results. However, noticeable differences between the two approaches are observed. On the coarsest mesh, the VMS-Gen-$\alpha$(0.5) scheme reproduces only the lowest portion of the DNS spectrum. In contrast, although the HERK$(3,3)$-VMS solution exhibits energy accumulation in the wavenumber range of ($20$,$40$), it still aligns closer with the DNS spectrum over a broader range of scales. With medium and fine meshes, the HERK$(3,3)$-VMS scheme captures a substantially larger portion of the DNS spectrum compared to the VMS-Gen-$\alpha$ scheme. In particular, on the finest mesh, the HERK$(3,3)$-VMS results remain in close agreement with DNS up to nearly 80\% of the resolved wavenumbers and exhibits a noticeably sharper spectral cutoff. By contrast, the VMS-Gen-$\alpha$ scheme exhibits excessive damping at high wavenumbers, resulting in significant underprediction of small-scale energy. Overall, the HERK$(3,3)$-VMS formulation demonstrates superior fidelity to the DNS spectrum across all resolvable scales.

Next, the influence of the time-step size on the TGV energy spectrum is examined. Figure \ref{fig:TGV_spectrum_dt_compare} presents the spectra obtained on the finest mesh for different time-step sizes. For the VMS-Gen-$\alpha$(0.5) scheme, instabilities at small time-step sizes are again observed. As the time-step size decreases from $1 \times 10^{-2}$ to $1 \times 10^{-3}$, deviations from the DNS spectrum become increasingly pronounced, particularly at high wavenumbers. In contrast, the HERK$(3,3)$+VMS scheme exhibits much lower sensitivity to the time-step size, with spectra obtained using the three time-step sizes nearly overlapping.

Finally, comparisons are made between HERK schemes of different orders of accuracy and generalized-$\alpha$ schemes with varying spectral radius $\rho_\infty$, using identical mesh resolution and time-step size. As shown in Figure \ref{fig:TGV_spectrum_RK_vs_GA}, the energy spectra obtained with different values of $\rho_\infty$ are nearly identical. This again confirms that, at least for the TGV problem, the results from the VMS-Gen-$\alpha$($\rho_{\infty}$) formulation are insensitive to this parameter, contrasting the numerical experiences in structural dynamics. For the HERK schemes, a marked improvement in the high wavenumber range is observed when the temporal accuracy is increased from first-order to second-order, yielding spectra that more closely follow the DNS data. When comparing the HERK and generalized-$\alpha$ schemes under identical spatial and temporal resolutions, both approaches reproduce the DNS spectra accurately at low wavenumbers. However, at high wavenumbers, even the first-order HERK scheme produces a noticeably sharper spectral cutoff and better alignment with the DNS results. The generalized-$\alpha$ scheme exhibits pronounced damping that limits its capability in resolving small-scale structures.

\paragraph{Instantaneous vortex visualization}
To further corroborate these findings, vortex structures obtained from the different solutions are compared in Figures \ref{fig:tgv_vortex_q_9} and \ref{fig:tgv_vortex_q_12}. The figures are based on the $Q$-criterion and depict isosurfaces of $Q = 20$ colored by vorticity magnitude at $t = 9$, the time of dissipation peak, and at $t = 12$, respectively. Compared with the HERK solutions, the generalized-$\alpha$ scheme tends to suppress the smallest turbulent structures, retaining only larger vortical features with relatively low vorticity. In contrast, higher-order HERK schemes ($p=2,3,4$) are capable of capturing finer vortex structures that are resolvable on the given mesh. These visual observations are consistent with the spectral analysis and the kinetic spectra behavior discussed earlier, confirming that higher-order HERK formulations preserve small-scale dynamics more effectively and yield a more accurate representation of the turbulent energy cascade.

\subsection{Open cavity flow}
The third numerical example considers flow over an open square cavity, a canonical configuration for studying supercritical Hopf bifurcation and global instability mechanisms in separated flows. At sufficiently large Reynolds numbers, the flow past the cavity produces a shear layer that rolls up into large-scale vortices. Their interaction with the downstream edge triggers pressure fluctuations, which propagate upstream through the recirculating region and destabilize the shear layer via the Kelvin-Helmholtz instability. The resulting feedback loop saturates into periodic limit cycle oscillations \cite{Barbagallo2009, Bengana2019, Meliga2017, Sipp2007}. The flow becomes linearly unstable via a Hopf bifurcation at Re $\approx$ 4140, and a second distinct instability arises via another Hopf bifurcation near Re $\approx$ 4350. If the Reynolds number range between 4455 and 4635, the two modes coexist and compete, and the final saturated flow state depends on the initial condition \cite{Meliga2019,Meliga2017}. For Re $>$ 4635, the flow ultimately reaches the higher-Reynolds-number periodic state. In this study, we select Re = 5000, which lies well within this regime. Previous studies have shown that excessive numerical dissipation in the time integration scheme leads to an incorrect limit-cycle selection, converging to an incorrect final periodic state \cite{Meliga2019}. Therefore, this example provides a stringent benchmark for evaluating the ability of numerical schemes to capture the correct oscillation frequency and amplitude.

\begin{figure}[htb]
\begin{center}
	\includegraphics[width=0.65\linewidth, trim=30 0 0 100, clip]{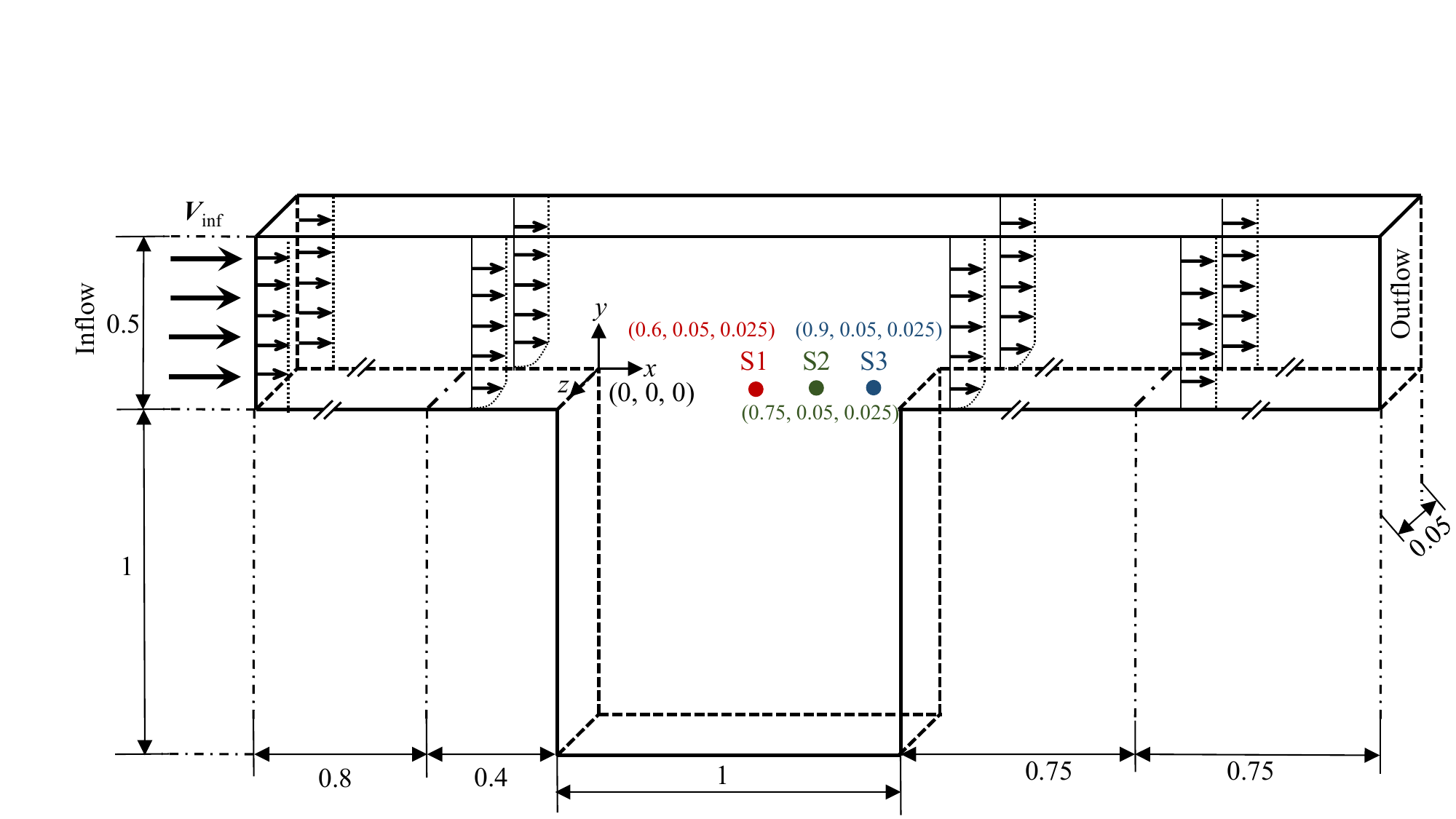}
\end{center}
\caption{Schematic of the open cavity flow.}
\label{fig:open_cavity_geo}
\end{figure}

The computational setup is illustrated in Figure \ref{fig:open_cavity_geo}. While the original open square cavity was designed as a two-dimensional benchmark, it is modeled here as a three-dimensional domain by extruding in the spanwise direction from $0$ to $0.05$. A Cartesian coordinate system is adopted, with the $x$-, $y$-, and $z$-axes aligned with the cavity wall, the wall-normal, and the extrusion directions, respectively. Periodic boundary conditions are applied on the planes at $z = 0$ and $z = 0.05$. The inflow boundary is located at $x=-1.2$, where a uniform velocity $\bm{V}_{\mathrm{inf}}=(1,0,0)$ is prescribed, and a traction-free condition is imposed at the outflow boundary at $x=2.5$. A free-slip condition is applied on the upper boundary at $y=0.5$ and on the cavity walls corresponding to $y=0$ and $x\in[-1.2,-0.4]\cup[1.75,2.5]$, whereas the remaining cavity walls are treated as no-slip surfaces.

The domain is discretized using $C^1$-continuous quadratic NURBS elements, comprising a total of 159200 elements with a single layer in the $z$-direction. Local refinement is applied near the junctions between the no-slip and free-slip boundaries, as well as within the adjacent boundary-layer regions, to ensure adequate resolution of shear layer development and wall-bounded gradients. The simulations are performed using the HERK$(2,2)$-VMS and HERK$(3,3)$-VMS schemes, with a time-step size of $\Delta t_n = 5\times10^{-4}$. The initial condition is constructed from a long-time simulation, from which the instantaneous velocity field is time-averaged to construct the steady base flow. A random perturbation, corresponding to 3\% Gaussian noise with zero mean, is superimposed on this base flow to initiate unsteady dynamics. All simulations are initialized from this identical perturbed state. To evaluate the accuracy of the schemes, the streamwise velocity is monitored at three locations within the shear layer, namely S1 = (0.60, 0.05, 0.025), S2 = (0.75, 0.05, 0.025), and S3 = (0.90, 0.05, 0.025), as indicated in Figure \ref{fig:open_cavity_geo}. 

\begin{figure}
	\begin{center}
		\begin{tabular}{cc}
			\multicolumn{2}{c}{\includegraphics[width=0.66\linewidth, trim=0 0 0 0]{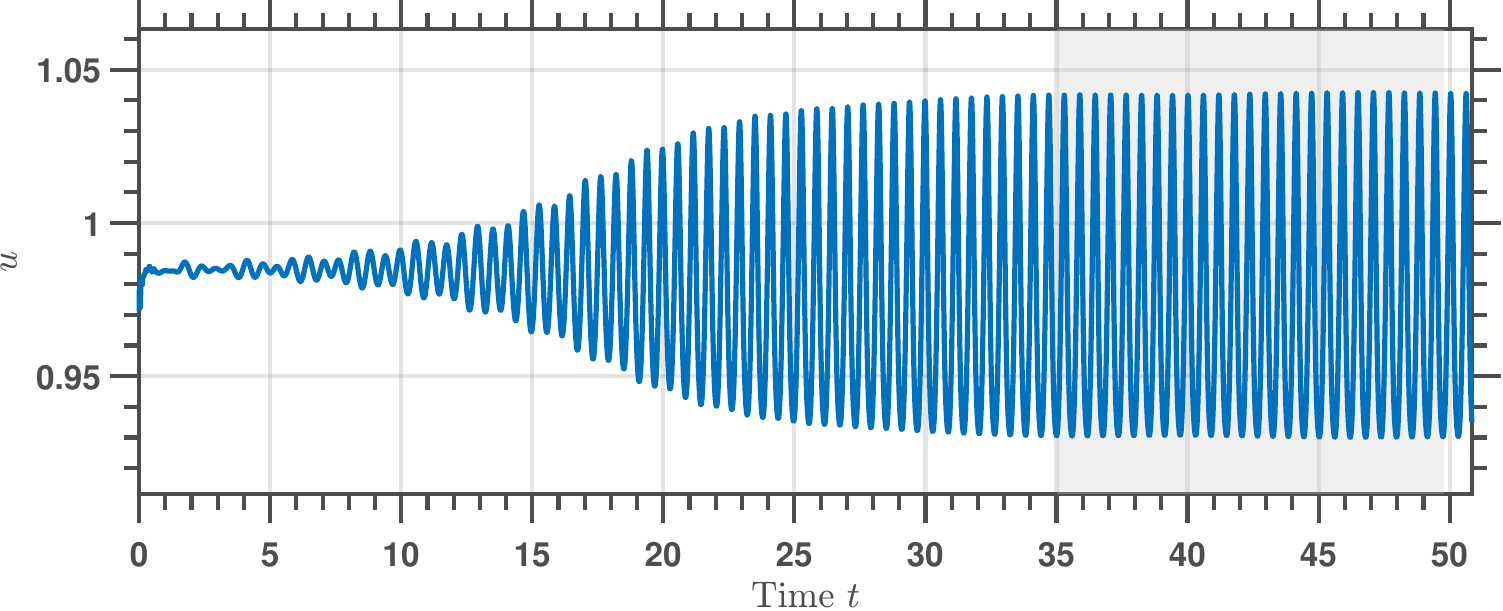}} \\
			\multicolumn{2}{c}{(a)} \\
			\includegraphics[width=0.33\linewidth, trim=0 0 0 0]{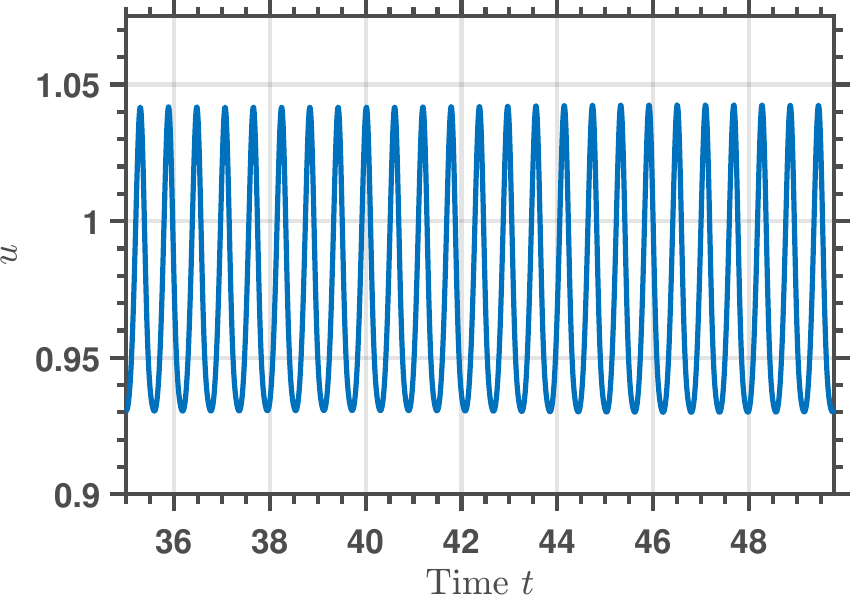} &
			\includegraphics[width=0.32\linewidth, trim=0 0 0 0]{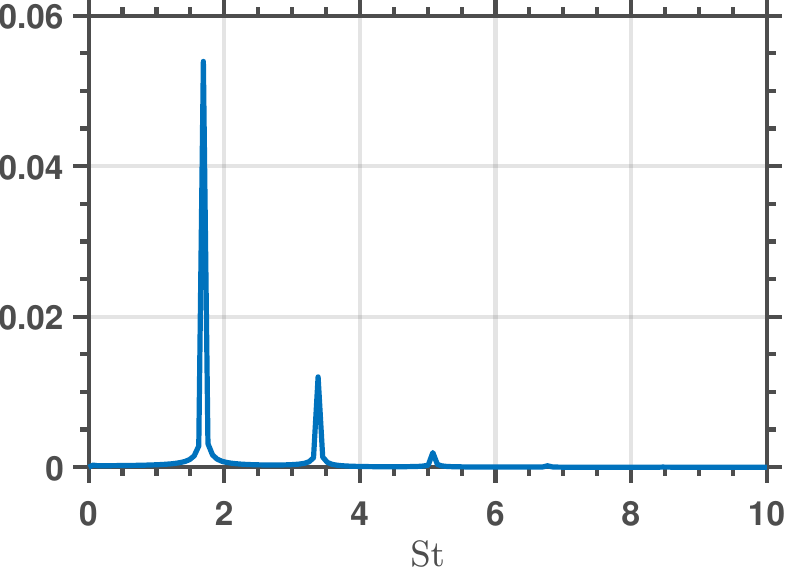}\\
			(b) & (c)
		\end{tabular}
	\end{center}
	\caption{Time evolution of the streamwise velocity at the S2 monitor for HERK$(2,2)$-VMS: (a) full time history; (b) close-up view of the shaded region in (a); and (c) the frequency spectrum corresponding to (b).}
	\label{fig:velo_S2_herk2}
\end{figure}

\begin{figure}
\begin{center}
	\begin{tabular}{cc}
		\multicolumn{2}{c}{\includegraphics[width=0.66\linewidth, trim=0 0 0 0]{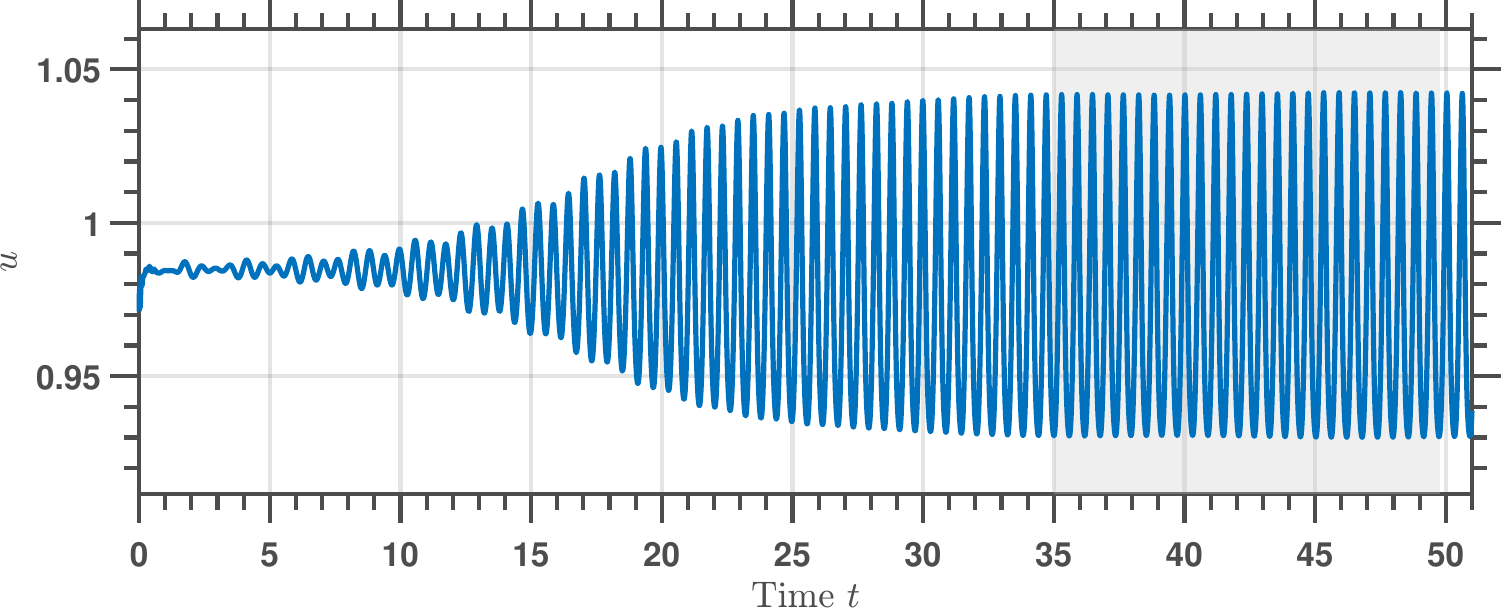}} \\
		\multicolumn{2}{c}{(a)} \\
		\includegraphics[width=0.33\linewidth, trim=0 0 0 0]{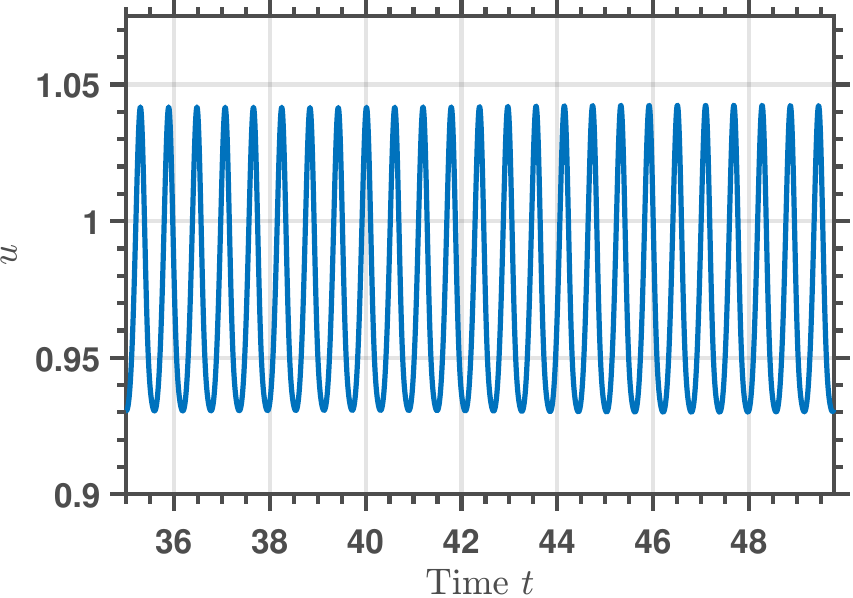} &
		\includegraphics[width=0.32\linewidth, trim=0 0 0 0]{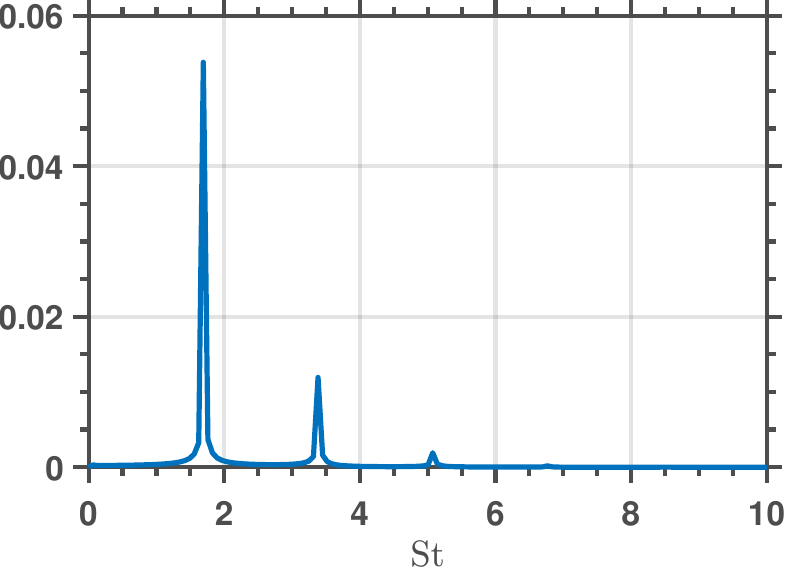}\\
		(b) & (c)
	\end{tabular}
\end{center}
\caption{Time evolution of the streamwise velocity at the S2 monitor for HERK$(3,3)$-VMS: (a) full time history; (b) close-up view of the shaded region in (a); and (c) the frequency spectrum corresponding to (b).}
\label{fig:velo_S2_herk3}
\end{figure}
	
\begin{figure}[htbp]
	\centering
	\begin{tabular}{cc}
 		\includegraphics[width=0.476\linewidth, trim=50 100 50 100, clip]{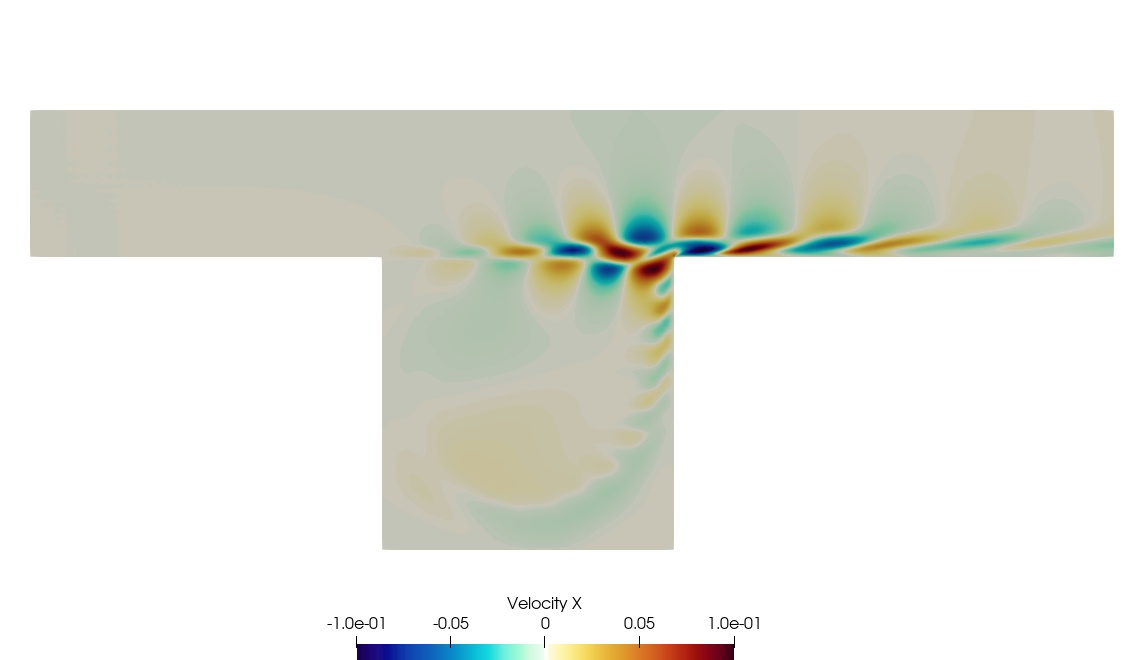} &
		\includegraphics[width=0.476\linewidth, trim=50 100 50 100, clip]{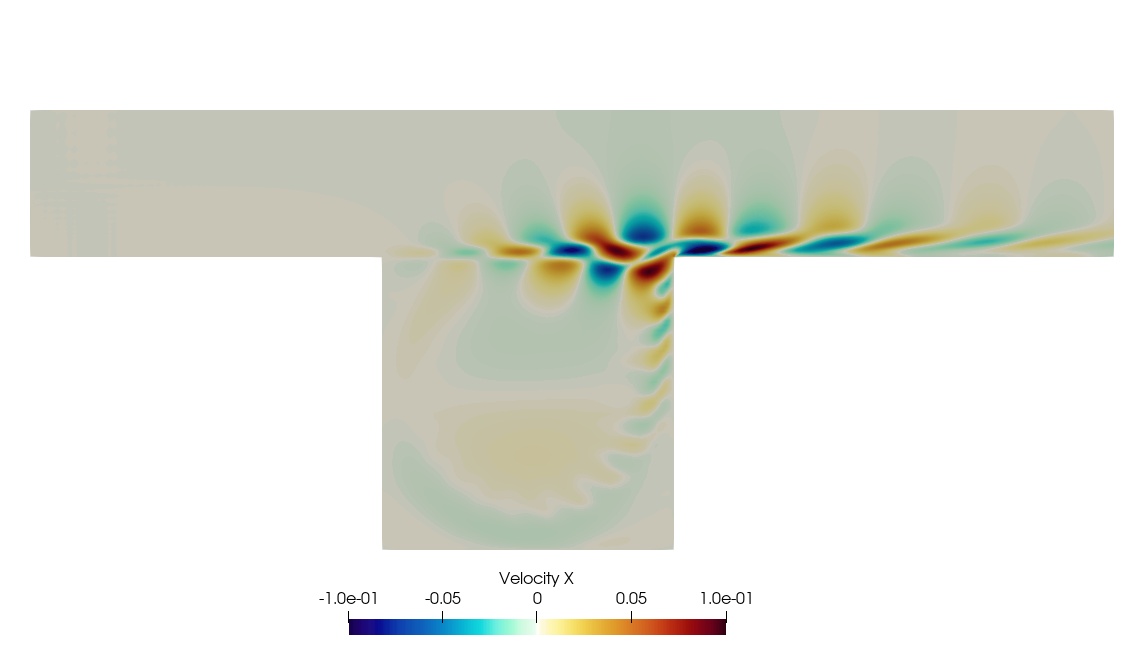} \\
		HERK$(2,2)$-VMS & HERK$(3,3)$-VMS \\
	\end{tabular}
	\vspace{0.5em}
	\includegraphics[width=0.55\linewidth, trim=200 250 200 300, clip]{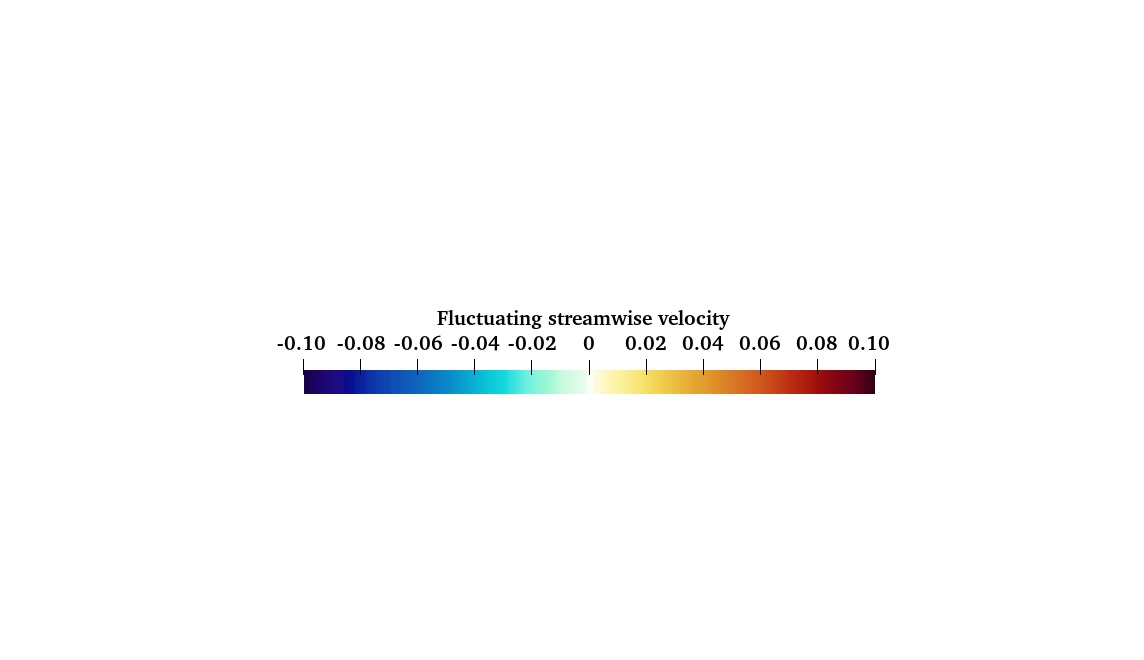} \\[-0.5em]
	\caption{Contours of the fluctuating streamwise velocity.}
	\label{fig:fluc_x-velo-peak}
\end{figure}

\begin{table}[htbp]
	\centering
	\tabcolsep=0.13cm
	\renewcommand{\arraystretch}{1.2}
	\caption{Statistics of the limit-cycle oscillation at Re = 5000.}
	\label{tab:limit_cycle_oscillation}
	\begin{tabular}{@{\extracolsep{4pt}}P{4.0cm} P{1.0cm} P{1.0cm} P{1.0cm} P{1.0cm} P{1.0cm} P{1.0cm} P{1.0cm} P{0.0cm} @{}}
 		\toprule
		\multirow{2}{*}{Case} & 
		\multicolumn{3}{c}{$u_a (\times 10^{-1})$} & 
		\multicolumn{3}{c}{$u_f (\times 10^{-2})$} & 
		\multirow{2}{*}{St} \\
		\cline{2-4} \cline{5-7}
		& S1 & S2 & S3 & S1 & S2 & S3 &\\
		\midrule
		HERK$(2,2)$-VMS & 9.970 & 9.740 & 9.341 & 2.272 & 3.932 & 4.345 & 1.695 \\
		HERK$(3,3)$-VMS & 9.971 & 9.739 & 9.342 & 2.272 & 3.930& 4.344& 1.692\\
		DNS \cite{Meliga2019} & 9.965 & 9.730 & 9.315 & 2.380 & 4.065 & 4.443 & 1.692 \\
		\bottomrule
	\end{tabular}
\end{table}

Figures \ref{fig:velo_S2_herk2} and \ref{fig:velo_S2_herk3} depict the time evolution of the streamwise velocity $u$ monitored at point S2, computed using the HERK$(2,2)$-VMS and HERK$(3,3)$-VMS schemes, respectively. Both schemes accurately capture the transient evolution at S2. After approximately 30 time units, the flow exhibits well-defined oscillations, and a limit-cycle behavior is established within about 50 time units. Table \ref{tab:limit_cycle_oscillation} summarizes the limit-cycle statistics obtained with the two schemes. The quantities reported are based on the streamwise velocity, where $u_a$ denotes its temporal mean and $u_f$ denotes the root-mean-square of its fluctuation. Time series at the three monitoring points are collected over 25 oscillation periods after the flow reaches the limit-cycle regime (around 35 time units). The fundamental Strouhal number St is also listed and found to be identical across all three monitoring points. For reference, Table \ref{tab:limit_cycle_oscillation} also includes the DNS data of \cite{Meliga2019}, obtained using P$_2$-P$_1$ Taylor-Hood elements and the Crank-Nicolson scheme. Overall, both HERK$(2,2)$-VMS and HERK$(3,3)$-VMS exhibit good agreement with the reference data, and they successfully capture the correct dynamics of the limit cycle. In particular, the St number predicted by the HERK$(3,3)$-VMS scheme matches the DNS value with remarkable accuracy. It demonstrates that the proposed HERK$(s,p)$-VMS schemes possess sufficient accuracy to ensure the correct nonlinear mode selection.

The fidelity of the HERK$(s,p)$-VMS schemes is further illustrated in Figure \ref{fig:fluc_x-velo-peak}, which shows the spatial distribution of the fluctuating streamwise velocity on the plane $z=0.025$ at a fixed phase of the limit-cycle, corresponding to the peak velocity recorded at the monitoring point S2. The fluctuating field is obtained by subtracting the time-averaged velocity field. As observed in Figure \ref{fig:fluc_x-velo-peak}, the roll-up of the shear layer generates counter-rotating vortical structures that propagate and intensify along the layer. Both schemes accurately capture the vortical structures developing inside the cavity and within the downstream boundary layer, while the HERK$(3,3)$-VMS scheme provides a slightly better resolution of the vortices near the downstream cavity edge.

\section{Conclusion}
In this work, we develop a mathematically consistent framework for residual-based VMS turbulence modeling, aiming to enable higher-order temporal discretizations. Guided by the Rothe method, the HERK scheme is employed for temporal discretization, which is particularly well suited for index-2 DAE systems. It preserves higher-order temporal accuracy attainable for the corresponding ODEs and avoids order reduction commonly observed in multi-stage methods. When applied to the NS equations, the temporal discretization yields, at each RK stage, a steady subproblem governed by a Darcy-like differential operator, for which a corresponding residual-based VMS formulation is constructed. Owing to the simpler form of the differential operator, the subgrid modeling becomes more straightforward and does not require the linearization procedures and extra assumptions invoked in the conventional VMS analysis. By applying the bubble enrichment technique, the parameter $\tau$ in the subgrid model is analytically derived and is found to be independent of the spatial discretization parameter. Interestingly, the resulting fully discrete system is represented by a symmetric definite matrix with a block structure. Moreover, the submatrices depend linearly on the RK coefficients and the time-step size, which enables efficient assembly of the matrix and the preconditioner through reuse of the submatrices. These features facilitate the development of an efficient linear solver strategy based on the CG algorithm with block preconditioning. A Fourier analysis is performed to examine the effect of switching the order of space-time discretization based on the RK schemes. The analysis demonstrates that the RK($s,p$)-VMS schemes, arising from the horizontal method of lines, attain the expected temporal accuracy and maintain the optimal spatial accuracy. They also exhibit better stability and smaller dissipation and dispersion errors at high wavenumbers, in comparison with the VMS-RK($s,p$) schemes obtained via the vertical method of lines. Additionally, the spectral accuracy of the VMS-RK($s,p$) schemes depends on the asymptotic behavior of the stabilization parameter and achieves comparable accuracy only under specific conditions.

The proposed framework is first assessed using the method of manufactured solutions, which confirms that the optimal temporal and spatial accuracies are achieved by the proposed schemes. In the second test, the TGV problem is employed to examine the proposed model as a tool for LES. The numerical results demonstrate that the proposed schemes outperform the conventional residual-based VMS schemes in capturing the energy spectrum and reproducing the dissipation behavior consistent with the reference data. In contrast, the conventional VMS schemes produce a spurious double peak in the temporal evolution of the dissipation, indicating that the dispersion error is excessive. The proposed schemes are also more robust with respect to time-step refinement and preserve small-scale turbulent features that are suppressed in the conventional approach. Finally, the open-cavity flow is investigated to evaluate the capability of the method in predicting the supercritical Hopf bifurcation and associated flow instabilities. The results demonstrate that the proposed schemes possess sufficient accuracy to capture the correct limit-cycle dynamics. These findings together demonstrate the effectiveness and fidelity of the proposed approach for simulating highly transient, globally unstable, and turbulent flows.

In the present scheme, each RK stage requires the solution of a fully coupled velocity-pressure linear system. Recent developments in low-cost RK strategies \cite{Karam2021,Aithal2023,Karam2023} are appealing in that one may avoid solving the Poisson-like equation for the pressure at the intermediate stages without losing accuracy. When combined with mass-lumping techniques for higher-order spline elements, such approaches may fully exploit the computational advantages of explicit schemes, thereby making large-scale unsteady flow simulations more efficient.

\section*{Acknowledgements}
This work is supported by the National Natural Science Foundation of China [Grant Numbers 12172160, 12472201], Shenzhen Science and Technology Program [Grant Number JCYJ20220818100600002], Southern University of Science and Technology [Grant Number Y01326127], and the Department of Science and Technology of Guangdong Province [2021QN020642]. Computational resources are provided by the Center for Computational Science and Engineering at the Southern University of Science and Technology.

\appendix
\section{Butcher tableaux of explicit RK schemes used in this work}
\label{sec:butcher-tableaux-explicit-RK-in-examples}
For completeness, we list the Butcher tableaux of the explicit RK schemes adopted in this work, including
\begin{itemize}
\item[-] the one-stage, first-order explicit Euler method,
\item[-] the two-stage, second-order Heun method,
\item[-] the three-stage, third-order strong-stability-preserving method,
\item[-] the classical four-stage, fourth-order method.
\end{itemize}
Their corresponding Butcher tableaux are given below.
\begin{table}[h!]
\centering
\begin{tabular}{c|c}
0   & 0   \\
\hline
& 1
\end{tabular}
\hspace{10mm}
\begin{tabular}{c|cc}
0   & 0    & 0    \\
1   & 1    & 0    \\
\hline
& 1/2     & 1/2
\end{tabular}
\caption{Butcher tableaux of the forward Euler (left) and Heun (right) schemes.}
\end{table}

\begin{table}[h!]
\centering
\begin{tabular}{c|ccc}
0   & 0    & 0   & 0  \\
1   & 1    & 0   & 0  \\
1/2 & 1/4  & 1/4   & 0 \\
\hline
& 1/6   & 1/6  &2/3
\end{tabular}
\hspace{10mm}
\begin{tabular}{c|cccc}
0   & 0    & 0   & 0  & 0 \\
1/2 & 1/2  & 0   & 0  & 0 \\
1/2 & 0    & 1/2 & 0  & 0 \\
1   & 0    & 0   & 1  & 0\\
\hline
& 1/6   & 1/3   &1/3  &1/6
\end{tabular}
\caption{Butcher tableaux of the strong stability preserving (left) and classical four-stage, fourth-order (right) schemes.}
\end{table}

\bibliographystyle{unsrt}
\bibliography{herk-vms}
	
\end{document}